\documentclass[a4paper,11pt]{article}
\usepackage{jheppub} 
 \usepackage{graphics}
 \usepackage{float}
 \usepackage{tikz-cd} 
\usepackage{url}
\usepackage{float}
\usepackage{standalone}

 \usepackage{caption, subcaption}
 
 \usepackage{tikz}

\def\a{\alpha}
\def\b{\beta}
\def\c{\varepsilon}
\def\d{\delta}
\def\e{\epsilon}
\def\f{\phi}
\def\g{\gamma}
\def\h{\theta}
\def\k{\kappa}
\def\l{\lambda}
\def\m{\mu}
\def\n{\nu}
\def\p{\psi}
\def\q{\partial}
\def\r{\rho}
\def\s{\sigma}
\def\t{\tau}
\def\u{\upsilon}
\def\v{\varphi}
\def\w{\omega}
\def\x{\xi}
\def\y{\eta}
\def\z{\zeta}
\def\D{\Delta}
\def\G{\Gamma}
\def\H{\Theta}
\def\L{\zeta}
\def\F{\Phi}
\def\P{\Psi}
\def\S{\Sigma}

\def\aa{{\dot \a}}
\def\bb{{\dot \b}}
\def\ss{{\bar \s}}
\def\hh{{\bar \h}}
\def\CA{{\cal A}}
\def\CB{{\cal B}}
\def\CC{{\cal C}}
\def\CD{{\cal D}}
\def\CE{{\cal E}}
\def\CG{{\cal G}}
\def\CH{{\cal H}}
\def\CI{{\cal I}}
\def\CK{{\cal K}}
\def\CL{{\cal L}}
\def\CR{{\cal R}}
\def\CM{{\cal M}}
\def\CN{{\cal N}}
\def\CO{{\cal O}}
\def\CP{{\cal P}}
\def\CQ{{\cal Q}}
\def\CW{{\cal W}}

\def\capcup{\phantom{.}^{\frown}_{\smile}\phantom{.}}

\DeclareMathOperator{\Tr}{Tr}
\newcommand{\Slash}[1]{{\ooalign{\hfil/\hfil\crcr$#1$}}}

\def\o{\over}
\newcommand{\gsim}{ \mathop{}_{\textstyle \sim}^{\textstyle >} }
\newcommand{\lsim}{ \mathop{}_{\textstyle \sim}^{\textstyle <} }
\newcommand{\vev}[1]{ \left\langle {#1} \right\rangle }
\newcommand{\bra}[1]{ \langle {#1} | }
\newcommand{\ket}[1]{ | {#1} \rangle }
\newcommand{\EV}{ {\rm eV} }
\newcommand{\KEV}{ {\rm keV} }
\newcommand{\MEV}{ {\rm MeV} }
\newcommand{\GEV}{ {\rm GeV} }
\newcommand{\TEV}{ {\rm TeV} }
\def\diag{\mathop{\rm diag}\nolimits}
\def\Spin{\mathop{\rm Spin}}
\def\SO{\mathop{\rm SO}}
\def\O{\mathop{\rm O}}
\def\SU{\mathop{\rm SU}}
\def\U{\mathrm{U}}
\def\Sp{\mathop{\rm Sp}}
\def\SL{\mathop{\rm SL}}
\def\tr{\mathop{\rm tr}}
\def\rank{\mathop{\rm rank}}

\def\beq#1\eeq{\begin{align}#1\end{align}}

\title{On rank two theories with eight supercharges part I: local singularities}

\author[a,b]{Dan Xie}

\affiliation[a]{Yau Mathematics Science Center, Tsinghua University, Beijing, 100084, China}
\affiliation[b]{Department of Mathematics, Tsinghua University, Beijing, 100084, China}

\abstract{A complete study of local singularities of rank two $\mathcal{N}=2$ Coulomb branch geometry is given.  Low energy theory associated with 
the local singularity is identified: it can be superconformal field theory (SCFT), or IR free gauge theory, or the combination of them. 
Various invariants for local singularity are also listed which are essential for the study of global Coulomb branch. As a first application, global Coulomb branch with only simplest local singularities in the bulk are 
given for 4d theories (including SCFTs and  asymptotical free theories), 5d KK theories, and 6d KK theories; those examples appear to cover all the findings in the literature and suggest there are more possibilities. More general global  
Coulomb branch geometry would be discussed in the sequel of  this paper.}

\begin{document} 
\maketitle
\flushbottom

\section{Introduction}
There are lots of interests in studying supersymmetric theory with eight supercharges in various dimensions, such as 3d $\mathcal{N}=4$ theory \cite{Intriligator:1996ex}, 4d $\mathcal{N}=2$ theory \cite{Seiberg:1994rs,Seiberg:1994aj},
5d $\mathcal{N}=1$ theory \cite{Seiberg:1996bd}, and 6d $(1,0)$ theory \cite{Seiberg:1996qx}. These theories have many physical applications, i.e. the studies of them help us understand strongly coupled quantum field theory, 
confinement, strong-weak duality, etc; They also have many applications in modern mathematics, such as geometric representation theory \cite{Braverman:2016pwk} and invariants for four manifolds \cite{Witten:1994cg},etc.

Most of those theories are strongly coupled and so conventional field theory tools are of little use, but the powerful geometric methods make it possible to study various deep properties of those theories. 
In particular, one can construct a large class of 3d $\mathcal{N}=4$ theory using type IIB branes \cite{Hanany:1996ie}, 4d $\mathcal{N}=2$ theory using M5 branes \cite{Witten:1997sc,Gaiotto:2009we,Xie:2012hs} and 3-fold canonical singularities \cite{Shapere:1999xr,Xie:2015rpa}, 5d $\mathcal{N}=1$ theory  using 5 brane webs \cite{Aharony:1997bh} 
and 3-fold canonical singularities \cite{Intriligator:1997pq,Xie:2017pfl}, 6d $(1,0)$ theory using F theory on local elliptic  fibered 3-folds \cite{Heckman:2013pva}. 

 Although there are now a quite large space of theories, it is  desirable to have 
a classification of theories with 8 supercharges \footnote{Theories with 16 supercharges have a  fairly simple classification and are closed related to the classification of simple  Lie algebra \cite{Seiberg:1997ax}.}. While the geometric constructions listed above are  quite  powerful, it is never clear whether
they would give a classification (in fact most of times they would miss some theories).

 Argyres and his collaborators \cite{Argyres:2015ffa,Argyres:2015gha,Argyres:2016xmc,Argyres:2016xua,Caorsi:2018ahl} have took a different approach in classifying 4d $\mathcal{N}=2$ SCFTs by classifying the rank one 
 Coulomb branch solution found in \cite{Seiberg:1994rs,Seiberg:1994aj}. The classification scheme is a lot more complicated than the geometric approach, but  it is more
 complete in the sense that they would generate theories which are not found using geometric tools. Their approach 
 is  implemented by the author in classifying rank one 5d $\mathcal{N}=1$ and 6d $(1,0)$ SCFTs \cite{Xie:2022lcm}.  
 The link of  Coulomb branch geometry with the rational elliptic surface \cite{schutt2019elliptic} plays a crucial role in the classification of rank one theory.
 
The purpose of this paper and the follow-ups \cite{Xie:ranktwob,Xie:ranktwoc} is to give a complete classification for \textbf{rank two} theories with eight 
supercharges in dimension $D\geq 4$ (see \cite{Argyres:2018zay,Bourget:2021csg,Kaidi:2021tgr,Martone:2021ixp,Argyres:2022lah, Argyres:2022puv,Argyres:2022fwy} for the attempt in classifying 4d rank two SCFTs). 
Now the general structure of rank two Coulomb branch and its underlying mathematical  structure (genus two fibered rational surface) 
have not been thoroughly studied, and our work will fill the gap.

\begin{figure}
\begin{center}

\tikzset{every picture/.style={line width=0.75pt}} 

\begin{tikzpicture}[x=0.55pt,y=0.55pt,yscale=-1,xscale=1]

\draw    (298,180) -- (309,191.72) ;
\draw    (298,192) -- (307,180.72) ;

\draw    (375,300) -- (386,311.72) ;
\draw    (375,312) -- (384,300.72) ;

\draw    (173,321) -- (184,332.72) ;
\draw    (173,333) -- (182,321.72) ;

\draw    (197.5,147.72) .. controls (177.5,170.72) and (243,281.72) .. (235,329.72) ;
\draw  [color={rgb, 255:red, 208; green, 2; blue, 27 }  ,draw opacity=1 ] (345,306.72) .. controls (345,295.67) and (361.57,286.72) .. (382,286.72) .. controls (402.43,286.72) and (419,295.67) .. (419,306.72) .. controls (419,317.76) and (402.43,326.72) .. (382,326.72) .. controls (361.57,326.72) and (345,317.76) .. (345,306.72) -- cycle ;
\draw    (357.5,281.72) .. controls (441.5,20.72) and (333.5,256.72) .. (382,306.72) ;
\draw  [color={rgb, 255:red, 208; green, 2; blue, 27 }  ,draw opacity=1 ] (266,184.75) .. controls (266,173.7) and (282.57,164.75) .. (303,164.75) .. controls (323.43,164.75) and (340,173.7) .. (340,184.75) .. controls (340,195.8) and (323.43,204.75) .. (303,204.75) .. controls (282.57,204.75) and (266,195.8) .. (266,184.75) -- cycle ;
\draw    (283,55.75) .. controls (223,47.75) and (298,138.75) .. (303,184.75) ;
\draw    (210,401) -- (221,412.72) ;
\draw    (210,413) -- (219,401.72) ;

\draw    (374,386) -- (385,397.72) ;
\draw    (374,398) -- (383,386.72) ;

\draw   (128.75,359) .. controls (128.75,262.76) and (206.76,184.75) .. (303,184.75) .. controls (399.24,184.75) and (477.25,262.76) .. (477.25,359) .. controls (477.25,455.24) and (399.24,533.25) .. (303,533.25) .. controls (206.76,533.25) and (128.75,455.24) .. (128.75,359) -- cycle ;

\draw (275,171.4) node [anchor=north west][inner sep=0.75pt]    {$\infty $};
\draw (114,115.4) node [anchor=north west][inner sep=0.75pt]    {$H\left( F^{\bullet } ,W^{\bullet },Q(\cdot,\cdot)\right)$};
\draw (361,100.4) node [anchor=north west][inner sep=0.75pt]    {$H_{s\ \ \ \ }\left( F^{\bullet } ,W^{\bullet }\right)$};
\draw (360,75.4) node [anchor=north west][inner sep=0.75pt]    {$H_{lim}\left( F^{\bullet } ,W^{\bullet }(N)\right)$};
\draw (360,125.4) node [anchor=north west][inner sep=0.75pt]    {$H_{van}\left( F^{\bullet } ,W^{\bullet }(N)\right)$};
\draw (427,305.4) node [anchor=north west][inner sep=0.75pt]    {$T$};
\draw (248,22.4) node [anchor=north west][inner sep=0.75pt]    {$H_{\infty }\left( F^{\bullet } ,W^{\bullet }(N_\infty)\right)$};
\draw (450,396.4) node [anchor=north west][inner sep=0.75pt]    {$u$};

\end{tikzpicture}
\end{center}
\caption{The structure of $\mathcal{N}=2$ Coulomb branch: 1): at generic point, one has a vector space with a mixed Hodge structure; 2): at special point, there are three vector spaces and all of them carry mixed Hodge structure; and there is a monodromy group $T$ acting on these vector spaces;  3): One can also have a vector space at $\infty$ of moduli space.}
\label{coulomb1}
\end{figure}
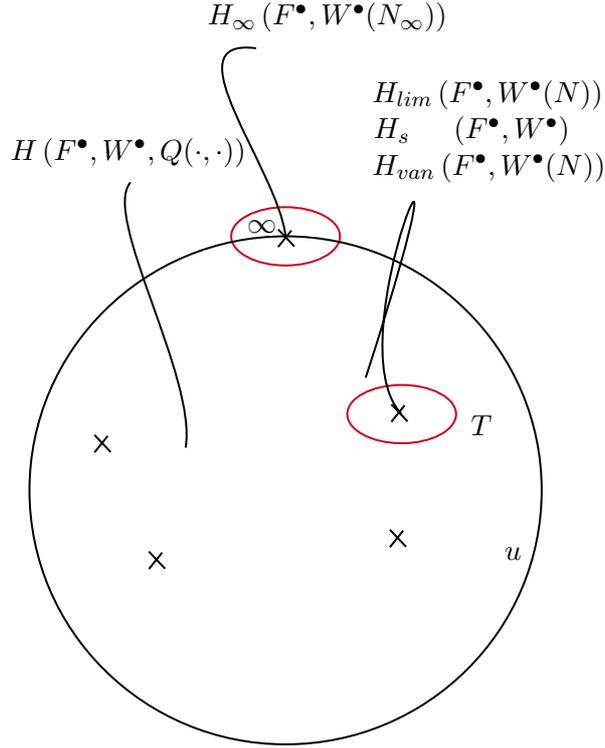

The basic idea of the classification for theories in $D\geq 4$ dimension has been described in \cite{Xie:2022lcm}.
Namely, one 
put 6d theory on $T^2$ and 5d theory on $S^1$ so that one get effective 4d $\mathcal{N}=2$ theory in the low energy (the resulting 4d effective theory is called KK theory). 
Then one use the general structure of  4d $\mathcal{N}=2$ Coulomb branch advocated in  \cite{Xie:2021hxd} to do the classification, see figure, \ref{coulomb1} for the illustration of Coulomb branch geometry. 

The crucial new ingredient in figure. \ref{coulomb1} is the object attached 
to the special points (We all them singularities.) at Coulomb branch. Those special points are the place where  new massless particles 
appear. The new object attached at singularity is 
defined by the \textbf{limiting} behavior of the structure attached to the nearby points. These  limiting 
objects are crucial in finding the low energy theory at special vacua.

Another crucial new ingredient is to compactify the Coulomb branch and so a limiting object can also be 
defined at the $\infty$ point of Coulomb branch. The limiting structure at $\infty$ can tell us the information 
of UV theory, such as which space-time dimension  the theory  lives. 

So the local behavior near the singularities (including $\infty$) is crucial in our classification program. 
For rank one theory,  the local singularity is classified by the conjugacy class of $SL(2,Z)$ group satisfying $(T^k-I)^2=0$,
 and there are  8 types which coincide with the Kodaira's list of singular fibers of elliptic surface. 
The associated low energy theory \footnote{Assuming the generic condition, namely, the singularities can always be split into a set of simplest $I_1$ singularities.} has been well studied \cite{Argyres:1995xn,Minahan:1996cj}.

The local singularities of rank two Coulomb branch is a lot more complicated. Firstly, there are many more 
possibilities for the choices of conjugacy classes of monodromy group; Secondly, unlike rank one case, the monodromy group  can not by itself decide the singularity. 
Now if one focus 
on a one  dimensional slice of Coulomb branch,  the local classification of singularities can be 
reduced to the study of degeneration of genus two pencils over the disk, which has been classified in \cite{namikawa1973complete}.
The classification is given by data $(M, \tau_{ij}(0), m)$, here $M$ is the monodromy group, $\tau_{ij}(0)$ is the fixed point of $M$, and $m$ is 
an invariant which is subtle to define. There are now a total of 120 classes. 

There are several quantities attach to local singularities: a) invariants $d_x, \delta_x, n_t$; b): a dual graph; c): monodromy group. 
All the data for these local singularities are listed in table. [\ref{elliptic1},\ref{elliptic2a},\ref{elliptic2b},\ref{parabolic3},\ref{parabolic4}], which 
are the main results of this paper. 
We then determine the low energy theory by using following two set of data: a): the possible scaling dimension inferred from the monodromy group \cite{Xie:2015xva,Caorsi:2018zsq}; b): the dual graph 
attached to the singularity, which can be used to find the 3d mirror for the low energy theory. 
There are a couple of new features for rank two local singularities:
\begin{enumerate}
\item One can define two invariants $d_x, \delta_x$ for the local singularity.  $\delta_x$ is called topological Euler number, and $d_x$  
is related to the minimal holomorphic family of  fibration. In the rank one case $d_x=\delta_x$. This is no longer the case for rank two cases: 
there are a few examples with $d_x \neq \delta_x$. 
\item In the rank one case, the local singularities can be split into  several $I_1$ singularity. In the rank two case, it is also possible to split  
the singularity into simpler ones, but now there are two kinds of atomic singularities  (those which can not be further splitted): $I_1$ and $\tilde{I}_1$, see figure. \ref{split} for the illustration.
The number of $I_1$ and $\tilde{I}_1$ singularities are determined by the number $d_x$ and $\delta_x$.
\item In the rank one case, the component $n_t$ in the dual graph is related to the rank of flavor symmetry $f$ as $f=n_t-1$.
Although the formula $f=n_t-1$ is still true for most of cases of rank two case, there are exceptions.
\end{enumerate}

As a first application of our approach, we construct some candidate global Coulomb branch geometries for 4d SCFTs (table. \ref{4ddeform} and \ref{4ddeform1}) and asymptotical 
free theories (table. \ref{af}), 5d $\mathcal{N}=1$ (table. \ref{5ddeform}) and 6d $(1,0)$ theories (table. \ref{6ddeform}). Here we only use simplest singularities 
in the bulk of the Coulomb branch. Our lists seem to be able to covers almost all the related results in the literature and actually suggest more possibilities.

Our approach is powerful in that it can deal with theories
in various dimensions simultaneously, and it can determine the low energy theory at \textbf{every} vacua of Coulomb branch. 
The results of this paper would be useful in studying the field theory dynamics, which we will explore elsewhere.

This paper is organized as follows: section 2 describes the local singularities 
of rank two theory, such as various invariants and the low energy theory attached to it; section 3 describes 
some global Coulomb branch geometries; section 4 gives a conclusion.

\section{Local singularities of rank two Coulomb branch geometry}
The basic properties of Coulomb branch of a 4d $\mathcal{N}=2$ theory are:
\begin{enumerate}
\item At a generic point, the low energy theory is described by $U(1)^r$ abelian gauge theory, free hypermultiplets, and possibly interacting SCFT whose 
Coulomb branch deformation is trivial. An important goal is to  determine the effective coupling for the $U(1)^r$ gauge theory, here $r$ is called the rank of the theory.
\item At a special point, new massless degrees of freedom appear.  The  low energy theory could be IR free gauge theory, or SCFT, or the direct sum of
them.
\item The new massless degrees of freedom at special point come from massive BPS particles at the generic point.
\end{enumerate}
The Coulomb branch solution for 4d $SU(2)$ gauge theory was solved in an elegant way by Seiberg and Witten \cite{Seiberg:1994rs,Seiberg:1994aj}.  They 
solved the theory by finding a family of algebraic curves $F(x,y, u,m, \Lambda)=0$ (Here $u$ parameterizes the Coulomb branch, $m$ the mass parameters,  and $\Lambda$ the dynamical generated scale.), and a SW differential $\lambda$ is also needed.

More generally, the Coulomb branch solution could be represented as a mixed Hodge module over the  generalized Coulomb branch (parameter space including Coulomb branch operators, masses,
relevant and marginal couplings).  Let's assume the rank of the theory is $r$ and the flavor symmetry has rank $f$ \footnote{This flavor number could be modified for 5d KK and 6d KK theories.}.

\textbf{At generic point}: There is a flat \footnote{The flat structure gives an integrable connection which is required for the definition of the mixed Hodge module.} holomorphic vector bundle whose rank is $2r+f$. To get the information of the low energy theory, two extra structures
are needed on the fiber $H$: 
\begin{itemize}
\item  A mixed Hodge structure, namely a Hodge filtration and a weight filtration; The weight filtration is an increasing filtration which takes the following form \footnote{If the SW geometry is given by a three dimensional variety \cite{Xie:2015rpa}, then the maximal weight is 4. These MHS could be brought to the form presented here by doing a Tate twist.}:
\begin{equation}
\{0\}=W^0\subset W^1\subset W^2= H;
\end{equation}
so we have two quotient spaces $Gr_1^W=W^1/W^0,~~Gr_2^W=W^2/W^1$, with dimension $dim(Gr_1^W)=2r,~dim(Gr_2^W)=f$. The weight filtration is needed so that we can separate the electric-magnetic part and the flavor part of the central charge: $Gr_1^W$ gives the
electric-magnetic charge, and $Gr_2^W$ gives the flavor charge. 
The Hodge filtration is a decreasing filtration and takes the following form
\begin{equation}
H=F^0\supset F^1;
\end{equation}
So in our case,  two holomorphic sub-bundles  $W^1$ and $F^1$ are needed.  The weight filtration and Hodge filtration together defines a so-called Mixed Hodge structure, and Hodge decomposition takes 
the form $Gr_1^W=H^{1,0}\oplus H^{0,1} $ and $Gr_2^W= H^{1,1}$, with dimension $h^{1,0}=h^{0,1}=r$ and $h^{1,1}=f$;

\item A polarization $Q(\cdot, \cdot)$ (which satisfies Riemann-Hodge bilinear relations on $Gr_1^W$ and acts trivially on $Gr_2^W$) on  $H$ so that  positive definite coupling constants can be defined. In fact, 
 a period matrix $Z_{ij},~i,j=1,\ldots, r$ which is symmetric and satisfies the condition $Im(Z)>0$ can be defined using the polarization.
\end{itemize}

\textbf{At  singular point \footnote{The Coulomb branch is not singular, but the physics is different from that of the generic point of the Coulomb branch.}}: There is also a vector space $H_s$ whose dimension is 
smaller than $H$, so the mathematical structure is not the vector bundle which is more familiar to physicists. The physics of the abelian gauge theory at singular point is described by $H_s$.
The crucial point of the mixed Hodge module is that one can define two more vector spaces at the singular point. The first 
is the so-called nearby cycle $H_{lim}$ which can be thought of as the limiting objects for the nearby vector spaces. There is a mixed Hodge structure on $H_{lim}$ which is quite different from that of the generic fiber described earlier:
The weight filtration is now determined by the nilpotent part $N$ of the monodromy group $T$ around the singularity. 

For the known solution, the monodromy group $T$ satisfies the following condition 
\begin{equation}
(T^k-1)^2=1.
\label{monodromytheorem}
\end{equation}
Namely the maximal size of the Jordan block is two, and the eigenvalue satisfies $\lambda^k=1$. We conjecture that this is true for the Coulomb branch solution of any $\mathcal{N}=2$ field theory. Furthermore, if we restrict the monodromy on the weight two part of the generic fiber, 
its action is trivial
\begin{equation}
T|_{Gr_2^W}=I.
\end{equation}
What this implies is that the monodromy matrix takes the form
\begin{equation*}
T=\left[\begin{array}{cc}
I&0\\
*&M\\
\end{array}\right].
\end{equation*}
and $M$ acts on weight one part. Finally, $H_s$ and $H_{lim}$ can be used to define a third vector space called vanishing cycle $H_{van}$. All of these three spaces carry mixed Hodge structure, and they form an exact sequence of mixed Hodge structure.  

Using the limit mixed Hodge structure (let's assume $H_s=0$), one can define 
a set of rational numbers $(\alpha_1,\alpha_2,\ldots, \alpha_s)$ called spectrum \cite{kulikov1998mixed}, and its relation to the eigenvalue of the monodromy group $T$ is given as 
\begin{equation*}
\lambda_i=\exp(2\pi i \alpha_i).
\end{equation*}
The monodromy group acts on the vector space $H_{lim}$, and so it has the decomposition $H_{lim}=\oplus_\lambda H^{lim}_\lambda$. The limit Hodge filtration defines a filtration on $H_\lambda^{lim}$: $F^0(H_\lambda^{lim})\supset F^{1}(H_\lambda^{lim})$. Now for
a basis element $e_i$ in $H_\lambda^{lim}$, a spectral number $\alpha$ is defined as
\begin{equation*}
\begin{cases}
& e_i\in F^1(H_\lambda^{lim}),~~~~~~~~~~~-1<\alpha\leq 0 \\
& e_i \in F^0(H_\lambda^{lim})/F^1(H_\lambda^{lim}),~~~~0<\alpha\leq 1
\end{cases}
\end{equation*}
here $\exp(2\pi \alpha)=\lambda$.
An important consistent condition is that the spectral numbers are in pair
\begin{equation*}
\alpha_i+\alpha_j=0.
\end{equation*}
One can find the Coulomb branch spectrum from the spectral numbers as follows. Let's denote the minimal spectrum number as $\alpha_{min}$, then one associate a Coulomb branch scaling dimension 
 for a spectral number $\alpha_i$ as follows \cite{Xie:2015xva}:
\begin{equation}
[u_i]={1+\alpha_{min}-\alpha_i\over 1+\alpha_{min}}.
\label{scale}
\end{equation}
So if $\alpha_i=0$, then $[u_i]=1$ which gives a mass parameter. The maximal scaling dimension is given as ${1\over 1+\alpha_{min}}$. 

\textbf{At Infinity}: While the above analysis is carried for the finite points of the Coulomb branch, it is possible to do the similar computation for the $\infty$ point on the moduli space, and the MHS at $\infty$ is useful to extract 
information for UV theory. 
The general structure of the Coulomb branch solution is summarized in figure. \ref{coulomb1}.

\subsection{Rank two Coulomb branch geometry and genus two pencils}
Let's now look at rank two Coulomb branch geometry: namely the dimension of $Gr_1^W$ is $4$. The generalized Coulomb branch parameters include 
expectation values of two Coulomb branch operators $u,v$ (here one assume the scaling dimension of $v$ is no less than $u$), the mass parameters,
and possibly relevant deformations for 4d theory.  There would be also dynamical generated scale or exact marginal deformation 
for 4d theory, but those parameters would be regarded as parameterizing the UV theory, and are not regarded as the Coulomb branch 
parameters of a given UV theory.

We now make a couple of simplifying assumption for the Coulomb branch geometry. First of all, since the monodromy group acts trivially on weight two part $Gr_2^W$, one might focus 
on weight one part, and now there is a rank four flat holomorphic bundle at generic points. One must be careful that lots of information such as the flavor group
is lost in this assumption, and the hope is to recover those information by looking at extra structure (such as the Mordell-Weil lattice) on the weight one part of  Coulomb branch geometry. 

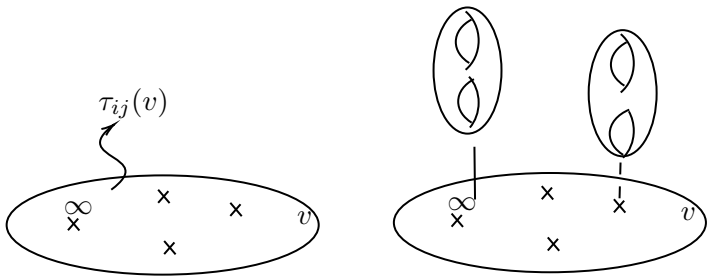
\begin{figure}[H]
\begin{center}

\tikzset{every picture/.style={line width=0.75pt}} 

\begin{tikzpicture}[x=0.45pt,y=0.45pt,yscale=-1,xscale=1]

\draw   (31,188.61) .. controls (31,165.63) and (89.09,147) .. (160.75,147) .. controls (232.41,147) and (290.5,165.63) .. (290.5,188.61) .. controls (290.5,211.59) and (232.41,230.22) .. (160.75,230.22) .. controls (89.09,230.22) and (31,211.59) .. (31,188.61) -- cycle ;
\draw    (82,182) -- (91.5,193.22) ;
\draw    (82,193) -- (87.23,186.98) -- (90.5,183.22) ;

\draw    (157,159) -- (166.5,170.22) ;
\draw    (157,170) -- (162.23,163.98) -- (165.5,160.22) ;

\draw    (217,170) -- (226.5,181.22) ;
\draw    (217,181) -- (222.23,174.98) -- (225.5,171.22) ;

\draw    (162,202) -- (171.5,213.22) ;
\draw    (162,213) -- (167.23,206.98) -- (170.5,203.22) ;

\draw    (118,159) .. controls (157.6,129.3) and (82.04,134.12) .. (119.34,105.11) ;
\draw [shift={(120.5,104.22)}, rotate = 143.13] [color={rgb, 255:red, 0; green, 0; blue, 0 }  ][line width=0.75]    (10.93,-3.29) .. controls (6.95,-1.4) and (3.31,-0.3) .. (0,0) .. controls (3.31,0.3) and (6.95,1.4) .. (10.93,3.29)   ;
\draw   (353,186.61) .. controls (353,163.63) and (411.09,145) .. (482.75,145) .. controls (554.41,145) and (612.5,163.63) .. (612.5,186.61) .. controls (612.5,209.59) and (554.41,228.22) .. (482.75,228.22) .. controls (411.09,228.22) and (353,209.59) .. (353,186.61) -- cycle ;
\draw    (401,179) -- (410.5,190.22) ;
\draw    (401,190) -- (406.23,183.98) -- (409.5,180.22) ;

\draw    (476,156) -- (485.5,167.22) ;
\draw    (476,167) -- (481.23,160.98) -- (484.5,157.22) ;

\draw    (536,167) -- (545.5,178.22) ;
\draw    (536,178) -- (541.23,171.98) -- (544.5,168.22) ;

\draw    (481,199) -- (490.5,210.22) ;
\draw    (481,210) -- (486.23,203.98) -- (489.5,200.22) ;

\draw   (386,59.11) .. controls (386,29.9) and (398.65,6.22) .. (414.25,6.22) .. controls (429.85,6.22) and (442.5,29.9) .. (442.5,59.11) .. controls (442.5,88.32) and (429.85,112) .. (414.25,112) .. controls (398.65,112) and (386,88.32) .. (386,59.11) -- cycle ;
\draw  [draw opacity=0] (415.7,52.78) .. controls (408.85,48.54) and (404.5,42.2) .. (404.5,35.11) .. controls (404.5,27.39) and (409.66,20.55) .. (417.6,16.35) -- (436,35.11) -- cycle ; \draw   (415.7,52.78) .. controls (408.85,48.54) and (404.5,42.2) .. (404.5,35.11) .. controls (404.5,27.39) and (409.66,20.55) .. (417.6,16.35) ;  
\draw  [draw opacity=0] (412.65,58.47) .. controls (419.22,53.03) and (423.5,44.1) .. (423.5,34) .. controls (423.5,23.92) and (419.23,15) .. (412.68,9.56) -- (397.75,34) -- cycle ; \draw   (412.65,58.47) .. controls (419.22,53.03) and (423.5,44.1) .. (423.5,34) .. controls (423.5,23.92) and (419.23,15) .. (412.68,9.56) ;  
\draw  [draw opacity=0] (417.7,102.78) .. controls (410.85,98.54) and (406.5,92.2) .. (406.5,85.11) .. controls (406.5,77.39) and (411.66,70.55) .. (419.6,66.35) -- (438,85.11) -- cycle ; \draw   (417.7,102.78) .. controls (410.85,98.54) and (406.5,92.2) .. (406.5,85.11) .. controls (406.5,77.39) and (411.66,70.55) .. (419.6,66.35) ;  
\draw  [draw opacity=0] (416.22,106.9) .. controls (420.73,101.5) and (423.5,94.13) .. (423.5,86) .. controls (423.5,77.44) and (420.42,69.71) .. (415.48,64.24) -- (397.75,86) -- cycle ; \draw   (416.22,106.9) .. controls (420.73,101.5) and (423.5,94.13) .. (423.5,86) .. controls (423.5,77.44) and (420.42,69.71) .. (415.48,64.24) ;  
\draw   (515,78.11) .. controls (515,48.9) and (527.65,25.22) .. (543.25,25.22) .. controls (558.85,25.22) and (571.5,48.9) .. (571.5,78.11) .. controls (571.5,107.32) and (558.85,131) .. (543.25,131) .. controls (527.65,131) and (515,107.32) .. (515,78.11) -- cycle ;
\draw  [draw opacity=0] (544.7,71.78) .. controls (537.85,67.54) and (533.5,61.2) .. (533.5,54.11) .. controls (533.5,46.39) and (538.66,39.55) .. (546.6,35.35) -- (565,54.11) -- cycle ; \draw   (544.7,71.78) .. controls (537.85,67.54) and (533.5,61.2) .. (533.5,54.11) .. controls (533.5,46.39) and (538.66,39.55) .. (546.6,35.35) ;  
\draw  [draw opacity=0] (541.65,77.47) .. controls (548.22,72.03) and (552.5,63.1) .. (552.5,53) .. controls (552.5,42.92) and (548.23,34) .. (541.68,28.56) -- (526.75,53) -- cycle ; \draw   (541.65,77.47) .. controls (548.22,72.03) and (552.5,63.1) .. (552.5,53) .. controls (552.5,42.92) and (548.23,34) .. (541.68,28.56) ;  
\draw  [draw opacity=0] (543.7,129.78) .. controls (536.85,125.54) and (532.5,119.2) .. (532.5,112.11) .. controls (532.5,103.41) and (539.06,95.83) .. (548.74,91.89) -- (564,112.11) -- cycle ; \draw   (543.7,129.78) .. controls (536.85,125.54) and (532.5,119.2) .. (532.5,112.11) .. controls (532.5,103.41) and (539.06,95.83) .. (548.74,91.89) ;  
\draw  [draw opacity=0] (543.52,132.21) .. controls (548.98,127.03) and (552.5,118.83) .. (552.5,109.61) .. controls (552.5,103.55) and (550.98,97.93) .. (548.38,93.32) -- (529.75,109.61) -- cycle ; \draw   (543.52,132.21) .. controls (548.98,127.03) and (552.5,118.83) .. (552.5,109.61) .. controls (552.5,103.55) and (550.98,97.93) .. (548.38,93.32) ;  
\draw    (420,123) -- (420.5,167.22) ;
\draw  [dash pattern={on 4.5pt off 4.5pt}]  (541.5,136.22) -- (540.5,172.22) ;

\draw (270,174.4) node [anchor=north west][inner sep=0.75pt]    {$v$};
\draw (76,166.4) node [anchor=north west][inner sep=0.75pt]    {$\infty $};
\draw (105,71.4) node [anchor=north west][inner sep=0.75pt]    {$\tau _{ij}( v)$};
\draw (589,171.4) node [anchor=north west][inner sep=0.75pt]    {$v$};
\draw (395,163.4) node [anchor=north west][inner sep=0.75pt]    {$\infty $};

\end{tikzpicture}

\end{center}
\caption{Left: The weight one part of Coulomb branch solution of a rank two theory: A multivalued holomorphic function $\tau_{ij}(v)$ is defined over the one dimensional slice of 
Coulomb branch parameterized by $v$; Right: One get a genus two fibration over $\mathbb{P}^1$ by identifying the complex structure of the genus two curve with $\tau_{ij}(v)$. }
\label{oneparameter}
\end{figure}

Secondly, one fix all the parameters but one (typically the parameter $v$, but it is not always possible to do it, then one need to look at a different family defined by the hypersurface  $f(u,v)=0$ in the $u,v$ plane), 
and a one parameter family of rank four flat bundles would be considered. See figure. \ref{oneparameter}. 

Now one has a  multi-valued holomorphic function $\tau_{ij}(v)$ at a generic point of the one family Coulomb branch geometry. The matrix $\tau_{ij}$ is a $2\times2$ symmetric matrix and satisfies following condition
\begin{equation*}
Im (\tau_{ij})>0;
\end{equation*}
We now attach a genus two curve at each base point on $v$ plane by identifying its complex structure with  $\tau_{ij}$. Therefore one has 
a family of genus two curves over $\mathbb{P}^1$, and the total space is  a surface $X$ with the map $f:X\to \mathbb{P}^1$.  At singularity, 
the genus two curves would degenerate, and 
the study of 
local singularities of rank two Coulomb branch (weight one part) is then reduced to the study of the one parameter
degeneration of genus two curves!

\subsection{Pencils of genus two curves}
The result of last subsection links the study of rank two Coulomb branch geometry to that of genus two pencils, and the degeneration of 
those pencils are completely classified in \cite{namikawa1973complete}. In this subsection, we will first review their classifications, 
and the local invariants, and finally the low energy theory would be given by physical input.

\subsubsection{Numerical type of  genus singular fibers}
In this subsection, the numerical type of a genus  two pencils will be reviewed.
Here the singular fiber $X_0$ is regarded as a divisor in $X$,
and $X_0$ does not have a component with self-intersection number  $-1$.

\begin{figure}[H]
\begin{center}
\tikzset{every picture/.style={line width=0.75pt}} 

\begin{tikzpicture}[x=0.45pt,y=0.45pt,yscale=-1,xscale=1]

\draw    (102,101.72) -- (189,101.45) ;
\draw   (110,96.72) .. controls (110,93.96) and (112.3,91.72) .. (115.14,91.72) .. controls (117.98,91.72) and (120.28,93.96) .. (120.28,96.72) .. controls (120.28,99.48) and (117.98,101.72) .. (115.14,101.72) .. controls (112.3,101.72) and (110,99.48) .. (110,96.72) -- cycle ;
\draw    (260,60) -- (260,180.72) ;
\draw    (272,161.72) .. controls (249,125.44) and (259,71.72) .. (316,81.72) ;
\draw    (401,108.72) -- (503,109.72) ;
\draw    (420,157.72) -- (480,61.72) ;
\draw    (481,161.72) -- (422,62.72) ;
\draw    (79,299.72) -- (199,300.72) ;
\draw    (80,359.72) -- (123,282.72) ;
\draw    (208,363.72) -- (163,283.72) ;
\draw    (122,396.72) -- (80,333.72) ;
\draw    (177,395.72) -- (217,329.72) ;
\draw    (128,397) -- (140,396.72) ;
\draw    (147,396) -- (159,395.72) ;
\draw    (164,396) -- (176,395.72) ;
\draw    (296,283) -- (398,281.72) ;
\draw    (462,280) -- (564,278.72) ;
\draw    (340,360) -- (340,260.72) ;
\draw    (520,360) -- (520,260.72) ;
\draw    (405,282) -- (417,281.72) ;
\draw    (427,282) -- (439,281.72) ;
\draw    (447,281) -- (459,280.72) ;
\draw    (319,318) -- (365,317.72) ;
\draw    (319,341) -- (365,340.72) ;
\draw    (499,340) -- (545,339.72) ;
\draw    (498,320) -- (544,319.72) ;
\draw    (26,541) -- (172,540.22) ;
\draw    (218,541.22) -- (409,539.22) ;
\draw    (445,540) -- (621,540.22) ;
\draw    (63,506) -- (36,580.22) ;
\draw    (105,508) -- (78,582.22) ;
\draw    (144,510) -- (117,584.22) ;
\draw    (41,551.22) -- (40,611.22) ;
\draw    (83,550.22) -- (82,610.22) ;
\draw    (123,555.22) -- (122,615.22) ;
\draw    (279,504) -- (280,583.22) ;
\draw    (321,501) -- (322,580.22) ;
\draw    (360,504) -- (361,583.22) ;
\draw    (245,569) -- (302,569.22) ;
\draw    (344,567) -- (401,567.22) ;
\draw    (379,552) -- (380,609.22) ;
\draw    (262,554) -- (264,613.22) ;
\draw    (540,501) -- (540,583.22) ;
\draw    (531,574) -- (584,574.22) ;
\draw    (569,561) -- (570,612.22) ;
\draw    (560,598) -- (597,598.22) ;
\draw    (588,623) -- (588,589.22) ;
\draw    (498,502) -- (498,584.22) ;
\draw    (458,505) -- (458,587.22) ;
\draw    (429,577) -- (482,577.22) ;

\draw (117,205.4) node [anchor=north west][inner sep=0.75pt]    {$II$};
\draw (254,205.4) node [anchor=north west][inner sep=0.75pt]    {$III$};
\draw (434,203.4) node [anchor=north west][inner sep=0.75pt]    {$IV$};
\draw (139,335.4) node [anchor=north west][inner sep=0.75pt]    {$n$};
\draw (336,296.4) node [anchor=north west][inner sep=0.75pt]    {$2$};
\draw (355,269.4) node [anchor=north west][inner sep=0.75pt]    {$2$};
\draw (490,267.4) node [anchor=north west][inner sep=0.75pt]    {$2$};
\draw (520,292.4) node [anchor=north west][inner sep=0.75pt]    {$2$};
\draw (439,405.4) node [anchor=north west][inner sep=0.75pt]    {$I_{n}^{*}$};
\draw (136,419.4) node [anchor=north west][inner sep=0.75pt]    {$I_{n}$};
\draw (401,288.4) node [anchor=north west][inner sep=0.75pt]    {$n-1$};
\draw (71,533.4) node [anchor=north west][inner sep=0.75pt]    {$3$};
\draw (45,545.4) node [anchor=north west][inner sep=0.75pt]    {$2$};
\draw (89,543.4) node [anchor=north west][inner sep=0.75pt]    {$2$};
\draw (129,543.4) node [anchor=north west][inner sep=0.75pt]    {$2$};
\draw (78,640.4) node [anchor=north west][inner sep=0.75pt]    {$IV^{*}$};
\draw (305,633.4) node [anchor=north west][inner sep=0.75pt]    {$III^{*}$};
\draw (510,628.4) node [anchor=north west][inner sep=0.75pt]    {$II^{*}$};
\draw (312,554.4) node [anchor=north west][inner sep=0.75pt]    {$2$};
\draw (351,547.4) node [anchor=north west][inner sep=0.75pt]    {$3$};
\draw (279,545.4) node [anchor=north west][inner sep=0.75pt]    {$3$};
\draw (268,568.4) node [anchor=north west][inner sep=0.75pt]    {$2$};
\draw (366,567.4) node [anchor=north west][inner sep=0.75pt]    {$2$};
\draw (509,522.4) node [anchor=north west][inner sep=0.75pt]    {$6$};
\draw (301,521.4) node [anchor=north west][inner sep=0.75pt]    {$4$};
\draw (529,542.4) node [anchor=north west][inner sep=0.75pt]    {$5$};
\draw (550,559.4) node [anchor=north west][inner sep=0.75pt]    {$4$};
\draw (570,579.4) node [anchor=north west][inner sep=0.75pt]    {$3$};
\draw (575,596.4) node [anchor=north west][inner sep=0.75pt]    {$2$};
\draw (446,548.4) node [anchor=north west][inner sep=0.75pt]    {$4$};
\draw (437,577.4) node [anchor=north west][inner sep=0.75pt]    {$2$};

\end{tikzpicture}
\caption{The dual graph for genus one degeneration, which is classified by Kodaira, see \cite{barth2015compact}. The data $(n_i,k_i, C_{ij})$ are read from the dual graph as follows: the number $n_i$ is listed for each component $C_i$ in the graph (if $n_i=1$, the number is omitted); 
$C_{ij}$ is read from the intersection number of component $C_i$ and $C_j$ in the graph; In genus one case, all the components are rational curve with $-2$ self-intersection number, and so $k_i=0$ for all the components.}
\label{kodaira}
\end{center}
\end{figure}
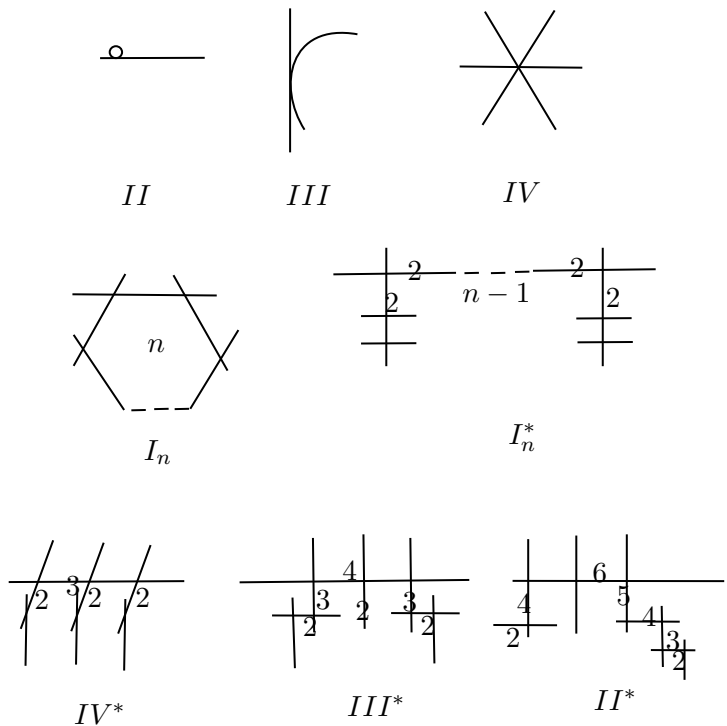

Let's consider a family of genus two fibration $\pi: X\to \Delta$, and assume the generic fiber on $b\in \Delta$ is a smooth 
genus two curve. Only the fiber $X_0$ at $b=0$ is a singular fiber.  $X_0$ is a reducible curve $X_0=\sum n_i C_i$, which satisfies following conditions: 
\begin{enumerate}
\item $X_0$ is connected, and $n_i>0$ for all $i$.
\item $C_i\cdot C_j=C_j\cdot C_i \geq 0$ if $i\neq j$, and $X\cdot C_i=0$.
\item $p(C_i)=\frac{1}{2}[C_i^2+C_i\cdot K)+1\geq 0$ for all $i$. Here $p(C_i)$ is the genus of curve $C_i$.
\end{enumerate}
$K$ is the canonical class of the compact surface $X$, which only plays a formal role here (all we need are  numbers $k_i=C_i \cdot K\geq 0$). Therefore a configuration of curves is 
specified by the data
\begin{equation}
(n_i,k_i, C_{ij}),~~~i=1,\ldots, n.
\end{equation}
The genus of $X_0$ is given by following formula:
\begin{equation}
g(X_0)=1+\frac{1}{2}(X_0\cdot K)=1+{1\over 2} \sum (n_i k_i),
\label{genus1}
\end{equation}
Which is determined by the data $n_i,k_i$ only. Since there is no $-1$ component (self-intersection number is $-1$) in $X_0$,  there is only a finite number of possibilities 
for a given genus \cite{artin1971degenerate}.  There is a complete classification of genus one pencils by Kodaira, and the configurations are represented by the dual graphs, see figure. \ref{kodaira}.

For genus two, the individual component of the singular fiber $\Gamma$ is one of the five types listed in table. \ref{component}. 
The proof is following: By using the formula \ref{genus1} for the genus, one has following solution for the components with non-zero $k_i$: a): there is one component with $k=2$ and $n=1$; 
and this component has genus $p(C)=\frac{1}{2} (C^2+2)+1$; Either $C^2=-4, p(C)=0$ (type D), or
$C^2=-2, p(C)=1$ ( type $C$). b): Two components with $k=1$; The genus of each component is given as $p(C)=\frac{1}{2} (C^2+1)+1$, and so $C^2=-3$ (type B) or $C^2=-1$ (type A). 
The genus formula does not put constraints on type $E$ component $(k_i=0)$, and they are always allowed.

\begin{table}[htp]
\begin{center}
\begin{tabular}{|c|c|c|c|}
\hline
Type & $k_i$ & $\Gamma^2$ & $p(\Gamma)$ \\ \hline
A: & $\Gamma\cdot K=1$ & $\Gamma^2=-1$ & $p(\Gamma)=1$ \\ \hline
B: & $\Gamma\cdot K=1$ & $\Gamma^2=-3$ & $p(\Gamma)=0$ \\ \hline
C: & $\Gamma\cdot K=2$ & $\Gamma^2=-2$ & $p(\Gamma)=1$ \\ \hline
D: & $\Gamma\cdot K=2$ & $\Gamma^2=-4$ & $p(\Gamma)=0$ \\ \hline
E: & $\Gamma\cdot K=0$ & $\Gamma^2=-2$ & $p(\Gamma)=0$ \\ \hline
\end{tabular}
\end{center}
\caption{The possible components for the genus two singular fiber.}
\label{component}
\end{table}%

Given the data in table. \ref{component}, there are following choices for components with $k_i>0$:
\begin{equation*}
C,~2A,~~(A,A),~~2B,~~(A,B),~~(B,B),~~~D.
\label{dualtype}
\end{equation*}
The dual graphs for all genus two singular fibers have been given in \cite{ogg1966pencils}, and there are a total of 44 types. 

\begin{figure}[htbp]
\begin{center}

\tikzset{every picture/.style={line width=0.75pt}} 

\begin{tikzpicture}[x=0.55pt,y=0.55pt,yscale=-1,xscale=1]

\draw    (194,218) -- (195,314.72) ;
\draw    (174,237) -- (314,237) ;
\draw    (213,220.28) -- (214,277) ;
\draw    (253,220.28) -- (254,277) ;
\draw    (293,220.28) -- (294,277) ;
\draw    (274,257) -- (334,257) ;
\draw    (314,246.72) -- (315,278.72) ;

\draw (173,293.4) node [anchor=north west][inner sep=0.75pt]    {$D$};
\draw (228,218.4) node [anchor=north west][inner sep=0.75pt]    {$4$};
\draw (218,251.4) node [anchor=north west][inner sep=0.75pt]    {$2$};
\draw (244,252.4) node [anchor=north west][inner sep=0.75pt]    {$2$};
\draw (299,247.4) node [anchor=north west][inner sep=0.75pt]    {$2$};
\draw (319,267.4) node [anchor=north west][inner sep=0.75pt]    {$1$};
\draw (281,236.4) node [anchor=north west][inner sep=0.75pt]    {$3$};

\end{tikzpicture}

\end{center}
\caption{A genus two singular fibers $X_0=\sum n_i C_i$: there is a component $D$, and the multiplicities $n_i$ are indicated. The type E component is not explicitly stated.}
\label{dual1}
\end{figure}
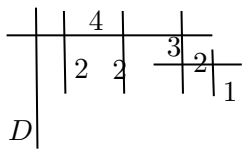

\textbf{Example}: Here we'd like to give an example to illustrate the idea of classification. 
 Let's assume that there is a type $D$ component, and so the singular fiber can be written as 
$X_0=D+\delta$. Now the condition $D\cdot X_0=0$ implies that $D^2+D\cdot \delta=0$, therefore
the following equation holds
\begin{equation*}
D\cdot \delta=4.
\end{equation*}
The above equation implies that $D$ intersects with the rest of $X_0$ (minus D) four times. The decomposition of $4$ is $4,3+1,2+2$.  Let's look at the decomposition $4$, and so 
$X_0=D+4C_1+\delta_1$, where $C_1$ is a $-2$ curve with $C_1\cdot D=4$ . We then use the condition $C_1\cdot X_0=0$, which then gives the equation
\begin{equation*}
C_1\cdot(D+4C_1+\delta_1)=0\to C_1\cdot \delta_1=-C_1\cdot D-4C_1^2=7.
\end{equation*}
So the total intersection number of $C_1$ with $\delta_1$ is 7. The legitimate decomposition of $7$ is $7=2+2+3,3+4,2+5,7$, and one of the configurations (with  decomposition  $7=2+2+3$)
is shown in figure. \ref{dual1}.  One can do  the similar computation to find all the other configurations listed in \cite{ogg1966pencils}.

\subsubsection{Classification of genus two singular fiber}
\begin{figure}[htbp]
\begin{center}

\tikzset{every picture/.style={line width=0.75pt}} 

\begin{tikzpicture}[x=0.55pt,y=0.55pt,yscale=-1,xscale=1]

\draw   (191,284.86) .. controls (191,270.97) and (222.79,259.72) .. (262,259.72) .. controls (301.21,259.72) and (333,270.97) .. (333,284.86) .. controls (333,298.74) and (301.21,310) .. (262,310) .. controls (222.79,310) and (191,298.74) .. (191,284.86) -- cycle ;
\draw  [fill={rgb, 255:red, 0; green, 0; blue, 0 }  ,fill opacity=1 ] (258,284.5) .. controls (258,283.12) and (259.12,282) .. (260.5,282) .. controls (261.88,282) and (263,283.12) .. (263,284.5) .. controls (263,285.88) and (261.88,287) .. (260.5,287) .. controls (259.12,287) and (258,285.88) .. (258,284.5) -- cycle ;
\draw    (259,233.72) .. controls (263.63,230.24) and (256,199.72) .. (260,194.72) .. controls (264,189.72) and (284.52,204.85) .. (280,195.72) .. controls (275.48,186.59) and (266.73,213.47) .. (264,205.72) .. controls (261.27,197.97) and (266.74,160.91) .. (275,154.72) ;
\draw    (209,234) .. controls (249,204) and (187,197.72) .. (227,167.72) ;
\draw  [dash pattern={on 4.5pt off 4.5pt}]  (213,242) -- (215,287.72) ;
\draw  [dash pattern={on 4.5pt off 4.5pt}]  (259,240) -- (261,285.72) ;

\draw (284,167.4) node [anchor=north west][inner sep=0.75pt]    {$X_{0}$};
\draw (184,164.4) node [anchor=north west][inner sep=0.75pt]    {$X_{t}$};

\end{tikzpicture}

\end{center}
\caption{A family of genus two curves over the disc. The fiber is smooth for $t\neq 0$, and the fiber degenerates at $t=0$}
\label{pens}
\end{figure}
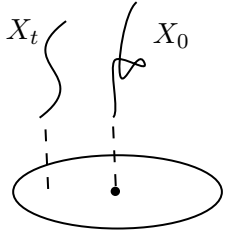
The numerical type does not completely determine the degeneration of the family of genus two curves. 
Mathematically, one consider a family of genus two curves over the disc $f:X\to \Delta$. One further 
assume that the fiber $X_t=f^{-1}(t)$ is smooth for $t\neq 0$. The only possible singular fiber is the fiber $X_0$
at $t=0$, see figure. \ref{pens}.  Once the above family is given, one can get a period mapping by looking at the first cohomology group of $X_t$:
the first cohomology group $H^1(X_t)$ has a Hodge decomposition $H^1(X_t)=H^{1,0}\otimes H^{0,1}$, and $H^{1,0}$ is 
spanned by the holomorphic differentials $w_1, w_2$. By choosing a symplectic basis of homology cycles $A_1, A_2, B_1, B_2$, one 
can form the period integrals and compute period matrix $\Omega$ whose entries are
\begin{equation*}
\int_{A_j} \omega_i,~~\int_{B_j} \omega_i,
\end{equation*}
The period matrix takes the form $\Omega=[I,\tau_{ij}]$
by choosing the basis of holomorphic differentials property. Here $\tau_{ij}$ is a two by two matrix satisfies following condition
\begin{equation*}
\tau=\tau^{T},~~Im(\tau)>0.
\end{equation*}
The entry of $\tau_{ij}(t)$ is moreover a multivalued holomorphic function of $t$, and so it has the following monodromy property
\begin{equation*}
\tau(e^{2\pi i}t)=M \tau (t).
\end{equation*}
 Here $M$ is the monodromy group. In fact one has a conjugacy class of $Sp(4,Z)$, as if the symplectic basis of 
 the homology group is changed, the monodromy group is also transformed by conjugating a $Sp(4,Z)$ matrix. The monodromy group satisfies 
 the relation
 \begin{equation*}
 (M^k-I)^2=0.
 \end{equation*}
 This means that $M$ is a quasi-unipotent matrix, namely, the eigenvalue is the roots of unity,
 and the maximal size of its Jordan block is two.  $M$ is called \textbf{elliptic} if it is semi-simple (finite order), otherwise it is 
 called \textbf{parabolic} (infinite order).

 It was shown in \cite{namikawa1973complete} that the isomorphic family of genus two pencil are classified by the following three types of data: 1) The conjugacy class of monodromy group $M$; 2) The limiting
 point $\tau(0)$ which can be represented by stable curves; 3): A third invariant $m$ which is more subtle to define.  In summary, a degeneration of genus 
 two family is classified by the data 
 \begin{equation}
 \boxed{(M, \tau(0), m)}.
 \end{equation}
  There are a total of around 120 types of degenerations for genus two curves (comparing 8 classes of 
 genus one degenerations).
 
\begin{figure}
  \begin{center}
\tikzset{every picture/.style={line width=0.75pt}} 

\begin{tikzpicture}[x=0.45pt,y=0.45pt,yscale=-1,xscale=1]

\draw   (126,137.22) .. controls (126,113.19) and (160.03,93.72) .. (202,93.72) .. controls (243.97,93.72) and (278,113.19) .. (278,137.22) .. controls (278,161.24) and (243.97,180.72) .. (202,180.72) .. controls (160.03,180.72) and (126,161.24) .. (126,137.22) -- cycle ;
\draw  [draw opacity=0] (207.96,141.92) .. controls (209.03,132.28) and (220.88,124.32) .. (235.65,123.74) .. controls (250.96,123.13) and (263.77,130.67) .. (264.85,140.73) -- (236.4,142.72) -- cycle ; \draw   (207.96,141.92) .. controls (209.03,132.28) and (220.88,124.32) .. (235.65,123.74) .. controls (250.96,123.13) and (263.77,130.67) .. (264.85,140.73) ;  
\draw  [draw opacity=0] (256,128.23) .. controls (256.81,130.36) and (257.21,132.68) .. (257.14,135.09) .. controls (256.83,145.57) and (247.56,153.98) .. (236.43,153.85) .. controls (225.31,153.73) and (216.54,145.12) .. (216.86,134.63) .. controls (216.93,132.22) and (217.47,129.92) .. (218.41,127.81) -- (237,134.86) -- cycle ; \draw   (256,128.23) .. controls (256.81,130.36) and (257.21,132.68) .. (257.14,135.09) .. controls (256.83,145.57) and (247.56,153.98) .. (236.43,153.85) .. controls (225.31,153.73) and (216.54,145.12) .. (216.86,134.63) .. controls (216.93,132.22) and (217.47,129.92) .. (218.41,127.81) ;  
\draw  [draw opacity=0] (136.2,143.2) .. controls (136.02,142.44) and (135.92,141.65) .. (135.89,140.85) .. controls (135.47,130.37) and (147.9,121.36) .. (163.65,120.74) .. controls (179.4,120.12) and (192.5,128.11) .. (192.92,138.59) .. controls (192.93,138.97) and (192.93,139.35) .. (192.91,139.72) -- (164.4,139.72) -- cycle ; \draw   (136.2,143.2) .. controls (136.02,142.44) and (135.92,141.65) .. (135.89,140.85) .. controls (135.47,130.37) and (147.9,121.36) .. (163.65,120.74) .. controls (179.4,120.12) and (192.5,128.11) .. (192.92,138.59) .. controls (192.93,138.97) and (192.93,139.35) .. (192.91,139.72) ;  
\draw  [draw opacity=0] (184,125.23) .. controls (184.81,127.36) and (185.21,129.68) .. (185.14,132.09) .. controls (184.83,142.57) and (175.56,150.98) .. (164.43,150.85) .. controls (153.31,150.73) and (144.54,142.12) .. (144.86,131.63) .. controls (144.93,129.22) and (145.47,126.92) .. (146.41,124.81) -- (165,131.86) -- cycle ; \draw   (184,125.23) .. controls (184.81,127.36) and (185.21,129.68) .. (185.14,132.09) .. controls (184.83,142.57) and (175.56,150.98) .. (164.43,150.85) .. controls (153.31,150.73) and (144.54,142.12) .. (144.86,131.63) .. controls (144.93,129.22) and (145.47,126.92) .. (146.41,124.81) ;  
\draw   (347,143.22) .. controls (347,127.48) and (366.7,114.72) .. (391,114.72) .. controls (415.3,114.72) and (435,127.48) .. (435,143.22) .. controls (435,158.96) and (415.3,171.72) .. (391,171.72) .. controls (366.7,171.72) and (347,158.96) .. (347,143.22) -- cycle ;
\draw  [draw opacity=0] (362.2,149.2) .. controls (362.02,148.44) and (361.92,147.65) .. (361.89,146.85) .. controls (361.47,136.37) and (373.9,127.36) .. (389.65,126.74) .. controls (405.4,126.12) and (418.5,134.11) .. (418.92,144.59) .. controls (418.93,144.97) and (418.93,145.35) .. (418.91,145.72) -- (390.4,145.72) -- cycle ; \draw   (362.2,149.2) .. controls (362.02,148.44) and (361.92,147.65) .. (361.89,146.85) .. controls (361.47,136.37) and (373.9,127.36) .. (389.65,126.74) .. controls (405.4,126.12) and (418.5,134.11) .. (418.92,144.59) .. controls (418.93,144.97) and (418.93,145.35) .. (418.91,145.72) ;  
\draw  [draw opacity=0] (410,131.23) .. controls (410.81,133.36) and (411.21,135.68) .. (411.14,138.09) .. controls (410.83,148.57) and (401.56,156.98) .. (390.43,156.85) .. controls (379.31,156.73) and (370.54,148.12) .. (370.86,137.63) .. controls (370.93,135.22) and (371.47,132.92) .. (372.41,130.81) -- (391,137.86) -- cycle ; \draw   (410,131.23) .. controls (410.81,133.36) and (411.21,135.68) .. (411.14,138.09) .. controls (410.83,148.57) and (401.56,156.98) .. (390.43,156.85) .. controls (379.31,156.73) and (370.54,148.12) .. (370.86,137.63) .. controls (370.93,135.22) and (371.47,132.92) .. (372.41,130.81) ;  
\draw   (435,142.22) .. controls (435,126.48) and (454.7,113.72) .. (479,113.72) .. controls (503.3,113.72) and (523,126.48) .. (523,142.22) .. controls (523,157.96) and (503.3,170.72) .. (479,170.72) .. controls (454.7,170.72) and (435,157.96) .. (435,142.22) -- cycle ;
\draw  [draw opacity=0] (450.2,148.2) .. controls (450.02,147.44) and (449.92,146.65) .. (449.89,145.85) .. controls (449.47,135.37) and (461.9,126.36) .. (477.65,125.74) .. controls (493.4,125.12) and (506.5,133.11) .. (506.92,143.59) .. controls (506.93,143.97) and (506.93,144.35) .. (506.91,144.72) -- (478.4,144.72) -- cycle ; \draw   (450.2,148.2) .. controls (450.02,147.44) and (449.92,146.65) .. (449.89,145.85) .. controls (449.47,135.37) and (461.9,126.36) .. (477.65,125.74) .. controls (493.4,125.12) and (506.5,133.11) .. (506.92,143.59) .. controls (506.93,143.97) and (506.93,144.35) .. (506.91,144.72) ;  
\draw  [draw opacity=0] (498,130.23) .. controls (498.81,132.36) and (499.21,134.68) .. (499.14,137.09) .. controls (498.83,147.57) and (489.56,155.98) .. (478.43,155.85) .. controls (467.31,155.73) and (458.54,147.12) .. (458.86,136.63) .. controls (458.93,134.22) and (459.47,131.92) .. (460.41,129.81) -- (479,136.86) -- cycle ; \draw   (498,130.23) .. controls (498.81,132.36) and (499.21,134.68) .. (499.14,137.09) .. controls (498.83,147.57) and (489.56,155.98) .. (478.43,155.85) .. controls (467.31,155.73) and (458.54,147.12) .. (458.86,136.63) .. controls (458.93,134.22) and (459.47,131.92) .. (460.41,129.81) ;  
\draw   (132,271.22) .. controls (132,255.48) and (151.7,242.72) .. (176,242.72) .. controls (200.3,242.72) and (220,255.48) .. (220,271.22) .. controls (220,286.96) and (200.3,299.72) .. (176,299.72) .. controls (151.7,299.72) and (132,286.96) .. (132,271.22) -- cycle ;
\draw  [draw opacity=0] (147.2,277.2) .. controls (147.02,276.44) and (146.92,275.65) .. (146.89,274.85) .. controls (146.47,264.37) and (158.9,255.36) .. (174.65,254.74) .. controls (190.4,254.12) and (203.5,262.11) .. (203.92,272.59) .. controls (203.93,272.97) and (203.93,273.35) .. (203.91,273.72) -- (175.4,273.72) -- cycle ; \draw   (147.2,277.2) .. controls (147.02,276.44) and (146.92,275.65) .. (146.89,274.85) .. controls (146.47,264.37) and (158.9,255.36) .. (174.65,254.74) .. controls (190.4,254.12) and (203.5,262.11) .. (203.92,272.59) .. controls (203.93,272.97) and (203.93,273.35) .. (203.91,273.72) ;  
\draw  [draw opacity=0] (195,259.23) .. controls (195.81,261.36) and (196.21,263.68) .. (196.14,266.09) .. controls (195.83,276.57) and (186.56,284.98) .. (175.43,284.85) .. controls (164.31,284.73) and (155.54,276.12) .. (155.86,265.63) .. controls (155.93,263.22) and (156.47,260.92) .. (157.41,258.81) -- (176,265.86) -- cycle ; \draw   (195,259.23) .. controls (195.81,261.36) and (196.21,263.68) .. (196.14,266.09) .. controls (195.83,276.57) and (186.56,284.98) .. (175.43,284.85) .. controls (164.31,284.73) and (155.54,276.12) .. (155.86,265.63) .. controls (155.93,263.22) and (156.47,260.92) .. (157.41,258.81) ;  
\draw [color={rgb, 255:red, 208; green, 2; blue, 27 }  ,draw opacity=1 ]   (128,281) -- (138,272.72) ;
\draw [color={rgb, 255:red, 208; green, 2; blue, 27 }  ,draw opacity=1 ]   (138,282) -- (130,272.72) ;
\draw   (338,269.22) .. controls (338,253.48) and (357.7,240.72) .. (382,240.72) .. controls (406.3,240.72) and (426,253.48) .. (426,269.22) .. controls (426,284.96) and (406.3,297.72) .. (382,297.72) .. controls (357.7,297.72) and (338,284.96) .. (338,269.22) -- cycle ;
\draw  [draw opacity=0] (353.2,275.2) .. controls (353.02,274.44) and (352.92,273.65) .. (352.89,272.85) .. controls (352.47,262.37) and (364.9,253.36) .. (380.65,252.74) .. controls (396.4,252.12) and (409.5,260.11) .. (409.92,270.59) .. controls (409.93,270.97) and (409.93,271.35) .. (409.91,271.72) -- (381.4,271.72) -- cycle ; \draw   (353.2,275.2) .. controls (353.02,274.44) and (352.92,273.65) .. (352.89,272.85) .. controls (352.47,262.37) and (364.9,253.36) .. (380.65,252.74) .. controls (396.4,252.12) and (409.5,260.11) .. (409.92,270.59) .. controls (409.93,270.97) and (409.93,271.35) .. (409.91,271.72) ;  
\draw  [draw opacity=0] (401,257.23) .. controls (401.81,259.36) and (402.21,261.68) .. (402.14,264.09) .. controls (401.83,274.57) and (392.56,282.98) .. (381.43,282.85) .. controls (370.31,282.73) and (361.54,274.12) .. (361.86,263.63) .. controls (361.93,261.22) and (362.47,258.92) .. (363.41,256.81) -- (382,263.86) -- cycle ; \draw   (401,257.23) .. controls (401.81,259.36) and (402.21,261.68) .. (402.14,264.09) .. controls (401.83,274.57) and (392.56,282.98) .. (381.43,282.85) .. controls (370.31,282.73) and (361.54,274.12) .. (361.86,263.63) .. controls (361.93,261.22) and (362.47,258.92) .. (363.41,256.81) ;  
\draw   (426,273) .. controls (426,261.95) and (438.09,253) .. (453,253) .. controls (467.91,253) and (480,261.95) .. (480,273) .. controls (480,284.05) and (467.91,293) .. (453,293) .. controls (438.09,293) and (426,284.05) .. (426,273) -- cycle ;
\draw [color={rgb, 255:red, 208; green, 2; blue, 27 }  ,draw opacity=1 ]   (483,281) -- (475,271.72) ;
\draw [color={rgb, 255:red, 208; green, 2; blue, 27 }  ,draw opacity=1 ]   (474,281) -- (484,272.72) ;
\draw   (143,403.36) .. controls (143,388.09) and (162.48,375.72) .. (186.5,375.72) .. controls (210.52,375.72) and (230,388.09) .. (230,403.36) .. controls (230,418.62) and (210.52,431) .. (186.5,431) .. controls (162.48,431) and (143,418.62) .. (143,403.36) -- cycle ;
\draw [color={rgb, 255:red, 208; green, 2; blue, 27 }  ,draw opacity=1 ]   (232,408) -- (224,398.72) ;
\draw [color={rgb, 255:red, 208; green, 2; blue, 27 }  ,draw opacity=1 ]   (224,408) -- (234,399.72) ;

\draw [color={rgb, 255:red, 208; green, 2; blue, 27 }  ,draw opacity=1 ]   (147,407) -- (139,397.72) ;
\draw [color={rgb, 255:red, 208; green, 2; blue, 27 }  ,draw opacity=1 ]   (139,407) -- (149,398.72) ;

\draw   (352,402.36) .. controls (352,387.09) and (371.48,374.72) .. (395.5,374.72) .. controls (419.52,374.72) and (439,387.09) .. (439,402.36) .. controls (439,417.62) and (419.52,430) .. (395.5,430) .. controls (371.48,430) and (352,417.62) .. (352,402.36) -- cycle ;
\draw [color={rgb, 255:red, 208; green, 2; blue, 27 }  ,draw opacity=1 ]   (527,409) -- (519,399.72) ;
\draw [color={rgb, 255:red, 208; green, 2; blue, 27 }  ,draw opacity=1 ]   (519,409) -- (529,400.72) ;

\draw [color={rgb, 255:red, 208; green, 2; blue, 27 }  ,draw opacity=1 ]   (356,406) -- (348,396.72) ;
\draw [color={rgb, 255:red, 208; green, 2; blue, 27 }  ,draw opacity=1 ]   (348,406) -- (358,397.72) ;

\draw   (439,401.36) .. controls (439,386.09) and (458.48,373.72) .. (482.5,373.72) .. controls (506.52,373.72) and (526,386.09) .. (526,401.36) .. controls (526,416.62) and (506.52,429) .. (482.5,429) .. controls (458.48,429) and (439,416.62) .. (439,401.36) -- cycle ;
\draw   (261,464.72) .. controls (281,454.72) and (405,454.72) .. (397,472.72) .. controls (389,490.72) and (385,486.72) .. (389,524.72) .. controls (393,562.72) and (360,561.72) .. (361,533.72) .. controls (362,505.72) and (323,483.72) .. (322,533.72) .. controls (321,583.72) and (302.57,521.8) .. (293,504.72) .. controls (283.43,487.63) and (268.02,558.02) .. (258,543) .. controls (247.98,527.98) and (241,474.72) .. (261,464.72) -- cycle ;
\draw   (317,552.72) .. controls (327,550.72) and (333.31,608.8) .. (355,607.72) .. controls (376.69,606.64) and (360.67,552.62) .. (377,553.72) .. controls (393.33,554.82) and (414.27,636.45) .. (406,644.72) .. controls (397.73,652.99) and (390,646.72) .. (353,662.72) .. controls (316,678.72) and (264,658.72) .. (246,647.72) .. controls (228,636.72) and (250,574.72) .. (258,549.72) .. controls (266,524.72) and (265.93,611.28) .. (286,606.72) .. controls (306.07,602.16) and (307,554.72) .. (317,552.72) -- cycle ;

\draw (81,125.4) node [anchor=north west][inner sep=0.75pt]    {$M$};
\draw (318,127.4) node [anchor=north west][inner sep=0.75pt]    {$N$};
\draw (74,264.4) node [anchor=north west][inner sep=0.75pt]    {$Ba$};
\draw (304,263.4) node [anchor=north west][inner sep=0.75pt]    {$Bb$};
\draw (79,405.4) node [anchor=north west][inner sep=0.75pt]    {$Ca$};
\draw (308,404.4) node [anchor=north west][inner sep=0.75pt]    {$Cb$};
\draw (148,533.4) node [anchor=north west][inner sep=0.75pt]    {$D$};

\end{tikzpicture}
\caption{Genus two stable curves. The red points are node where the local equation is $xy=0$ for the singularity.} 
\label{stable}
  \end{center}
  \end{figure}
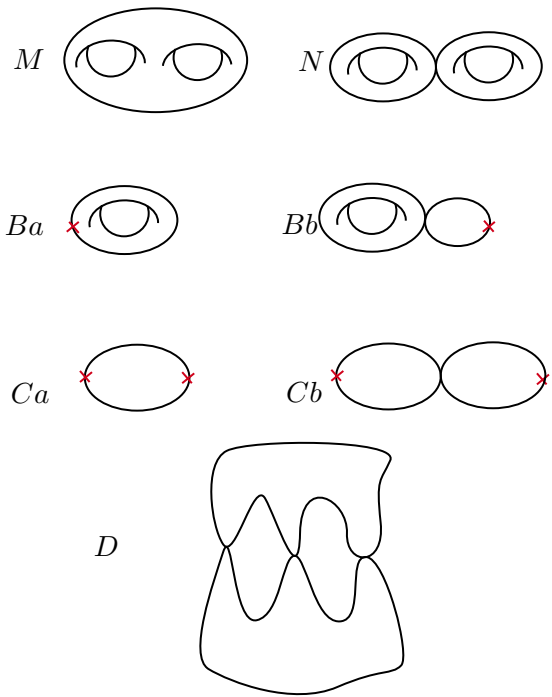

 The study of the conjugacy class of monodromy groups is left for section. \ref{monodromy}. Here let's explain the data $\tau(0)$ which is the fixed point of monodromy group. 
 $\tau(0)$ is represented by the genus two stable curves. A stable curve is a curve whose only singularities are nodes and whose smooth rational components 
  contain at least three singular points of the curve. For genus two, there are following classes of stable curves: M: an irreducible non-singular hypelliptic curve;
  N: two elliptic curves meeting at one point transversally; Ba): an elliptic curve with one ordinary double point; Bb): an elliptic curve and a rational curve with ordinary double point,
  the above two components intersect transversely; Ca): A rational curve with two ordinary double points; Cb): Two rational curves with one ordinary double points intersecting 
  transversely; D): Two non-singular rational curves meeting at three points transversally. See figure. \ref{stable} for the configuration. 
  
There are a total of five classes of genus two degenerations, see table. \ref{fivetypes}.
The meaning of the classification is following: a) First the conjugacy class of monodromy group is classified as elliptic if it is semi-simple (finite order, namely $M^k=I$ for some k), and 
 parabolic if it is of infinite order; b): The limiting point of the period matrix  $\tau_{ij}(0)$ correspond to a type of stable curve, see figure. \ref{stable}; c): the invariant $m$ is the most subtle one which 
can be defined using the limiting behavior of the period map and its intersection with the divisor $\bar{N}$ ($=N\cup Bb\cup Cb$ \footnote{These three types have separating node: namely one get two disconnected components by taking off
the node.}) in the space of stable curves, see \cite{namikawa1973complete} for further explanation of this number.

 \begin{table}[htp]
\begin{center}
\begin{tabular}{|c|c|c|c|}
\hline
Type & Monodromy group & Fixed point & degree \\ \hline
Elliptic [1]& Elliptic & M &$m=0$ \\ \hline
Elliptic [2] & Elliptic &N & $m\geq 0$ \\ \hline
Parabolic [3]& Infinite & B & $m\geq 0$ \\ \hline
Parabolic [4] & Infinite & C & $m\geq 0$ \\ \hline
Parabolic [5] & Infinite &D &$m= 0$ \\ \hline
\end{tabular}
\end{center}
\caption{Five classes of genus two singular fibers.}
\label{fivetypes}
\end{table}%

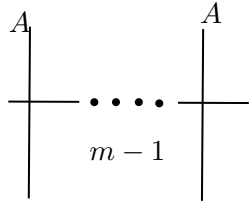
\begin{figure}[H]
\begin{center}

\tikzset{every picture/.style={line width=0.75pt}} 

\begin{tikzpicture}[x=0.45pt,y=0.45pt,yscale=-1,xscale=1]

\draw    (212,298.22) -- (213,154.22) ;
\draw    (195,217) -- (254,217.22) ;
\draw  [fill={rgb, 255:red, 0; green, 0; blue, 0 }  ,fill opacity=1 ] (264,217.5) .. controls (264,216.12) and (265.12,215) .. (266.5,215) .. controls (267.88,215) and (269,216.12) .. (269,217.5) .. controls (269,218.88) and (267.88,220) .. (266.5,220) .. controls (265.12,220) and (264,218.88) .. (264,217.5) -- cycle ;
\draw  [fill={rgb, 255:red, 0; green, 0; blue, 0 }  ,fill opacity=1 ] (282,217.5) .. controls (282,216.12) and (283.12,215) .. (284.5,215) .. controls (285.88,215) and (287,216.12) .. (287,217.5) .. controls (287,218.88) and (285.88,220) .. (284.5,220) .. controls (283.12,220) and (282,218.88) .. (282,217.5) -- cycle ;
\draw  [fill={rgb, 255:red, 0; green, 0; blue, 0 }  ,fill opacity=1 ] (301,217.5) .. controls (301,216.12) and (302.12,215) .. (303.5,215) .. controls (304.88,215) and (306,216.12) .. (306,217.5) .. controls (306,218.88) and (304.88,220) .. (303.5,220) .. controls (302.12,220) and (301,218.88) .. (301,217.5) -- cycle ;
\draw  [fill={rgb, 255:red, 0; green, 0; blue, 0 }  ,fill opacity=1 ] (318,218.5) .. controls (318,217.12) and (319.12,216) .. (320.5,216) .. controls (321.88,216) and (323,217.12) .. (323,218.5) .. controls (323,219.88) and (321.88,221) .. (320.5,221) .. controls (319.12,221) and (318,219.88) .. (318,218.5) -- cycle ;
\draw    (336,218) -- (395,218.22) ;
\draw    (356,300.22) -- (357,156.22) ;

\draw (193,139.4) node [anchor=north west][inner sep=0.75pt]    {$A$};
\draw (352,132.4) node [anchor=north west][inner sep=0.75pt]    {$A$};
\draw (260,247.4) node [anchor=north west][inner sep=0.75pt]    {$m-1$};

\end{tikzpicture}

\end{center}
\caption{The dual graph for $[2]-I_0-I_0-m$ singular fiber.  The invariant $m$ for the  singular fiber is given  by the   $A_{m-1}$ chain of rational curves with self-intersection  number  $-2$.}
\label{deg}
\end{figure}

 The invariant $m$ deserves further explanation.  First of all, let's look the degeneration of type $[2]-I_0-I_0-m$ (We use the label used in \cite{namikawa1973complete}). The period mapping has the following form
 \begin{equation*}
 \tau =\left( \begin{array}{cc}
 a(t)&t^m\\
 t^m& b(t) \\
 \end{array}\right),~~Im(a(0))>0,~~Im(b(0))>0.
 \end{equation*}
Here the number $m$ is just the degree for the degeneration. We notice that the number $m$ does not affect the limiting value of the period matrix, which is determined by $a(0)$ and $b(0)$ only. The dual graph of this singular fiber 
is shown in figure. \ref{deg}, and the number $m$ is related to the number of $-2$ rational curves in between two elliptic curves (they form a chain of $A_{m-1}$ Dynkin diagram). $a(0)$ and $b(0)$ gives the moduli of two elliptic curves. 

\subsection{Monodromy group and Coulomb branch spectrum}
\label{monodromy}
We'd like to derive the possible scaling dimension from the data of monodromy group of a singular fiber.
The conjugacy class of $Sp(4,Z)$ has been classified, see \cite{eie1984dimensions}. Here we'd like to focus on the so-called elliptic conjugacy class, i.e. the 
conjugacy class with the property $M^k=I$; they are also called conjugacy class with finite order. The eigenvalues for 
these conjugacy classes are roots of unity. The characteristic polynomial and the minimal polynomial for them are listed in table. \ref{character}.
\begin{table}
\begin{center}
\begin{tabular}{|c|c|}
\hline
Characteristic polynomial $\Psi(x)$ & minimal polynomial $P(x)$ \\ \hline
$(x\pm 1)^4$ & $(x\pm1)$ \\ \hline
$(x-1)^2(x+1)^2$ & $x^2-1$ \\ \hline
$ (x^2+1)^2$ & $x^2+1$ \\ \hline
$(x^2\pm x+1)^2$ & $x^2\pm x+1$ \\ \hline
 $(x^2+1)(x\pm1)^2$ & $(x^2+1)(x\pm 1)$ \\ \hline
 $(x\pm1)^2(x^2+x+1)$ & $(x\pm 1) (x^2+x+1)$ \\ \hline
 $(x\pm1)^2(x^2-x+1)$ & $(x\pm 1) (x^2-x+1)$ \\ \hline
$x^4+1,~x^4\pm x^2+1,~(x^2+1)(x^2\pm x+1),~~x^4\pm x^3 +x^2 \pm x +1$ & $~~$ \\ \hline
\end{tabular}
\end{center}
\caption{Characteristic polynomial and its minimal polynomial for elliptic conjugacy class of $Sp(4,Z)$ group.}
\label{character}
\end{table}%
The Coulomb branch spectrum are related to singularity spectrum by the formula \ref{scale}.
Using the limit mixed Hodge structure, one can define 
a set of rational numbers $(\alpha_1,\alpha_2,\alpha_3, \alpha_4)$ called spectrum \cite{kulikov1998mixed}, and its relation to the eigenvalue of the monodromy group is 
\begin{equation}
\lambda_i=\exp(2\pi i \alpha_i).
\label{spec1}
\end{equation}
An important consistent condition is that the spectral numbers are in pair
\begin{equation}
\alpha_i+\alpha_j=0.
\label{spec2}
\end{equation}

Even without knowing the exact limit mixed Hodge structure, one can get possible scaling dimension by using the information of eigenvalues and the two constraints \ref{spec1}, \ref{spec2}.
The data is listed in table. \ref{scalingdimension} (We assume that there is no eigenvalue one.), where  the exponent $e_i=\alpha_i+1$ is shown. The pairing for the exponent takes the form $e_i+e_j=2$. The formula
for scaling dimension is  changed 
\begin{equation*}
[u_i]={1+e_{min}-e_{i}\over e_{min}}.
\end{equation*}
and the pairing for $[u_i]$ is $[u_i]+[u_j]=2$. Only the scaling dimension with $[u_i]>1$ is listed in table.  \ref{scalingdimension}, and others can be found by using the pairing  symmetry.

\begin{table}
\begin{center}
\resizebox{5.5in}{!}{\begin{tabular}{|l|p{22mm}|p{15mm}|c|}
\hline
 Eigenvalues & Possible exponents & Scaling dimension &Characteristic polynomial \\ \hline
  $(\exp(\pi i),\exp(\pi i),\exp(\pi i),\exp(\pi i))$ & $[\frac{1}{2}, \frac{1}{2}, \frac{3}{2}, \frac{3}{2}]$ & $[2,2]$ & $(x+1)^4$  \\ \hline
 $(\exp(\pi i/2),\exp(\pi i/2),\exp(3\pi i/2),\exp(3\pi i/2))$ & $[\frac{1}{4},\frac{1}{4},\frac{7}{4},\frac{7}{4}]$ \newline $[\frac{1}{4},\frac{3}{4},\frac{5}{4},\frac{7}{4}]$ \newline $[\frac{3}{4},\frac{3}{4},\frac{5}{4},\frac{5}{4}]$ 
   & $[4,4]$ \newline $[4,2]$ \newline $[\frac{4}{3}, \frac{4}{3}]$ & $(x^2+1)^2$
     \\ \hline
 $(\exp(2\pi i/3),\exp(2\pi i/3),\exp(4\pi i/3),\exp(4\pi i/3))$ & $[\frac{1}{3},\frac{1}{3},\frac{5}{3},\frac{5}{3}]$ \newline $[\frac{1}{3},\frac{2}{3},\frac{4}{3},\frac{5}{3}]$ \newline $[\frac{2}{3},\frac{2}{3},\frac{4}{3},\frac{4}{3}]$ 
   & $[3,3]$ \newline $[3,2]$ \newline $[\frac{3}{2}, \frac{3}{2}]$ & $(x^2+x+1)^2$
     \\ \hline
 $(\exp(2\pi i/6),\exp(2\pi i/6),\exp(10\pi i/6),\exp(10\pi i/6))$ & $[\frac{1}{6},\frac{1}{6},\frac{11}{6},\frac{11}{6}]$ \newline $[\frac{1}{6},\frac{5}{6},\frac{7}{6},\frac{11}{6}]$ \newline $[\frac{5}{6},\frac{5}{6},\frac{7}{6},\frac{7}{6}]$ 
   & $[6,6]$ \newline $[6,2]$ \newline $[\frac{6}{5}, \frac{6}{5}]$ &$(x^2-x+1)^2$
     \\ \hline    
       $(\exp(2\pi i/4),\exp(6\pi i/4),\exp(\pi i),\exp(\pi i))$ & $[\frac{1}{4},\frac{1}{2},\frac{3}{2},\frac{7}{4}]$ \newline $[\frac{1}{2},\frac{3}{4},\frac{5}{4},\frac{3}{2}]$  
   & $[4,3]$ \newline $[2,\frac{3}{2}]$ & $(x+1)^2(x^2+1)$
     \\ \hline 
      $(\exp(2\pi i/3),\exp(4\pi i/3),\exp(\pi i),\exp(\pi i))$ & $[\frac{1}{3},\frac{1}{2},\frac{3}{2},\frac{5}{3}]$ \newline $[\frac{1}{2},\frac{2}{3},\frac{4}{3},\frac{3}{2}]$  
   & $[3,\frac{5}{2}]$ \newline $[2,\frac{5}{3}]$ & $(x+1)^2(x^2+x+1)$
     \\ \hline      
  $(\exp(2\pi i/6),\exp(10\pi i/6),\exp(\pi i),\exp(\pi i))$ & $[\frac{1}{6},\frac{1}{2},\frac{3}{2},\frac{11}{6}]$ \newline $[\frac{1}{2},\frac{5}{6},\frac{7}{6},\frac{3}{2}]$ 
   & $[6,4]$ \newline $[2,\frac{4}{3}]$ & $(x+1)^2(x^2-x+1)$
     \\ \hline  
       $(\exp(2\pi i/8),\exp(6\pi i/8),\exp(10\pi i/8),\exp(14 \pi i/8))$ & $[\frac{1}{8},\frac{3}{8},\frac{13}{8},\frac{15}{8}]$ \newline $[\frac{1}{8},\frac{5}{8},\frac{11}{8},\frac{15}{8}]$  \newline 
       $[\frac{3}{8},\frac{7}{8},\frac{9}{8},\frac{13}{8}]$ \newline $[\frac{5}{8},\frac{7}{8},\frac{9}{8},\frac{11}{8}]$
   & $[8,6]$ \newline $[8,4]$ \newline $[\frac{8}{3},\frac{4}{3}]$  \newline $[\frac{8}{5},\frac{6}{5}]$& $x^4+1$
     \\ \hline  
      $(\exp(2\pi i/12),\exp(10\pi i/12),\exp(22\pi i/12),\exp(14 \pi i/12))$ & $[\frac{1}{12},\frac{5}{12},\frac{19}{12},\frac{23}{12}]$ \newline $[\frac{1}{12},\frac{7}{12},\frac{17}{12},\frac{23}{12}]$  \newline 
       $[\frac{5}{12},\frac{11}{12},\frac{13}{12},\frac{19}{12}]$ \newline $[\frac{7}{12},\frac{11}{12},\frac{13}{12},\frac{17}{12}]$
   & $[12,8]$ \newline $[12,6]$ \newline $[\frac{12}{5},\frac{6}{5}]$  \newline $[\frac{12}{7},\frac{8}{7}]$& $x^4-x^2+1$
     \\ \hline   
       $(\exp(2\pi i/3),\exp(4\pi i/3),\exp(2\pi i/6),\exp(10 \pi i/6))$ & $[\frac{1}{6},\frac{1}{3},\frac{5}{3},\frac{11}{6}]$ \newline $[\frac{1}{6},\frac{2}{3},\frac{4}{3},\frac{11}{6}]$  \newline 
       $[\frac{1}{3},\frac{5}{6},\frac{7}{6},\frac{5}{3}]$ \newline $[\frac{2}{3},\frac{5}{6},\frac{7}{6},\frac{4}{3}]$
   & $[6,5]$ \newline $[6,3]$ \newline $[3,\frac{3}{2}]$  \newline $[\frac{3}{2},\frac{5}{4}]$& $x^4+x^2+1$
     \\ \hline  
       $(\exp(2\pi i/4),\exp(6\pi i/4),\exp(2\pi i/3),\exp(4 \pi i/3))$ & $[\frac{1}{4},\frac{1}{3},\frac{5}{3},\frac{7}{4}]$ \newline $[\frac{1}{4},\frac{2}{3},\frac{4}{3},\frac{7}{4}]$  \newline 
       $[\frac{1}{3},\frac{3}{4},\frac{5}{4},\frac{5}{3}]$ \newline $[\frac{2}{3},\frac{3}{4},\frac{5}{4},\frac{4}{3}]$
   & $[4,\frac{11}{3}]$ \newline $[4,\frac{7}{3}]$ \newline $[3,\frac{7}{4}]$  \newline $[\frac{3}{2},\frac{23}{8}]$& $(x^2+1)(x^2+x+1)$
     \\ \hline 
       $(\exp(2\pi i/6),\exp(10\pi i/6),\exp(2\pi i/4),\exp(6 \pi i/4))$ & $[\frac{1}{6},\frac{1}{4},\frac{7}{4},\frac{11}{6}]$ \newline $[\frac{1}{6},\frac{3}{4},\frac{5}{4},\frac{11}{6}]$  \newline 
       $[\frac{1}{4},\frac{5}{6},\frac{7}{6},\frac{7}{4}]$ \newline $[\frac{3}{4},\frac{5}{6},\frac{7}{6},\frac{5}{4}]$
   & $[6,\frac{11}{2}]$ \newline $[6,\frac{5}{2}]$ \newline $[4,\frac{5}{3}]$  \newline $[\frac{4}{3},\frac{11}{9}]$& $(x^2+1)(x^2-x+1)$
     \\ \hline    
     $(\exp(2\pi i/5),\exp(4\pi i/5),\exp(8\pi i/5),\exp(6 \pi i/5))$ & $[\frac{1}{5},\frac{2}{5},\frac{8}{5},\frac{9}{5}]$ \newline $[\frac{1}{5},\frac{3}{5},\frac{7}{5},\frac{9}{5}]$  \newline
     $[\frac{2}{5},\frac{4}{5},\frac{6}{5},\frac{8}{5}]$ \newline $[\frac{3}{5},\frac{4}{5},\frac{6}{5},\frac{7}{5}]$ 
        & $[5,4]$ \newline $[5,3]$ \newline $[\frac{5}{2},\frac{3}{2}]$ \newline $[\frac{5}{3},\frac{4}{3}]$& $x^4+x^3+x^2+x+1$
     \\ \hline  
     $(\exp(2\pi i/10),\exp(6\pi i/10),\exp(14\pi i/10),\exp(18 \pi i/10))$ & $[\frac{1}{10},\frac{3}{10},\frac{17}{10},\frac{19}{10}]$ \newline $[\frac{1}{10},\frac{7}{10},\frac{13}{10},\frac{19}{10}]$  \newline 
       $[\frac{3}{10},\frac{9}{10},\frac{11}{10},\frac{17}{10}]$ \newline $[\frac{7}{10},\frac{9}{10},\frac{11}{10},\frac{13}{10}]$
   & $[10,8]$ \newline $[10,4]$ \newline $[\frac{10}{3},\frac{4}{3}]$  \newline $[\frac{10}{7},\frac{8}{7}]$& $x^4-x^3+x^2-x+1$
     \\ \hline  
\end{tabular}}
\end{center}
\caption{The possible scaling dimension given the characteristic polynomial of the elliptic conjugacy class. }
\label{scalingdimension}
\end{table}%

In the above computations, we assume that the  limit mixed Hodge structure is irreducible. For some cases, the limit is reducible and so the monodromy group is 
the direct sum of two $SL(2,Z)$ groups. The characteristic polynomial is then the product of two degree two polynomials. One can get 
the scaling dimension by simply using the result of rank one case, see \cite{Xie:2022lcm}.

For the parabolic conjugacy class, one can also find the possible scaling dimension by looking at the eigenvalues. The method is completely  same.
The dual graph is useful in finding the precise scaling dimension, see the discussion in the section. \ref{data}.

\textbf{Isolated fixed points of monodromy group}: A conjugacy class of $Sp(4,Z)$ is called regular elliptic if it has
isolated fixed points. There are only six type of fixed points: 
\begin{align*}
&Z_1=\diag[i,i],~~Z_2=\diag[\rho,\rho],~~\rho=\exp(\pi i/3) \\ \nonumber\\
& Z_3=\diag[i,\rho],~~Z_4={i\over \sqrt{3}}\left(\begin{array}{cc} 2&1\\1&2 \end{array} \right) \nonumber\\
&Z_5=\left(\begin{array}{cc} \eta&(\eta-1)/2\\(\eta-1)/2 & \eta \end{array} \right),~~\eta=\frac{1}{3}+{2\sqrt{2} i\over 3} \nonumber\\
& Z_6=\left(\begin{array}{cc} w&w+w^{-2}\\w+w^{-2} & -w^{-1} \end{array} \right),~~w=\exp(2\pi i/5)
\end{align*}
The isotropy subgroups at $Z_i$ are groups of order $16, 36, 12,12,24,5$. These subgroups are useful for the studies of discrete gauging of 4d $\mathcal{N}=2$ SCFTs.

\subsection{Local invariants}
One can define a set of local invariant
$ n_t, \delta_x, d_x$ for a singular fiber. These invariants are very important for extracting the low energy theory associated with the singular fibers. 
They are also crucial in constructing global Coulomb branch geometry.

\textbf{The number $n_t$}: This number is given by the number of irreducible components in the dual graph, and  it can be easily computed from the dual graph.

\textbf{The number $\delta_x$}: This
number $\delta_x$ is defined as
\begin{equation*}
\delta_x=2+\epsilon(F).
\end{equation*}
here $\epsilon(F)$ is  the topological Euler number of the dual graph $\Gamma=\sum n_iC_i$. It can be computed as follows: first, the Euler number of a component $C_i$ is given as $2-2g$ ($g$ is the genus), and 
there is a minus one contribution for every double point. The total Euler number is the sum of the contributions from each component and the double points. 

\textit{Example}: Let's compute $\delta_x$ for the dual graph in figure. \ref{dual1}. There are seven component with genus zero and each contributes $2$ to the total Euler number, and there are six intersection points, so $\epsilon(F)=14-6=8$, and
$\delta_x=10$.

\textbf{The number $d_x$}:  This number is the most complicated one to define, see \cite{kenji1988discriminants} for details. To begin with, we notice that an arbitrary  one parameter genus two curves can be represented by an equation
\begin{equation*}
y^2=f(x,t).
\end{equation*}
with $f(x,t)$ a polynomial of degree $5$ or $6$ in $x$. The discriminant for a sextic $g(x)=u_0\prod_{i=1}^6(x-a_i)$ is defined as 
\begin{equation}
\Delta=u_0^{10} \prod_{i<j}(a_i-a_j)^2.
\end{equation}
Notice that there are many equations which would give the same surface. Now the basis for the holomorphic differential for a genus two hyper-elliptic curve is taken as $w_1=dx/y$ and $w_2=xdx/y$, and 
$\Delta(w_1\wedge w_2)^{10}$ is actually independent of the defining equation for the given surface \footnote{At a fixed point $t$ on the base $\mathbb{P}^1$, there is  fiber of a  rank two holomorphic vector bundle $V$  spanned by 
the differential $w_1$ and $w_2$. The antisymmetric tensor product $\Lambda^2 V$ (the basis is $w_1\wedge w_2$) gives a line bundle over  $\mathbb{P}^1$.}. Now the invariant $d_x$ is defined as 
\begin{equation*}
d_x=ord_t\Delta(w_1\wedge w_2)^{10}.
\end{equation*}
Here $ord_t$ means the order of the zero of $\Delta(w_1\wedge w_2)^{10}$ at $t=0$. A hypelliptic equation is called \textbf{minimal} equation if $ord_t\Delta =d_x$, here $\Delta$ is 
the discriminant of the defining equation.

The invariants $d_x$ and $\delta_x$ can also be similarly defined for genus one fibration, and  it can be shown  $d_x=\delta_x$. The equality does not hold
for genus two case. The difference of $d_x$ and $\delta_x$ is  given by the following equation (for all the cases except eight types):
\begin{equation*}
d_x=\delta_x+l+1.
\end{equation*}
here $l$ is a simple function of $m$ (in most cases $l=m-1$). Let's recall that, $m$ is the degree needed for the classification of degenerations.
 For the following eight types: $[1]-II,~[1]-IV,~[1]-VIII-3,~[3]-II_{n-0}^*,~[3]-IV^*-II_0,~[3]-III^*-II_0,~[5]-II_{n-p}^*,~[5]-III_n^*$, the formula is changed to
\begin{equation*}
d_x=\delta_x+l+2.
\end{equation*}

\textbf{Gauge algebra}: One can also associate a gauge algebra on the singular fiber, which can  be read from the dual graph  $F$.  The crucial idea is the balanced chain 
in the dual graph. Let's first review the notion of a balanced chain.  Given a dual graph $F=\sum n_i C_i$, a node $C_i$ is called balanced if $\sum_{j\neq i} C_{ij} n_j=2n_i$, and one can verify that 
only type $E$ component is balanced. \textbf{Proof}: Let's write the dual graph as $F=n_i C_i+\delta$, and since $C_i\cdot F=0$, we have 
\begin{align*}
& C_i\cdot (n_iC_i+\delta)=0\to (C_i\cdot C_i)n_i+C_i  \cdot \delta=0, \nonumber  \\
&\to \sum_{j\neq i} C_{ij} n_j=-(C_i\cdot C_i)n_i.
\end{align*}
so the balance condition only holds for type $E$ component (with $C_i\cdot C_i=-2$).

A connected subgraph in $F$ is called balanced chain if all of the components in this subgraph is balanced, and it can be proven that
the balanced chain form a $ADE$ or affine $ADE$ Dynkin diagram. Now the gauge algebra is read from the dual graph as follows : each balanced ADE chain (or affine ADE chain) gives a 
corresponding ADE algebra. 

\textit{Example}: Let's look at the example in figure. \ref{dual1}. There is a balanced chain forming a $D_6$ Dynkin diagram (only D component is not balanced), and so the gauge algebra of this singular fiber is $D_6$.

Physically, the gauge algebra gives the flavor symmetry of low energy theory.

\subsubsection{Local splitting of singularities}
\begin{figure}[H]
\begin{center}

\tikzset{every picture/.style={line width=0.75pt}} 

\begin{tikzpicture}[x=0.40pt,y=0.40pt,yscale=-1,xscale=1]

\draw   (93.99,409.85) .. controls (94.04,388.3) and (130.81,370.91) .. (176.12,371) .. controls (221.43,371.1) and (258.13,388.65) .. (258.08,410.2) .. controls (258.04,431.75) and (221.27,449.14) .. (175.96,449.05) .. controls (130.64,448.95) and (93.95,431.4) .. (93.99,409.85) -- cycle ;
\draw  [draw opacity=0] (228.2,406.9) .. controls (222.57,414.58) and (213.14,419.94) .. (202.17,420.67) .. controls (189.83,421.48) and (178.66,416.24) .. (172.26,407.75) -- (200.19,390.73) -- cycle ; \draw   (228.2,406.9) .. controls (222.57,414.58) and (213.14,419.94) .. (202.17,420.67) .. controls (189.83,421.48) and (178.66,416.24) .. (172.26,407.75) ;  
\draw  [draw opacity=0] (168.43,415.76) .. controls (170.4,402.97) and (182.83,392.69) .. (198.39,391.88) .. controls (215.82,390.96) and (230.58,402.27) .. (231.41,417.14) -- (199.81,418.87) -- cycle ; \draw   (168.43,415.76) .. controls (170.4,402.97) and (182.83,392.69) .. (198.39,391.88) .. controls (215.82,390.96) and (230.58,402.27) .. (231.41,417.14) ;  
\draw  [draw opacity=0] (96.81,419.35) .. controls (95.57,417.66) and (94.87,415.74) .. (94.87,413.69) .. controls (94.89,406.88) and (102.76,401.3) .. (112.44,401.24) .. controls (121.28,401.18) and (128.58,405.73) .. (129.78,411.7) -- (112.41,413.58) -- cycle ; \draw   (96.81,419.35) .. controls (95.57,417.66) and (94.87,415.74) .. (94.87,413.69) .. controls (94.89,406.88) and (102.76,401.3) .. (112.44,401.24) .. controls (121.28,401.18) and (128.58,405.73) .. (129.78,411.7) ;  
\draw  [draw opacity=0] (128.33,407.75) .. controls (129.47,409.51) and (130.06,411.47) .. (129.93,413.51) .. controls (129.5,420.32) and (121.31,425.41) .. (111.64,424.89) .. controls (102.82,424.42) and (95.8,419.44) .. (94.96,413.41) -- (112.41,412.58) -- cycle ; \draw   (128.33,407.75) .. controls (129.47,409.51) and (130.06,411.47) .. (129.93,413.51) .. controls (129.5,420.32) and (121.31,425.41) .. (111.64,424.89) .. controls (102.82,424.42) and (95.8,419.44) .. (94.96,413.41) ;  
\draw   (369,406.22) .. controls (369,390.48) and (388.7,377.72) .. (413,377.72) .. controls (437.3,377.72) and (457,390.48) .. (457,406.22) .. controls (457,421.96) and (437.3,434.72) .. (413,434.72) .. controls (388.7,434.72) and (369,421.96) .. (369,406.22) -- cycle ;
\draw  [draw opacity=0] (384.2,412.2) .. controls (384.02,411.44) and (383.92,410.65) .. (383.89,409.85) .. controls (383.47,399.37) and (395.9,390.36) .. (411.65,389.74) .. controls (427.4,389.12) and (440.5,397.11) .. (440.92,407.59) .. controls (440.93,407.97) and (440.93,408.35) .. (440.91,408.72) -- (412.4,408.72) -- cycle ; \draw   (384.2,412.2) .. controls (384.02,411.44) and (383.92,410.65) .. (383.89,409.85) .. controls (383.47,399.37) and (395.9,390.36) .. (411.65,389.74) .. controls (427.4,389.12) and (440.5,397.11) .. (440.92,407.59) .. controls (440.93,407.97) and (440.93,408.35) .. (440.91,408.72) ;  
\draw  [draw opacity=0] (432,394.23) .. controls (432.81,396.36) and (433.21,398.68) .. (433.14,401.09) .. controls (432.83,411.57) and (423.56,419.98) .. (412.43,419.85) .. controls (401.31,419.73) and (392.54,411.12) .. (392.86,400.63) .. controls (392.93,398.22) and (393.47,395.92) .. (394.41,393.81) -- (413,400.86) -- cycle ; \draw   (432,394.23) .. controls (432.81,396.36) and (433.21,398.68) .. (433.14,401.09) .. controls (432.83,411.57) and (423.56,419.98) .. (412.43,419.85) .. controls (401.31,419.73) and (392.54,411.12) .. (392.86,400.63) .. controls (392.93,398.22) and (393.47,395.92) .. (394.41,393.81) ;  
\draw   (457,405.22) .. controls (457,389.48) and (476.7,376.72) .. (501,376.72) .. controls (525.3,376.72) and (545,389.48) .. (545,405.22) .. controls (545,420.96) and (525.3,433.72) .. (501,433.72) .. controls (476.7,433.72) and (457,420.96) .. (457,405.22) -- cycle ;
\draw  [draw opacity=0] (472.2,411.2) .. controls (472.02,410.44) and (471.92,409.65) .. (471.89,408.85) .. controls (471.47,398.37) and (483.9,389.36) .. (499.65,388.74) .. controls (515.4,388.12) and (528.5,396.11) .. (528.92,406.59) .. controls (528.93,406.97) and (528.93,407.35) .. (528.91,407.72) -- (500.4,407.72) -- cycle ; \draw   (472.2,411.2) .. controls (472.02,410.44) and (471.92,409.65) .. (471.89,408.85) .. controls (471.47,398.37) and (483.9,389.36) .. (499.65,388.74) .. controls (515.4,388.12) and (528.5,396.11) .. (528.92,406.59) .. controls (528.93,406.97) and (528.93,407.35) .. (528.91,407.72) ;  
\draw  [draw opacity=0] (520,393.23) .. controls (520.81,395.36) and (521.21,397.68) .. (521.14,400.09) .. controls (520.83,410.57) and (511.56,418.98) .. (500.43,418.85) .. controls (489.31,418.73) and (480.54,410.12) .. (480.86,399.63) .. controls (480.93,397.22) and (481.47,394.92) .. (482.41,392.81) -- (501,399.86) -- cycle ; \draw   (520,393.23) .. controls (520.81,395.36) and (521.21,397.68) .. (521.14,400.09) .. controls (520.83,410.57) and (511.56,418.98) .. (500.43,418.85) .. controls (489.31,418.73) and (480.54,410.12) .. (480.86,399.63) .. controls (480.93,397.22) and (481.47,394.92) .. (482.41,392.81) ;  
\draw  [fill={rgb, 255:red, 208; green, 2; blue, 27 }  ,fill opacity=1 ] (455,405.5) .. controls (455,404.12) and (456.12,403) .. (457.5,403) .. controls (458.88,403) and (460,404.12) .. (460,405.5) .. controls (460,406.88) and (458.88,408) .. (457.5,408) .. controls (456.12,408) and (455,406.88) .. (455,405.5) -- cycle ;
\draw  [fill={rgb, 255:red, 208; green, 2; blue, 27 }  ,fill opacity=1 ] (93,411.5) .. controls (93,410.12) and (94.12,409) .. (95.5,409) .. controls (96.88,409) and (98,410.12) .. (98,411.5) .. controls (98,412.88) and (96.88,414) .. (95.5,414) .. controls (94.12,414) and (93,412.88) .. (93,411.5) -- cycle ;
\draw   (100,141.61) .. controls (100,122.5) and (134.03,107) .. (176,107) .. controls (217.97,107) and (252,122.5) .. (252,141.61) .. controls (252,160.72) and (217.97,176.22) .. (176,176.22) .. controls (134.03,176.22) and (100,160.72) .. (100,141.61) -- cycle ;
\draw    (173,134) -- (186,149.22) ;
\draw    (174,147) -- (184,133.22) ;

\draw    (288,141) -- (378,142.19) ;
\draw [shift={(380,142.22)}, rotate = 180.76] [color={rgb, 255:red, 0; green, 0; blue, 0 }  ][line width=0.75]    (10.93,-3.29) .. controls (6.95,-1.4) and (3.31,-0.3) .. (0,0) .. controls (3.31,0.3) and (6.95,1.4) .. (10.93,3.29)   ;
\draw   (414,141.61) .. controls (414,122.5) and (448.03,107) .. (490,107) .. controls (531.97,107) and (566,122.5) .. (566,141.61) .. controls (566,160.72) and (531.97,176.22) .. (490,176.22) .. controls (448.03,176.22) and (414,160.72) .. (414,141.61) -- cycle ;
\draw  [fill={rgb, 255:red, 208; green, 2; blue, 27 }  ,fill opacity=1 ] (444,128.11) .. controls (444,126.51) and (445.29,125.22) .. (446.89,125.22) .. controls (448.49,125.22) and (449.78,126.51) .. (449.78,128.11) .. controls (449.78,129.71) and (448.49,131) .. (446.89,131) .. controls (445.29,131) and (444,129.71) .. (444,128.11) -- cycle ;
\draw  [fill={rgb, 255:red, 208; green, 2; blue, 27 }  ,fill opacity=1 ] (471,122.11) .. controls (471,120.51) and (472.29,119.22) .. (473.89,119.22) .. controls (475.49,119.22) and (476.78,120.51) .. (476.78,122.11) .. controls (476.78,123.71) and (475.49,125) .. (473.89,125) .. controls (472.29,125) and (471,123.71) .. (471,122.11) -- cycle ;
\draw  [fill={rgb, 255:red, 208; green, 2; blue, 27 }  ,fill opacity=1 ] (463,152.11) .. controls (463,150.51) and (464.29,149.22) .. (465.89,149.22) .. controls (467.49,149.22) and (468.78,150.51) .. (468.78,152.11) .. controls (468.78,153.71) and (467.49,155) .. (465.89,155) .. controls (464.29,155) and (463,153.71) .. (463,152.11) -- cycle ;
\draw  [fill={rgb, 255:red, 208; green, 2; blue, 27 }  ,fill opacity=1 ] (486,144.11) .. controls (486,142.51) and (487.29,141.22) .. (488.89,141.22) .. controls (490.49,141.22) and (491.78,142.51) .. (491.78,144.11) .. controls (491.78,145.71) and (490.49,147) .. (488.89,147) .. controls (487.29,147) and (486,145.71) .. (486,144.11) -- cycle ;
\draw  [fill={rgb, 255:red, 208; green, 2; blue, 27 }  ,fill opacity=1 ] (503,127.11) .. controls (503,125.51) and (504.29,124.22) .. (505.89,124.22) .. controls (507.49,124.22) and (508.78,125.51) .. (508.78,127.11) .. controls (508.78,128.71) and (507.49,130) .. (505.89,130) .. controls (504.29,130) and (503,128.71) .. (503,127.11) -- cycle ;
\draw  [fill={rgb, 255:red, 0; green, 0; blue, 0 }  ,fill opacity=1 ] (523,147.11) .. controls (523,145.51) and (524.29,144.22) .. (525.89,144.22) .. controls (527.49,144.22) and (528.78,145.51) .. (528.78,147.11) .. controls (528.78,148.71) and (527.49,150) .. (525.89,150) .. controls (524.29,150) and (523,148.71) .. (523,147.11) -- cycle ;

\draw (444,474.4) node [anchor=north west][inner sep=0.75pt] [font=\tiny]   {$\widetilde{I_{1}}$};
\draw (164,478.4) node [anchor=north west][inner sep=0.75pt]  [font=\tiny]  {$I_{1}$};
\draw (298,112) node [anchor=north west][inner sep=0.75pt]   [align=left] [font=\tiny]  {Splitting};
\draw (467,128.4) node [anchor=north west][inner sep=0.75pt]   [font=\tiny] {$I_{1}$};
\draw (528,131.4) node [anchor=north west][inner sep=0.75pt]   [font=\tiny] {$\widetilde{I_{1}}$};

\end{tikzpicture}

\end{center}
\caption{Up: Local splitting of a singularity: the complicated singularity is split into several $I_1$ and $\tilde{I}_1$ singularities. Bottom: The geometric representation of $I_1$ and $\tilde{I}_1$ singularity; $I_1$ singularity has 
a  non-separating  node while $\tilde{I}_1$ singularity has a separating node.}
\label{split}
\end{figure}
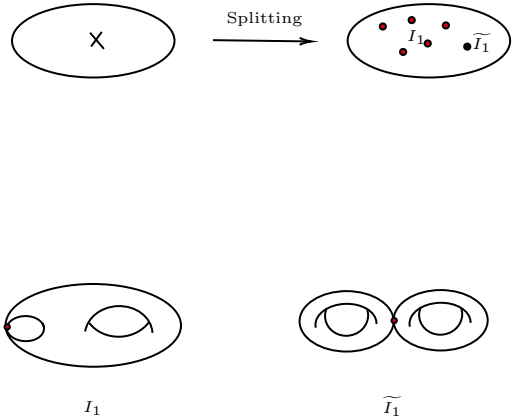

It is possible to deform a local singularity $F$ by changing the defining family. In fact, it was shown in \cite{ashikaga2000global} that each local singularity can be deformed into some number of $a_x$ $I_1$ (type $I_{[1-0-0]}$ in cite{}) singularity  and $t_x$ $\tilde{I}_1$ (type $I_{[0-0-1]}$ in \cite{namikawa1973complete}) singularity, see figure.\ref{split}.  Those numbers $a_x, t_x$ are related to the local invariants $d_x,\delta_x$ as follows:
\begin{equation}
\boxed{t_x=d_x-\delta_x,~~a_x=2\delta_x-d_x}.
\end{equation}
The dual graph for $I_1$ and $\tilde{I_1}$ singularity are shown in figure. \ref{split}, and they are called atomic singular fiber \cite{takamura2004towards}.   The local data $(d_x, \delta_x)$ for $I_1$ ($\tilde{I}_1$) singularities are $I_1=(1,1)$ ($\tilde{I}_1=(2,1)$), and it
can be easily verified the data for the original singularity $F$ is equal to the sum of the local data of atomic singularity ($I_1$ and $\tilde{I}_1$ type). 

\subsection{Low energy theory}
\label{data}
The important data for genus two singular fibers are summarized in following tables. \ref{elliptic1}, \ref{elliptic2a},\ref{elliptic2b},\ref{parabolic3},\ref{parabolic4} .  The invariants 
$d_x, \delta_x, n_t$ are included, and so the invariant $t_x=d_x-\delta_x$ can be computed directly. $n_t$ and $\delta_x$ can be found from 
the dual graph listed in \cite{namikawa1973complete}. $d_x$ has been computed in \cite{kenji1988discriminants}. 

\begin{figure}
\begin{center}

\tikzset{every picture/.style={line width=0.75pt}} 

\begin{tikzpicture}[x=0.55pt,y=0.55pt,yscale=-1,xscale=1]

\draw    (192,200) -- (387.5,201.22) ;
\draw    (203,187) .. controls (243,157) and (221.5,317.22) .. (261.5,287.22) ;
\draw    (314,177) .. controls (354,147) and (332.5,307.22) .. (372.5,277.22) ;
\draw    (187,385) -- (382.5,386.22) ;
\draw    (260,368) .. controls (300,338) and (278.5,498.22) .. (318.5,468.22) ;

\draw (200,274.4) node [anchor=north west][inner sep=0.75pt] [font=\tiny]   {$Kod_{1}$};
\draw (370,282.4) node [anchor=north west][inner sep=0.75pt]    [font=\tiny]  {$Kod_{2}$};
\draw (167,193.4) node [anchor=north west][inner sep=0.75pt]   [font=\tiny]   {$D$};
\draw (238,447.4) node [anchor=north west][inner sep=0.75pt]   [font=\tiny]   {$Kod$};
\draw (162,378.4) node [anchor=north west][inner sep=0.75pt]   [font=\tiny]   {$C$};

\end{tikzpicture}

\end{center}
\caption{The dual graph for a genus two singular fiber: it is given by two Kodaira dual graphs intersecting with a type $D$ component or type $C$ component. Kodaira's dual graph always 
has a component with multiplicity one, which is replaced by $C$  or $D$ component in the figure. }
\label{twokoda}
\end{figure}
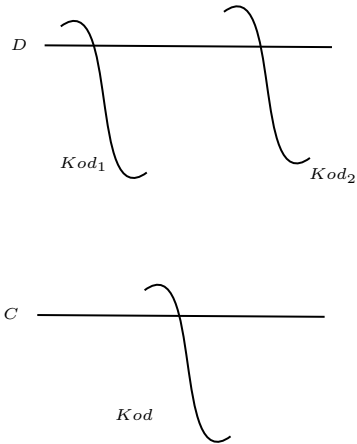

The gauge algebra is also computed using the dual graph, which should be part of the flavor symmetry of the low energy 
theory. The scaling dimension of the Coulomb branch operator is constrained by the eigenvalues of the monodromy group,
and there are actually several choices for a given set of eigenvalues, see table. \ref{scalingdimension}. 
The dual graph plays an essential role for us to find the exact scaling dimension and the low energy theory. First of all, if the dual graph is formed by 
two Kodaira's graphs intersected with a type $D$ or $C$ component (see figure. \ref{twokoda}), it is natural to guess the low energy theory 
is just the direct sum of two rank one theories.  The map between Kodaira's singularity and low energy theory is listed in table. \ref{simple}.
More generally, one can find a 3d mirror from the dual graph, which is then 
used to find the low energy theory.

\begin{table}[htp]
\begin{center}
\resizebox{2.5in}{!}{
\begin{tabular}{|c|c|}
\hline 
Kodaira's list &Theory  \\ \hline
$II^*$ & $E_8$ SCFT\\ \hline
$III^*$ & $E_7$ SCFT \\ \hline
$IV^*$ & $E_6$ SCFT \\ \hline
$II$ & $H_0$ SCFT \\ \hline
$III$ &  $H_1$ SCFT \\ \hline
$IV$ &  $H_2$ SCFT \\ \hline
$I_n$ & $U(1)-n$ \\ \hline
$I_n^*$ & $SU(2)-(n+4)$\\ \hline
\end{tabular}}
\end{center}
\caption{The theory associated with Kodaira's list of rank one singular fiber. The $E_n$ type theories were found in \cite{Minahan:1996cj}, and the $H_i$ type theories were found in \cite{Argyres:1995xn}.}
\label{simple}
\end{table}

\subsubsection{3d mirror from dual graph}
Let's first review 3d mirror symmetry relevant for our context. One can compactify 4d $\mathcal{N}=2$ theory on a circle to get a 3d $\mathcal{N}=4$ theory $A$. 3d $\mathcal{N}=4$ theory has 
interesting mirror symmetry property: there is a mirror theory $B$ such that the Higgs branch of $B$ is identified with the Coulomb branch
of $A$ and vice versa. In our case, the hyperkahler dimension of the theory $A$ is two, and so the Higgs branch of the mirror theory  $B$
should  have dimension 2.  

The way of attaching a quiver gauge theory for a singular fiber $F=\sum n_i C_i$ is following: a): attach a quiver node for each component $C_i$, and the
gauge group is $U(n_i)$; b): the number of bi-fundamental hypermultiplets between two quiver nodes $i,j$ is  $C_{ij}$; c): There is one adjoint hypermultiplet for genus one component. 
Let's now compute the Higgs branch of the constructed quiver (there is an added one because the overall $U(1)$ gauge group is decoupled; we also leave the extra adjoint hypermultiplet aside for a while.):
\begin{align*}
& dim Higgs=  {1\over 2} \sum_{i\neq j} n_i n_j C_{ij}-\sum n_i^2+1 \nonumber\\
& ={1\over2} (\sum n_i C_i)\cdot (\sum n_i C_i)+{1\over2}\sum(-C_i^2 n_i^2-2n_i^2 )+1 \nonumber\\
&={1\over2}\sum_{C_i \neq E}(-C_i^2 n_i^2-2n_i^2 )+1.
\end{align*}
Here the fact $C_i\cdot F=0$ is used. By looking at the data listed in table. \ref{component}, one find (one need to add the contribution of adjoint hypermultiplet for the $2A$ case):
\begin{align*}
& dim Higgs=3,~~\text{for type}~~2B \nonumber\\
&dim Higgs=2,~~\text{for type}~~D,~(B,B), \nonumber\\
& dim Higgs=1,~~\text{for type}~~C,~(A,B) \nonumber\\
& dim Higgs=0,~~\text{for type}~~(A,A) \nonumber\\
& dim Higgs=1,~~\text{for type}~~2A\nonumber\\
\end{align*}
After some experiments, the  following rules for finding the  3d mirror are found:
\begin{enumerate}
\item For type $D$ and $(B,B)$, the 3d mirror is just the quiver gauge theory found from the dual graph.
\item For type $2B$, if there is a type $E$ component with multiplicity one which intersects  $2B$ component, then the 3d mirror is found by removing the $U(1)$ node corresponding to $E$.
One can easily see that the  Higgs branch of the new quiver has dimension two. One can do similar thing for type $2A$.
\item For type $C$, the 3d mirror is just that of a rank one theory represented by Kodaira dual graph.
\end{enumerate}
We do not know how to extract 3d mirror for other type of dual graphs.

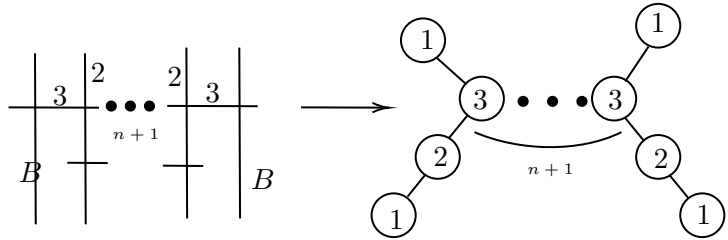
\begin{figure}[H]
\begin{center}

\tikzset{every picture/.style={line width=0.75pt}} 

\begin{tikzpicture}[x=0.55pt,y=0.55pt,yscale=-1,xscale=1]

\draw    (63,201) -- (124.5,201.22) ;
\draw    (81,164) -- (81.5,281.22) ;
\draw    (220,160) -- (220.5,277.22) ;
\draw    (115,164) -- (115.5,281.22) ;
\draw    (184,160) -- (184.5,277.22) ;
\draw    (171,200) -- (232.5,200.22) ;
\draw    (103,239) -- (130.5,239.22) ;
\draw    (168,241) -- (195.5,241.22) ;
\draw  [fill={rgb, 255:red, 0; green, 0; blue, 0 }  ,fill opacity=1 ] (129.78,200.11) .. controls (129.78,198.39) and (131.17,197) .. (132.89,197) .. controls (134.61,197) and (136,198.39) .. (136,200.11) .. controls (136,201.83) and (134.61,203.22) .. (132.89,203.22) .. controls (131.17,203.22) and (129.78,201.83) .. (129.78,200.11) -- cycle ;
\draw  [fill={rgb, 255:red, 0; green, 0; blue, 0 }  ,fill opacity=1 ] (142.78,200.11) .. controls (142.78,198.39) and (144.17,197) .. (145.89,197) .. controls (147.61,197) and (149,198.39) .. (149,200.11) .. controls (149,201.83) and (147.61,203.22) .. (145.89,203.22) .. controls (144.17,203.22) and (142.78,201.83) .. (142.78,200.11) -- cycle ;
\draw  [fill={rgb, 255:red, 0; green, 0; blue, 0 }  ,fill opacity=1 ] (155.78,200.11) .. controls (155.78,198.39) and (157.17,197) .. (158.89,197) .. controls (160.61,197) and (162,198.39) .. (162,200.11) .. controls (162,201.83) and (160.61,203.22) .. (158.89,203.22) .. controls (157.17,203.22) and (155.78,201.83) .. (155.78,200.11) -- cycle ;
\draw    (262,201) -- (319.5,201.21) ;
\draw [shift={(321.5,201.22)}, rotate = 180.21] [color={rgb, 255:red, 0; green, 0; blue, 0 }  ][line width=0.75]    (10.93,-3.29) .. controls (6.95,-1.4) and (3.31,-0.3) .. (0,0) .. controls (3.31,0.3) and (6.95,1.4) .. (10.93,3.29)   ;
\draw   (370,195) .. controls (370,186.72) and (376.72,180) .. (385,180) .. controls (393.28,180) and (400,186.72) .. (400,195) .. controls (400,203.28) and (393.28,210) .. (385,210) .. controls (376.72,210) and (370,203.28) .. (370,195) -- cycle ;
\draw   (460,195) .. controls (460,186.72) and (466.72,180) .. (475,180) .. controls (483.28,180) and (490,186.72) .. (490,195) .. controls (490,203.28) and (483.28,210) .. (475,210) .. controls (466.72,210) and (460,203.28) .. (460,195) -- cycle ;
\draw   (490,235) .. controls (490,226.72) and (496.72,220) .. (505,220) .. controls (513.28,220) and (520,226.72) .. (520,235) .. controls (520,243.28) and (513.28,250) .. (505,250) .. controls (496.72,250) and (490,243.28) .. (490,235) -- cycle ;
\draw   (520,275) .. controls (520,266.72) and (526.72,260) .. (535,260) .. controls (543.28,260) and (550,266.72) .. (550,275) .. controls (550,283.28) and (543.28,290) .. (535,290) .. controls (526.72,290) and (520,283.28) .. (520,275) -- cycle ;
\draw   (490,145) .. controls (490,136.72) and (496.72,130) .. (505,130) .. controls (513.28,130) and (520,136.72) .. (520,145) .. controls (520,153.28) and (513.28,160) .. (505,160) .. controls (496.72,160) and (490,153.28) .. (490,145) -- cycle ;
\draw   (330,155) .. controls (330,146.72) and (336.72,140) .. (345,140) .. controls (353.28,140) and (360,146.72) .. (360,155) .. controls (360,163.28) and (353.28,170) .. (345,170) .. controls (336.72,170) and (330,163.28) .. (330,155) -- cycle ;
\draw   (340,235) .. controls (340,226.72) and (346.72,220) .. (355,220) .. controls (363.28,220) and (370,226.72) .. (370,235) .. controls (370,243.28) and (363.28,250) .. (355,250) .. controls (346.72,250) and (340,243.28) .. (340,235) -- cycle ;
\draw   (310,275) .. controls (310,266.72) and (316.72,260) .. (325,260) .. controls (333.28,260) and (340,266.72) .. (340,275) .. controls (340,283.28) and (333.28,290) .. (325,290) .. controls (316.72,290) and (310,283.28) .. (310,275) -- cycle ;
\draw  [fill={rgb, 255:red, 0; green, 0; blue, 0 }  ,fill opacity=1 ] (410,196.89) .. controls (410,195.17) and (411.39,193.78) .. (413.11,193.78) .. controls (414.83,193.78) and (416.22,195.17) .. (416.22,196.89) .. controls (416.22,198.61) and (414.83,200) .. (413.11,200) .. controls (411.39,200) and (410,198.61) .. (410,196.89) -- cycle ;
\draw  [fill={rgb, 255:red, 0; green, 0; blue, 0 }  ,fill opacity=1 ] (430,196.89) .. controls (430,195.17) and (431.39,193.78) .. (433.11,193.78) .. controls (434.83,193.78) and (436.22,195.17) .. (436.22,196.89) .. controls (436.22,198.61) and (434.83,200) .. (433.11,200) .. controls (431.39,200) and (430,198.61) .. (430,196.89) -- cycle ;
\draw  [fill={rgb, 255:red, 0; green, 0; blue, 0 }  ,fill opacity=1 ] (450,196.89) .. controls (450,195.17) and (451.39,193.78) .. (453.11,193.78) .. controls (454.83,193.78) and (456.22,195.17) .. (456.22,196.89) .. controls (456.22,198.61) and (454.83,200) .. (453.11,200) .. controls (451.39,200) and (450,198.61) .. (450,196.89) -- cycle ;
\draw    (482.5,208.22) -- (495.5,224.22) ;
\draw    (512.5,247.22) -- (525.5,263.22) ;
\draw    (482.5,183.22) -- (498.5,159.22) ;
\draw    (362.5,222.22) -- (375.5,206.22) ;
\draw    (334.5,262.22) -- (346.5,247.22) ;
\draw    (374.5,185.22) -- (354.5,167.22) ;
\draw    (380,218) .. controls (410.5,231.22) and (451.5,232.22) .. (480,218) ;

\draw (226,241.4) node [anchor=north west][inner sep=0.75pt]    {$B$};
\draw (68,236.4) node [anchor=north west][inner sep=0.75pt]    {$B$};
\draw (117,168.4) node [anchor=north west][inner sep=0.75pt]    {$2$};
\draw (169,171.4) node [anchor=north west][inner sep=0.75pt]    {$2$};
\draw (91,184.4) node [anchor=north west][inner sep=0.75pt]    {$3$};
\draw (195,182.4) node [anchor=north west][inner sep=0.75pt]    {$3$};
\draw (132,213.4) node [anchor=north west][inner sep=0.75pt]  [font=\tiny]  {$n+1$};
\draw (350,226.4) node [anchor=north west][inner sep=0.75pt]    {$2$};
\draw (500,227.4) node [anchor=north west][inner sep=0.75pt]    {$2$};
\draw (529,266.4) node [anchor=north west][inner sep=0.75pt]    {$1$};
\draw (320,269.4) node [anchor=north west][inner sep=0.75pt]    {$1$};
\draw (469,187.4) node [anchor=north west][inner sep=0.75pt]    {$3$};
\draw (414,237.4) node [anchor=north west][inner sep=0.75pt]  [font=\tiny]  {$n+1$};
\draw (341,146.4) node [anchor=north west][inner sep=0.75pt]    {$1$};
\draw (498,134.4) node [anchor=north west][inner sep=0.75pt]    {$1$};
\draw (377,187.4) node [anchor=north west][inner sep=0.75pt]    {$3$};

\end{tikzpicture}

\end{center}
\caption{Left: the dual graph for singularity $III_n$; right: the 3d mirror for the low energy theory associated with the singular fiber $III_n$.}
\label{su3mirror}
\end{figure}

\textbf{Example 2}: The dual graph for $III_n$ type singular fiber is shown in figure. \ref{su3mirror}. We recognize that it is the 3d mirror
for $SU(3)$ gauge theory coupled with $n+6$ fundamental flavors \cite{Nanopoulos:2010bv}, which is therefore the low energy theory associated with 
the singularity $III_n$.

\subsubsection{List of tables}
The monodromy group, invariants $d_x, \delta_x,  n_t$ for every genus two singular fibers are included. The scaling dimension, gauge algebra, and the theory 
are included. The fixed points and dual graph for each singular fiber can be found in \cite{namikawa1973complete}. We only show  data for  singular fiber with $m=0$, with the only 
exception $I_0-I_0-1$ ($\tilde{I}_1$) fiber.

\begin{table}[H]
\begin{center}
\caption{Elliptic fiber [1].}
\label{elliptic1}
\resizebox{4.0in}{!}{\begin{tabular}{|c|c|c|c|c|c|c|c|c|} \hline
Type &  Monodromy & $d_x$ & $\delta_x$ & $l$ & Theory & Scaling dimension & Gauge algebra &$n_t$ (components)\\  \hline 
$I_{0-0-0}$& $ \left(\begin{array}{cccc} 1&0&0&0\\0&1&0&0\\0&0&1&0\\0&0&0&1\end{array}\right)$ & 0 & 0 & -1 & $U(1)^2$ & $(1,1)$ & $\emptyset$ &1 \\ \hline
$I^*_{0-0-0}$ &$ \left(\begin{array}{cccc} -1&0&0&0\\0&-1&0&0\\0&0&-1&0\\0&0&0&-1\end{array}\right)$& $10$ & $10$ & $-1$ & $2-SU(2)-SU(2)-2$ & $(2,2)$ & $A_1^6$&7 \\ \hline
\textcolor{red}{$II$} &$ \left(\begin{array}{cccc} 0&1&0&0\\1&0&0&0\\0&0&0&1\\0&0&1&0\end{array}\right)$& $5$ & $4$ & $-1$ & $?$ & $(2,1)$ & $A_1^2$&3  \\ \hline
$III$ &$ \left(\begin{array}{cccc} 0&-1&0&0\\1&-1&0&0\\0&0&-1&-1\\0&0&1&0\end{array}\right)$& $10$ & $10$ & $-1$ & $SU(3)-6$ & $(3,2)$ & $A_5$ &7 \\ \hline
\textcolor{red}{$IV$}& $ \left(\begin{array}{cccc} 0&1&0&0\\-1&1&0&0\\0&0&1&1\\0&0&-1&0\end{array}\right)$& $10$ & $9$ & $-1$ & $?$ & $(6,2)$ & $D_5$&6 \\ \hline
$V$ & $ \left(\begin{array}{cccc} 0&0&1&0\\0&0&1&1\\-1&1&0&0\\0&-1&0&0\end{array}\right)$& $5$ & $5$ & $-1$ & SCFT & $(\frac{6}{4}, \frac{5}{4})$ & $~$ &2 \\ \hline
\textcolor{red}{$V^*$ }& $ \left(\begin{array}{cccc} 0&0&-1&0\\0&0&-1&-1\\1&-1&0&0\\0&1&0&0\end{array}\right)$& $15$ & $15$ & $-1$ & SCFT & $(6,5)$ & $A_{11}$ &12\\ \hline
$VI$ &$ \left(\begin{array}{cccc} 0&0&-1&0\\0&0&-1&-1\\1&-1&0&0\\0&1&0&0\end{array}\right)$& $10$ & $10$ & $-1$ & $Sp(4)-6(\textbf{F})$ & $(4,2)$ & $D_6$ & 7 \\ \hline
$VII$ & $ \left(\begin{array}{cccc} 0&1&1&0\\1&0&0&-1\\0&0&0&-1\\0&0&1&0\end{array}\right)$ & $5$ & $5$ & $-1$ & $SCFT$ & $(\frac{8}{5}, \frac{6}{5})$ & $A_1$ &2 \\ \hline
$VII^*$ & $ \left(\begin{array}{cccc} 0&-1&-1&0\\1&0&0&1\\0&0&0&1\\0&0&-1&0\end{array}\right)$& $15$ & $15$ & $-1$ & $SCFT$ & $(8, 6)$ & $ D_{10}\oplus A_1$ &12 \\ \hline
$VIII-1$ & $ \left(\begin{array}{cccc} 0&1&1&0\\1&0&0&1\\-1&1&1&0\\0&-1&0&0\end{array}\right)$& $4$ & $4$ & $-1$ & $SCFT$ & $(\frac{10}{7},\frac{8}{7})$ & $\emptyset$&1 \\ \hline
$VIII-2$ & $ \left(\begin{array}{cccc} 0&0&1&1\\-1&1&1&0\\0&-1&0&0\\0&0&-1&0\end{array}\right)$& $12$ & $12$ & $-1$ & $SCFT$ & $(10,4)$ & $E_8$ &9 \\ \hline
\textcolor{red}{$VIII-3$ }&$ \left(\begin{array}{cccc} 0&-1&-1&-1\\0&0&-1&0\\0&0&0&-1\\1&0&0&1\end{array}\right)$& $8$ & $7$ & $-1$ & $SCFT$ & $(\frac{10}{3}, \frac{4}{3})$ & $A_3$&4 \\ \hline
\textcolor{red}{$VIII-4$ }& $ \left(\begin{array}{cccc} 1&0&-1&0\\0&0&0&-1\\1&0&0&1\\-1&1&1&0\end{array}\right)$&$16$ & $16$ & $-1$ & $SCFT$ & $(10,8)$ & $D_{12}$ & $13$ \\ \hline
$I_X-1$ &  $ \left(\begin{array}{cccc} 0&1&1&1\\0&0&1&0\\0&0&0&1\\-1&0&0&-1\end{array}\right)$ &$8$ & $8$ & $-1$  & $D_2(SU(5))$ & $(\frac{5}{2},\frac{3}{2})$ & $A_4$ &5 \\ \hline
$I_X-2$ &  $ \left(\begin{array}{cccc} -1&0&1&0\\0&0&0&1\\-1&0&0&-1\\1&-1&-1&0\end{array}\right)$ &$6$ & $6$ & $-1$  & $SCFT$ & $(\frac{5}{3},\frac{4}{3})$ & $A_1$ &3   \\ \hline
$I_X-3$ & $ \left(\begin{array}{cccc} 0&-1&-1&0\\-1&0&0&-1\\1&-1&-1&0\\0&1&0&0\end{array}\right)$& $14$ & $14$ &$-1$ &$SCFT$ & $(5,4)$ & $A_9\oplus A_1$  &11 \\ \hline
$I_X-4$ &$ \left(\begin{array}{cccc} 0&0&-1&-1\\1&-1&-1&0\\0&1&0&0\\0&0&1&0\end{array}\right)$& $12$ & $12$ & $-1$ & $SCFT$ & $(5,3)$ & $D_7\oplus A_1$ & 9 \\ \hline
\end{tabular}}
\end{center}
\end{table}%

\begin{table}[H]
\begin{center}
\caption{Elliptic fiber [2].}
\label{elliptic2a}
\resizebox{4.0in}{!}{\begin{tabular}{|c|c|c|c|c|c|c|c|c|} \hline
Type &  Monodromy & $d_x$ & $\delta_x$ & $l$ & Theory & Scaling dimension & Gauge algebra & $n_t$ (components)\\  \hline 
$I_0-I_0-1$& $ \left(\begin{array}{cccc} 1&0&0&0\\0&1&0&0\\0&0&1&0\\0&0&0&1\end{array}\right)$ & 2 & 1 & 0 & $U(1)^2$ & $(1,1)$ & $\emptyset$  &2 \\ \hline
$I^*_0-I_0^*-0$ &$ \left(\begin{array}{cccc} -1&0&0&0\\0&-1&0&0\\0&0&-1&0\\0&0&0&-1\end{array}\right)$& $12$ & $12$ & $-1$ & $D_4 \oplus D_4$ & $(2,2)$ & $D_4\oplus D_4$ &9 \\ \hline
$I_0-I_0^*-0$  &$ \left(\begin{array}{cccc} 1&0&0&0\\0&-1&0&0\\0&0&1&0\\0&0&0&-1\end{array}\right)$& $6$ & $6$ & $-1$ & $D_4\oplus U(1)$ & $(2,1)$ & $D_4$ &5 \\ \hline
$2I_0-0$ &$ \left(\begin{array}{cccc} 0&1&0&0\\1&0&0&0\\0&0&0&1\\0&0&1&0\end{array}\right)$& $7$ & $5$ & $1$ & $?$ & $(2,1)$ & $D_3$ &4 \\ \hline
\textcolor{red}{$2I_0^*-0$}& $ \left(\begin{array}{cccc} 0&-1&0&0\\1&0&0&0\\0&0&0&-1\\0&0&1&0\end{array}\right)$& $11$ & $10$ & $0$ & $SCFT$ & $(4,2)$ & $D_4\oplus A_1^2$ &7\\ \hline
$I_0-II-0$ & $ \left(\begin{array}{cccc} 1&0&1&0\\0&1&0&0\\-1&0&0&0\\0&0&0&1\end{array}\right)$& $2$ & $2$ & $-1$ & $H_0\oplus U(1)$ & $(\frac{6}{5}, 1)$ & $\emptyset$ &1 \\ \hline
$I_0-II^*-0$ & $ \left(\begin{array}{cccc} 0&0&-1&0\\0&1&0&0\\1&0&1&0\\0&0&0&1\end{array}\right)$& $10$ & $10$ & $-1$ & $E_8\oplus U(1)$ & $(6,1)$ & $E_8$ &10 \\ \hline
$I_0-IV-0$ &$ \left(\begin{array}{cccc} 0&0&-1&0\\0&0&-1&-1\\1&-1&0&0\\0&1&0&0\end{array}\right)$& $4$ & $4$ & $-1$ & $H_2\oplus U(1)$ & $(\frac{3}{2},1)$ & $A_2$&4 \\ \hline
$I_0-IV^*-0$ & $ \left(\begin{array}{cccc} -1&0&-1&0\\0&1&0&0\\-1&0&-1&0\\0&0&0&1\end{array}\right)$ & $8$ & $8$ & $-1$ & $E_6\oplus U(1)$ & $(3, 1)$ & $E_6$ &8\\ \hline
$I_0^*-II-0$ & $ \left(\begin{array}{cccc}1&0&1&0\\0&-1&0&0\\-1&0&0&0\\0&0&0&-1\end{array}\right)$& $8$ & $8$ & $-1$ & $D_4\oplus  H_0$ & $(2, \frac{6}{5})$ & $ D_4$&5 \\ \hline
$I_0^*-II^*-0$ & $ \left(\begin{array}{cccc} 0&0&-1&0\\0&-1&0&0\\1&0&1&0\\0&0&0&-1\end{array}\right)$& $16$ & $16$ & $-1$ & $D_4\oplus  E_8$ & $(6,2)$ & $E_8\oplus  D_4$&14 \\ \hline
\textcolor{red}{$I_0^*-II^*-\alpha$} & $ ~$& $14$ & $14$ & $-1$ & $?$ & $(6,2)$? & $E_7\oplus A_1^3$ &11  \\ \hline
$I_0^*-IV-0$ &$ \left(\begin{array}{cccc} 0&0&1&0\\0&-1&0&0\\-1&0&-1&0\\0&0&0&-1\end{array}\right)$& $10$ & $10$ & $-1$ & $D_4\oplus  H_2$ & $(2, \frac{3}{2})$ & $D_4\oplus A_2$ &7\\ \hline
$I_0^*-IV^*-0$ & $ \left(\begin{array}{cccc} -1&0&-1&0\\0&-1&0&0\\1&0&0&0\\0&0&0&-1\end{array}\right)$&$14$ & $14$ & $-1$ & $D_4\oplus  E_6$ & $(2,3)$ & $D_4\oplus E_6$ &12 \\ \hline
\textcolor{red}{$I_0^*-IV^*-\alpha$ }&  $ ~$ &$12$ & $12$ & $-1$  & $?$ & $(3,2)$? & $A_5\oplus A_1^3$ &9 \\ \hline
$I_0-III-0$ &  $ \left(\begin{array}{cccc} 0&0&1&0\\0&1&0&0\\-1&0&0&0\\0&0&0&1\end{array}\right)$ &$3$ & $3$ & $-1$  & $U(1)\oplus H_1$ & $(\frac{4}{3},1)$ & $ A_1$ &3  \\ \hline
$I_0-III^*-0$ & $ \left(\begin{array}{cccc} 0&0&-1&0\\0&1&0&0\\1&0&0&0\\0&0&0&1\end{array}\right)$& $9$ & $9$ &$-1$ &$U(1)\oplus E_7$ & $(4,1)$ & $E_7$ & 9   \\ \hline
$I_0^*-III-0$ &$ \left(\begin{array}{cccc} 0&0&1&0\\0&-1&0&0\\-1&0&0&0\\0&0&0&-1\end{array}\right)$& $9$ & $9$ & $-1$ & $D_4\oplus  H_1$ & $(2,\frac{4}{3})$ & $D_4\oplus A_1$ & 6 \\ \hline
$I_0^*-III^*-0$ &$ \left(\begin{array}{cccc} 0&0&-1&0\\0&-1&0&0\\1&0&0&0\\0&0&0&-1\end{array}\right)$& $15$ & $15$ & $-1$ & $D_4\oplus  E_7$ & $(4,2)$ & $D_4\oplus E_7$  &12\\ \hline
\textcolor{red}{$I_0^*-III^*-\alpha$} &$ ~$& $13$ & $13$ & $-1$ & $?$ & $(4,2)$? & $D_6\oplus A_1^3$ &10  \\ \hline
\textcolor{red}{$2II-0$}& $ \left(\begin{array}{cccc} 0&1&0&1\\1&0&0&0\\0&-1&0&0\\0&0&1&0\end{array}\right)$ & 7 & 6 & 0 & $SCFT$ & $(\frac{12}{5},\frac{6}{5})$ & $A_1^2$ &3  \\ \hline
\textcolor{red}{$2II^*-0$} &$ \left(\begin{array}{cccc} 0&0&0&-1\\1&0&0&0\\0&1&0&1\\0&0&1&0\end{array}\right)$& $15$ & $14$ & $0$ & $SCFT$ & $(12,6)$ & $E_8\oplus A_1^2$ &11 \\ \hline
$II-II-0$  &$ \left(\begin{array}{cccc} 1&0&1&0\\0&1&0&1\\-1&0&0&0\\0&-1&0&0\end{array}\right)$& 4 & 4 & $-1$ & $H_0\oplus H_0$ & $(\frac{6}{5},\frac{6}{5})$ & $\emptyset$ &1 \\ \hline
$II-II^*-0$  &$ \left(\begin{array}{cccc} 1&0&1&0\\0&0&0&-1\\-1&0&0&0\\0&1&0&1\end{array}\right)$& $12$ & $12$ & $-1$ & $H_0\oplus E_8$ & $(6,\frac{6}{5})$ & $E_8$ &12 \\ \hline
$II^*-II^*-0$& $ \left(\begin{array}{cccc} 0&0&-1&0\\0&0&0&-1\\1&0&1&0\\0&1&0&1\end{array}\right)$& $20$ & $20$ & $-1$ & $E_8\oplus E_8$ & $(6,6)$ & $E_8\oplus E_8$ &19 \\ \hline
\textcolor{red}{$II^*-II^*-\alpha$} & ~& $18$ & $18$ & $-1$ & $?$ & $?$ & $E_7\oplus E_7$ &15 \\ \hline
$II-IV-0$ & $ \left(\begin{array}{cccc} 0&0&1&0\\0&1&0&1\\-1&0&-1&0\\0&-1&0&0\end{array}\right)$& $6$ & $6$ & $-1$ & $H_0\oplus H_2$ & $(\frac{3}{2},\frac{6}{5})$ & $A_2$ &3 \\ \hline
\end{tabular}}
\end{center}
\end{table}%

\begin{table}[H]
\begin{center}
\caption{Elliptic fiber [2]-continued.}
\label{elliptic2b}
\resizebox{4.0in}{!}{\begin{tabular}{|c|c|c|c|c|c|c|c|c|} \hline
Type &  Monodromy & $d_x$ & $\delta_x$ & $l$ & Theory & Scaling dimension & Gauge algebra & $n_t$ (components)\\  \hline 
$II-IV^*-0$ &$ \left(\begin{array}{cccc} 1&0&1&0\\0&-1&0&-1\\-1&0&0&0\\0&1&0&0\end{array}\right)$& $10$ & $10$ & $-1$ & $H_0\oplus E_6$ & $(3,\frac{6}{5})$ & $E_6$ &8 \\ \hline
$II^*-IV-0$ & $ \left(\begin{array}{cccc} -1&0&-1&0\\0&1&0&0\\-1&0&-1&0\\0&0&0&1\end{array}\right)$ & $14$ & $14$ & $-1$ & $H_2\oplus E_8$ & $(6, \frac{3}{2})$ & $E_8\oplus A_2$ &11\\ \hline
\textcolor{red}{$II^*-IV-\alpha$} & ~& $12$ & $12$ & $-1$ & $?$ & $?$ & $ E_7$ &9  \\ \hline
$II^*-IV^*-0$ & $ \left(\begin{array}{cccc} 0&0&-1&0\\0&-1&0&-1\\1&0&1&0\\0&1&0&0\end{array}\right)$& $18$ & $18$ & $-1$ & $E_6\oplus  E_8$ & $(6,3)$ & $E_8\oplus E_6$ & 17 \\ \hline
\textcolor{red}{$II^*-IV^*-\alpha$ }& $ ~$& $16$ & $16$ & $-1$ & $?$ & $?$ & $E_7\oplus A_5$ & 13\\ \hline
\textcolor{red}{$2IV-0$} &$ \left(\begin{array}{cccc} 0&0&0&1\\1&0&0&0\\0&-1&0&-1\\0&0&1&0\end{array}\right)$& $9$ & $8$ & $0$ & $SCFT$ & $(3, \frac{3}{2})$ & $A_2\oplus A_1^2$ &5 \\ \hline
\textcolor{red}{$2IV^*-0$ }& $ \left(\begin{array}{cccc} 0&-1&0&-1\\1&0&0&0\\0&1&0&0\\0&0&1&0\end{array}\right)$&$13$ & $12$ & $0$ & $SCFT$ & $(6,3)$ & $ E_6 \oplus A_1^2$ & 9 \\ \hline
$IV-IV-0$ &   $ \left(\begin{array}{cccc} 0&0&1&0\\0&0&0&1\\-1&0&-1&0\\0&-1&0&-1\end{array}\right)$&$8$ & $8$ & $-1$  & $H_2\oplus H_2$ & $(\frac{3}{2},\frac{3}{2})$ & $A_2\oplus A_2$ & 5 \\ \hline
$IV-IV^*-0$ &  $ \left(\begin{array}{cccc} 0&0&1&0\\0&-1&0&-1\\-1&0&-1&0\\0&1&0&0\end{array}\right)$ &$12$ & $12$ & $-1$  & $E_6\oplus H_2$ & $(3,\frac{3}{2})$ & $ A_1$   &10 \\ \hline
$IV^*-IV^*-0$ & $ \left(\begin{array}{cccc} -1&0&-1&0\\0&-1&0&-1\\1&0&0&0\\0&1&0&0\end{array}\right)$& $16$ & $16$ &$-1$ &$E_6\oplus E_6$ & $(3,3)$ & $E_6\oplus E_6$  & 15  \\ \hline
\textcolor{red}{$IV^*-IV^*-\alpha$} &~& $14$ & $14$ & $-1$ & $?$ & $?$ & $A_5\oplus A_5$ &11 \\ \hline
$II-III-0$ &$ \left(\begin{array}{cccc} 1&0&1&0\\0&0&0&-1\\-1&0&0&0\\0&1&0&0\end{array}\right)$& $5$ & $5$ & $-1$ & $H_0\oplus  H_1$ & $(\frac{4}{3},\frac{6}{5})$ & $A_1$ &2 \\ \hline
$II-III^*-0$ &$  \left(\begin{array}{cccc} 1&0&1&0\\0&0&0&1\\-1&0&0&0\\0&-1&0&0\end{array}\right)$& $11$ & $11$ & $-1$ & $H_0\oplus E_7$ & $(4,\frac{6}{5})$ & $E_7$ & 9 \\ \hline
$II^*-III-0$ &$  \left(\begin{array}{cccc} 0&0&-1&0\\0&0&0&1\\1&0&1&0\\0&-1&0&0\end{array}\right)$& $13$ & $13$ & $-1$ & $E_8\times H_1$ & $(6,\frac{4}{3})$ & $E_8\oplus A_1$ &11  \\ \hline
\textcolor{red}{$II^*-III-\alpha$ }&$ ~$& $11$ & $11$ & $-1$ & $?$ & $?$ & $E_7$ &8 \\ \hline
$II^*-III^*-0$& $ \left(\begin{array}{cccc} 0&0&-1&0\\0&0&0&-1\\1&0&1&0\\0&1&0&0\end{array}\right)$ & 19 & 19 & -1 & $E_8\oplus E_7$ & $(6,4)$ & $E_8\oplus E_7$ &17  \\ \hline
\textcolor{red}{$II^*-III^*-\alpha$ }&$ ~$ & $17$ & $17$  &$-1$ & $?$ & $(6,4)$? & $E_7\oplus D_6$ & 14 \\ \hline
$IV-III-0$  &$ \left(\begin{array}{cccc} 0&0&1&0\\0&0&0&1\\-1&0&-1&0\\0&-1&0&0\end{array}\right)$& $7$ & $7$ & $-1$ & $H_1\oplus H_2$ & $(\frac{3}{2},\frac{4}{3})$ & $A_2\oplus A_1$ &4  \\ \hline
$IV-III^*-0$  &$ \left(\begin{array}{cccc} 0&0&1&0\\0&0&0&-1\\-1&0&-1&0\\0&1&0&0\end{array}\right)$& $13$ & $13$ & $1$ & $H_1\oplus E_7$ & $(\frac{3}{2},4)$ & $A_2\times E_7$ &11 \\ \hline
\textcolor{red}{$IV-III^*-\alpha$}& $ ~$& $11$ & $11$ & $0$ & $?$ & $(4,3)$? & $D_6$ & 8 \\ \hline
$IV^*-III-0$ & $ \left(\begin{array}{cccc} -1&0&-1&0\\0&0&0&1\\1&0&0&0\\0&-1&0&0\end{array}\right)$& $11$ & $11$ & $-1$ & $H_1\oplus E_6$ & $(3,\frac{4}{3})$ & $E_6\oplus A_1$ &9 \\ \hline
$IV^*-III^*-0$ & $ \left(\begin{array}{cccc} -1&0&-1&0\\0&0&0&-1\\1&0&0&0\\0&1&0&0\end{array}\right)$& $17$ & $17$ & $-1$ & $E_7\oplus E_6$ & $(4,3)$ & $E_7\oplus E_6$ & 16 \\ \hline
\textcolor{red}{$IV^*-III^*-\alpha$} &$~$& $15$ & $15$ & $-1$ & $?$ & $?$ & $A_5\oplus D_6$& 12  \\ \hline
\textcolor{red}{$2III-0$} & $ \left(\begin{array}{cccc} 0&0&0&1\\1&0&0&0\\0&-1&0&0\\0&0&1&0\end{array}\right)$ & $8$ & $7$ & $0$ & $SCFT$ & $(\frac{8}{3}, \frac{4}{3})$ & $A_1\oplus A_1^2$ & 4\\ \hline
\textcolor{red}{$2III^*-0$} & $ \left(\begin{array}{cccc}0&0&0&-1\\1&0&0&0\\0&1&0&0\\0&0&1&0\end{array}\right)$& $14$ & $13$ & $0$ & $SCFT$ & $(8, 4)$ & $ E_7\oplus A_1^2$ & 10 \\ \hline
$III-III-0$ & $ \left(\begin{array}{cccc} 0&0&1&0\\0&0&0&1\\-1&0&0&0\\0&-1&0&0\end{array}\right)$& $6$ & $6$ & $-1$ & $H_1\oplus  H_1$ & $(\frac{4}{3},\frac{4}{3})$ & $A_1\oplus A_1$ & 3\\ \hline
$III-III^*-0$ & $ \left(\begin{array}{cccc} 0&0&1&0\\0&0&0&-1\\-1&0&0&0\\0&1&0&0\end{array}\right)$& $12$ & $12$ & $-1$ & $E_7\oplus  H_1$ & $(4,\frac{4}{3})$ & $E_7\oplus  A_1$ & 10 \\ \hline
$III^*-III^*-0$ &$ \left(\begin{array}{cccc} 0&0&-1&0\\0&0&0&-1\\1&0&0&0\\0&1&0&0\end{array}\right)$& $18$ & $18$ & $-1$ & $E_7\oplus E_7$ & $(4,4)$ & $E_7\oplus E_7$ & 17 \\ \hline
\textcolor{red}{$III^*-III^*-\alpha$ }&$~$& $16$ & $16$ & $-1$ & $?$ & $?$ & $D_6\oplus D_6$ & 13 \\ \hline
\end{tabular}}
\end{center}
\end{table}%

\begin{table}[H]
\begin{center}
\caption{Parabolic fiber [3].}
\label{parabolic3}
\resizebox{4.5in}{!}{\begin{tabular}{|c|c|c|c|c|c|c|c|c|} \hline
Type &  Monodromy & $d_x$ & $\delta_x$ & $l$ & Theory & Scaling dimension & Gauge algebra & $n_t$(components)\\  \hline 
$I_{n-0-0}$& $ \left(\begin{array}{cccc} 1&0&0&0\\0&1&0&n\\0&0&1&0\\0&0&0&1\end{array}\right)$ & $n $& $n$ & -1 & $U(1)-n\oplus U(1)$ & $(1,1)$ & $A_{n-1}$ & $n$  \\ \hline
$I_0-I_n^*-0$  &$ \left(\begin{array}{cccc} 1&0&0&0\\0&-1&0&-n\\0&0&1&0\\0&0&0&-1\end{array}\right)$& $n+6$ & $n+6$ & $-1$ & $SU(2)-(n+4)\oplus U(1)$ & $(2,1)$ & $D_{n+4}$ & $n+5$  \\ \hline
$I_n-I_0^*-0$  &$ \left(\begin{array}{cccc} -1&0&0&0\\0&1&0&n\\0&0&-1&0\\0&0&0&1\end{array}\right)$& $n+6$ & $n+6$ & $-1$ & $SU(2)-(4)\oplus U(1)-n$ & $(2,1)$ & $D_4\oplus A_{n-1}$ & $n+4$  \\ \hline
$I_{n-0-0}^*$  &$ \left(\begin{array}{cccc} -1&0&0&0\\0&-1&0&-n\\0&0&-1&0\\0&0&0&-1\end{array}\right)$&$n+10$ & $n+10$ & $-1$ & $(n+2)-SU(2)-SU(2)-2$ & $(2,2)$ & $D_{n+2} \times A_1^3$ & $n+7$ \\ \hline
$I_{0}^*-I_n^*-0$  &$ \left(\begin{array}{cccc}  -1&0&0&0\\0&-1&0&-n\\0&0&-1&0\\0&0&0&-1\end{array}\right)$&$n+12$ & $n+12$ & $-1$ & $SU(2)-4 \oplus SU(2)-(n+4)$ & $(2,2)$ & $D_4\times D_{n+4} $ & $n+9$ \\ \hline
$II_{n-0}$ & $ \left(\begin{array}{cccc} -1&0&0&0\\-1&1&0&n\\0&0&-1&-1\\0&0&0&1\end{array}\right)$& $n+5$ & $n+5$ & $-1$ & $(n-1)-U(1)-SU(2)-3$ & $(2,1)$ & $A_{n-2}\oplus D_3$ & $n+3$ \\ \hline
$II_{n-0}^*$ & $ \left(\begin{array}{cccc} 1&0&0&0\\1&-1&0&-n\\0&0&1&1\\0&0&0&-1\end{array}\right)$& $n+5$ & $n+4$ & $-1$ & $?$ & $(2,1)$ & $D_{n+2}$ & $n+3$ \\ \hline
$II-I_n-0$&$ \left(\begin{array}{cccc} 1&0&1&0\\0&1&0&n\\-1&0&0&0\\0&0&0&1\end{array}\right)$& $n+2$ & $n+2$ & $-1$ & $H_0\oplus I_n$ & $(\frac{6}{5},1)$ & $A_{n-1}$ & $n$ \\ \hline
$II^*-I_n-0$ & $ \left(\begin{array}{cccc} 0&0&-1&0\\0&1&0&n\\1&0&1&0\\0&0&0&1\end{array}\right)$ & $n+10$ & $n+10$ & $-1$ & $E_8\oplus I_n$ & $(6, 1)$ & $E_8\oplus A_{n-1}$ & $n+9$ \\ \hline
$IV-I_n-0$ & $ \left(\begin{array}{cccc}0&0&1&0\\0&1&0&n\\-1&0&-1&0\\0&0&0&1\end{array}\right)$& $n+4$ & $n+4$ & $-1$ & $H_2\oplus I_n$ & $(\frac{3}{2}, 1)$ & $ A_2\oplus A_{n-1}$ & $n+2$ \\ \hline
$IV^*-I_n-0$ & $ \left(\begin{array}{cccc} -1&0&-1&0\\0&1&0&n\\1&0&0&0\\0&0&0&1\end{array}\right)$& $n+8$ & $n+8$ & $-1$ & $E_6\oplus  I_n$ & $(3,1)$ & $E_6\oplus A_{n-1}$ & $n+7$\\ \hline
$II-I_n^*-0$ & $ \left(\begin{array}{cccc} 1&0&1&0\\0&-1&0&-n\\-1&0&0&0\\0&0&0&-1\end{array}\right)$& $n+8$ & $n+8$ & $-1$ & $H_0\oplus  I_n^*$ & $(2,\frac{6}{5})$ & $ D_{n+4}$ & $n+5$ \\ \hline
$II^*-I_n^*-0$ &$ \left(\begin{array}{cccc} 0&0&-1&0\\0&-1&0&-n\\1&0&1&0\\0&0&0&-1\end{array}\right)$& $n+16$ & $n+16$ & $-1$ & $E_8 \oplus I_n^*$ & $(6,2)$ & $E_8\oplus D_{n+4}$ & $n+14$ \\ \hline
$II^*-I_n^*-\alpha$ &$~$& $n+14$ & $n+14$ & $-1$ & $(n+2)-SU(2)-E_8$ & $(6,2)$ & $D_{n+2}\oplus E_7 \oplus A_1$ & $n+11$ \\ \hline
$IV-I_n^*-0$ &$ \left(\begin{array}{cccc} 0&0&1&0\\0&-1&0&-n\\-1&0&-1&0\\0&0&0&-1\end{array}\right)$& $n+10$ & $n+10$ & $-1$ & $H_2\oplus I_n^*$ & $(\frac{3}{2},2)$ & $E_6\oplus A_2$ & $n+7$ \\ \hline
$IV^*-I_n^*-0$ &$ \left(\begin{array}{cccc} -1&0&-1&0\\0&-1&0&-n\\1&0&0&0\\0&0&0&-1\end{array}\right)$& $n+14$ & $n+14$ & $-1$ & $E_6\oplus I_n^*$ & $(3,2)$ & $E_6\oplus D_{n+4}$ & $n+12$ \\ \hline
$IV^*-I_n^*-\alpha$ &$ ~$& $n+12$ & $n+12$ & $-1$ & $(n+2)-SU(2)-E_6$ & $(3,2)$ & $A_5\oplus D_{n+2} \oplus A_1$ & $n+9$ \\ \hline
$IV-II_n$ &$ \left(\begin{array}{cccc} 0&0&1&0\\0&1&-1&n+1\\-1&0&-1&1\\0&0&0&1\end{array}\right)$& $n+4$ & $n+4$ & $-1$ & $n-U(1)-H_2$ & $(\frac{3}{2},1)$ & $A_{n-1}$ & $n+2$ \\ \hline
$IV^*-II_0$ &~& $7$ & $6$ & $-1$ & $U(1)\capcup E_6$ & $(3,1)$ & $A_3$ & $4$ \\ \hline
$IV^*-II_n$ &$ \left(\begin{array}{cccc} -1&0&-1&-1\\-1&1&0&n\\1&0&0&0\\0&0&0&1\end{array}\right)$& $n+7$ & $n+7$ & $-1$ & $(n-1)-U(1) - E_6$ & $(3,1)$ & $ D_5 \oplus A_{n-2}$ & $n+5$ \\ \hline
$II-II^*_n$ &$ \left(\begin{array}{cccc} 1&0&1&1\\1&-1&0&-n\\-1&0&0&0\\0&0&0&-1\end{array}\right)$& $n+7$ & $n+7$ & $-1$ & $(n+3)-SU(2)-H_1$ & $(2,4/3)$ & $D_{n+3}$ & $n+4$ \\ \hline
$II^*-II_n^*$ &$ \left(\begin{array}{cccc} 0&0&-1&0\\0&-1&1&-n\\1&0&1&-1\\0&0&0&1\end{array}\right)$& $n+14$ & $n+14$ & $-1$ & $(n+2)-SU(2)-E_8$ $(E_8=D_7\oplus A_1)$ & $(6,4)$? & $A_1\oplus D_7\oplus D_{n+2}$ & $n+11$ \\ \hline
$III-I_n-0$ &$ \left(\begin{array}{cccc} 0&0&1&0\\0&1&0&n\\-1&0&0&0\\0&0&0&1\end{array}\right)$& $n+3$ & $n+3$ & $-1$ & $H_1\oplus I_n$ & $(\frac{4}{3},1)$ & $A_1\oplus A_{n-1}$ & $n+1$ \\ \hline
$III^*-I_n-0$ &$ \left(\begin{array}{cccc} 0&0&-1&0\\0&1&0&n\\1&0&0&0\\0&0&0&1\end{array}\right)$& $n+9$ & $n+9$ & $-1$ & $E_7\oplus I_n$ & $(4,1)$ & $E_7\oplus A_{n-1}$ & $n+8$ \\ \hline
$III-I_n^*-0$ &$ \left(\begin{array}{cccc} 0&0&1&0\\0&-1&0&-n\\-1&0&0&0\\0&0&0&-1\end{array}\right)$& $n+9$ & $n+9$ & $-1$ & $H_1\oplus I_n^*$ & $(\frac{4}{3},2)$ & $A_1\oplus A_{n+4}$ & $n+6$ \\ \hline
\end{tabular}}
\end{center}
\end{table}%

\begin{table}[H]
\begin{center}
\caption{Parabolic fiber [3] [4] [5].}
\label{parabolic4}
\resizebox{5.5in}{!}{\begin{tabular}{|c|c|c|c|c|c|c|c|c|} \hline
Type &  Monodromy & $d_x$ & $\delta_x$ & $l$ & Theory & Scaling dimension & Gauge algebra & $n_t$ (components)\\  \hline 
$III^*-I_n^*-0$ &$ \left(\begin{array}{cccc} 0&0&-1&0\\0&-1&0&-n\\1&0&0&0\\0&0&0&-1\end{array}\right)$& $n+15$ & $n+15$ & $-1$ & $E_7 \oplus I_n^*$ & $(4,2)$ & $E_7\oplus D_{n+4}$ & $n+13$ \\ \hline
$III^*-I_n^*-\alpha$ &$~$& $n+13$ & $n+13$ & $-1$ & $(n+2)-SU(2)-E_7$ & $(4,2)$ & $D_6\oplus D_{n+2}\oplus A_1$ & $n+9$\\ \hline
$III-II_n$ &$ \left(\begin{array}{cccc} 0&0&1&0\\0&1&1&n+1\\-1&0&0&-1\\0&0&0&1\end{array}\right)$& $n+3$ & $n+3$ & $-1$ & $n-U(1)-H_1$ & $(\frac{4}{3},1)$ & $A_{n-1}$ & $n+1$ \\ \hline
$III^*-II_n$& $ \left(\begin{array}{cccc} 0&0&-1&1\\1&1&0&n\\1&0&0&0\\0&0&0&1\end{array}\right)$ & $n+8 $& $n+8$ & -1 & $(n-1)-U(1)-E_7$ & $(4,1)$ & $A_{n-2}\oplus E_6$ & $n+6$  \\ \hline
$III-II_n^*$&$ \left(\begin{array}{cccc} 0&0&1&-1\\-1&-1&0&-n\\-1&0&0&0\\0&0&0&-1\end{array}\right)$& $n+8$ & $n+8$  &$-1$ & $(n+3)-SU(2)-H_2$ & $(\frac{3}{2},2)$ & $D_{n+3}$& $n+5$ \\ \hline
$III^*-II_n^*$  &$ \left(\begin{array}{cccc} 0&0&-1&0\\0&-1&-1&-n-1\\1&0&0&1\\0&0&0&-1\end{array}\right)$& $n+13$ & $n+13$ & $-1$ & $(n+2)-SU(2)-E_7$ $(E_7=SU(2)\oplus A_6)$& $(4,2)$? & $D_{n+2}\oplus A_6\oplus A_1$ & $n+10$ \\ \hline
\textcolor{red}{Parabolic[4]}&~&~&~&~&~&~&~\\ \hline
$I_{n-p-0}$  &$ \left(\begin{array}{cccc} 1&0&p&0\\0&1&0&n\\0&0&1&0\\0&0&0&1\end{array}\right)$& $n+p$ & $n+p$ & $1$ & $I_n\oplus I_p$ & $(1,1)$ & $A_{n-1}\oplus A_{p-1}$ & $n+p-1$ \\ \hline
$I_{n-p-0}^*$& $ \left(\begin{array}{cccc} -1&0&-p&0\\0&-1&0&-n\\0&0&-1&0\\0&0&0&-1\end{array}\right)$& $n+p+10$ & $n+p+10$ & $0$ & $(n+2)-SU(2)-SU(2)-(p+2)$ & $(2,2)$ & $D_{n+2}\oplus A_1^2 \oplus D_{p+2}$ & $n+p+7$ \\ \hline
$I_n^*-I_p^*-0$ &  $\left(\begin{array}{cccc} -1&0&-p&0\\0&-1&0&-n\\0&0&-1&0\\0&0&0&-1\end{array}\right)$& $n+p+12$ & $n+p+12$ & $-1$ & $I_n^*\oplus I_p^*$ & $(2,2)$ & $D_{n+4}\oplus D_{p+4}$ & $n+p+9$\\ \hline
$I_n-I_p^*-0$ & $ \left(\begin{array}{cccc} -1&0&-p&0\\0&1&0&n\\0&0&-1&0\\0&0&0&1\end{array}\right)$& $n+p+6$ & $n+p+6$ & $-1$ & $I_n\oplus I_p^*$ & $(2,1)$ & $A_{n-1}\oplus D_{p+4}$ & $n+p+4$\\ \hline
$2I_n-0$ & $ \left(\begin{array}{cccc} 0&1&0&n\\1&0&0&0\\0&0&0&1\\0&0&1&0\end{array}\right)$& $n+5$ & $n+4$ & $0$ & $?$ & $(2,1)$ & $A_{n-1}\oplus A_1^2$ & $n+2$ \\ \hline
$2I_n^*-0$ & $ \left(\begin{array}{cccc} 0&-1&0&-n\\1&0&0&0\\0&0&0&-1\\0&0&1&0\end{array}\right)$& $n+11$ & $n+10$ & $0$ & $V-Sp(4)-(n+4)(\textbf{4})$ & $(4,2)$ & $D_{n+4}\oplus A_1^2$ & $n+7$\\ \hline
$II_{n-p}$&$ \left(\begin{array}{cccc} -1&0&-p&-1\\0&1&1&n\\0&0&-1&0\\0&0&0&1\end{array}\right)$& $n+p+5$ & $n+p+5$ & $-1$ & $(n-1)-U(1)-SU(2)-(p+3)$? & $(2,1)$ & $A_{n-2}\oplus  D_{p+3}$ & $n+p+3$ \\ \hline
$\tilde{III}_n$ & $ \left(\begin{array}{cccc} 0&-1&1&0\\1&0&n&-1\\0&0&0&-1\\0&0&1&0\end{array}\right)$ & $n+10$ & $n+10$ & $-1$ & $Sp(4)-(n+6)(\textbf{4})$ & $(4, 2)$ & $D_{n+6}$ & $n+7$ \\ \hline
\textcolor{red}{Parabolic[5]}&~&~&~&~&~&~&~\\ \hline
$I_{n-p-q}$ & $ \left(\begin{array}{cccc}1&0&p+q&-q\\0&1&-q&n+q\\0&0&1&0\\0&0&0&1\end{array}\right)$& $n+p+q$ & $n+p+q$ & $-1$ & $n-U(1)-q-U(1)-p$? & $(1, 1)$ & $A_{n-1}\oplus A_{p-1}\oplus A_{q-1}$ & $n+p+q-1$ \\ \hline
$I_{n-p-q}^*$ & $ \left(\begin{array}{cccc} -1&0&-p-q&q\\0&-1&q&-n-q\\0&0&-1&0\\0&0&0&-1\end{array}\right)$& $n+p+q+10$ & $n+p+q+10$ & $-1$ & $(n+1)-SU(2)-q+1-SU(2)-(p+1)$?& $(2,2)$ & $D_{n+2}\oplus D_{p+2}\oplus D_{q+2}$ & $n+p+q+7$ \\ \hline
$\tilde{II}_{n-p}$ & $ \left(\begin{array}{cccc} -1&0&-p&0\\1&1&p&n\\0&0&-1&1\\0&0&0&1\end{array}\right)$& $n+p+5$ & $n+p+5$ & $-1$ &$?$ & $(2,1)$ & $A_{n-2}\oplus  A_{p+3}$ & $n+p+3$ \\ \hline
$II_{n-p}^*$ &$ \left(\begin{array}{cccc} 1&0&-p&0\\-1&-1&p&-n\\0&0&1&-1\\0&0&0&-1\end{array}\right)$& $n+p+5$ & $n+p+4$ & $-1$ & $?$ & $(2,1)$ & $A_{p-1}\oplus D_{n+2}$ & $n+p+2$ \\ \hline
$III_n$ &$ \left(\begin{array}{cccc} 0&1&-n&n\\-1&-1&0&0\\0&0&-1&1\\0&0&-1&0\end{array}\right)$& $n+10$ & $n+10$ & $-1$ & $SU(3)-(n+6)$ & $(3,2)$ & $A_{n+5}$ & $n+7$ \\ \hline
$III_n^*$ &$ \left(\begin{array}{cccc} 0&-1&n&-n\\1&1&0&0\\0&0&1&-1\\0&0&1&0\end{array}\right)$& $n+10$ & $n+9$ & $-1$ & \textcolor{red}{$G_2-(n+4)\textbf{7}$} & $(6,2)$ & \textcolor{red}{$D_{n+5}$} & $n+6$\\ \hline
\end{tabular}}
\end{center}
\end{table}%

\newpage
\section{Global coulomb branch geometries }
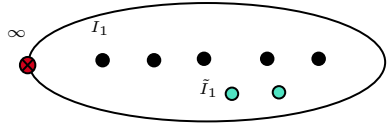
\begin{figure}[htbp]
\begin{center}

\tikzset{every picture/.style={line width=0.75pt}} 

\begin{tikzpicture}[x=0.55pt,y=0.55pt,yscale=-1,xscale=1]

\draw   (185,138.61) .. controls (185,116.18) and (239.29,98) .. (306.25,98) .. controls (373.21,98) and (427.5,116.18) .. (427.5,138.61) .. controls (427.5,161.04) and (373.21,179.22) .. (306.25,179.22) .. controls (239.29,179.22) and (185,161.04) .. (185,138.61) -- cycle ;
\draw  [fill={rgb, 255:red, 208; green, 2; blue, 27 }  ,fill opacity=1 ] (179,140.61) .. controls (179,137.51) and (181.35,135) .. (184.25,135) .. controls (187.15,135) and (189.5,137.51) .. (189.5,140.61) .. controls (189.5,143.71) and (187.15,146.22) .. (184.25,146.22) .. controls (181.35,146.22) and (179,143.71) .. (179,140.61) -- cycle ; \draw   (180.54,136.64) -- (187.96,144.58) ; \draw   (187.96,136.64) -- (180.54,144.58) ;
\draw  [fill={rgb, 255:red, 0; green, 0; blue, 0 }  ,fill opacity=1 ] (231,137.25) .. controls (231,134.9) and (232.9,133) .. (235.25,133) .. controls (237.6,133) and (239.5,134.9) .. (239.5,137.25) .. controls (239.5,139.6) and (237.6,141.5) .. (235.25,141.5) .. controls (232.9,141.5) and (231,139.6) .. (231,137.25) -- cycle ;
\draw  [fill={rgb, 255:red, 0; green, 0; blue, 0 }  ,fill opacity=1 ] (266,137.25) .. controls (266,134.9) and (267.9,133) .. (270.25,133) .. controls (272.6,133) and (274.5,134.9) .. (274.5,137.25) .. controls (274.5,139.6) and (272.6,141.5) .. (270.25,141.5) .. controls (267.9,141.5) and (266,139.6) .. (266,137.25) -- cycle ;
\draw  [fill={rgb, 255:red, 0; green, 0; blue, 0 }  ,fill opacity=1 ] (343,136.25) .. controls (343,133.9) and (344.9,132) .. (347.25,132) .. controls (349.6,132) and (351.5,133.9) .. (351.5,136.25) .. controls (351.5,138.6) and (349.6,140.5) .. (347.25,140.5) .. controls (344.9,140.5) and (343,138.6) .. (343,136.25) -- cycle ;
\draw  [fill={rgb, 255:red, 0; green, 0; blue, 0 }  ,fill opacity=1 ] (300,136.25) .. controls (300,133.9) and (301.9,132) .. (304.25,132) .. controls (306.6,132) and (308.5,133.9) .. (308.5,136.25) .. controls (308.5,138.6) and (306.6,140.5) .. (304.25,140.5) .. controls (301.9,140.5) and (300,138.6) .. (300,136.25) -- cycle ;
\draw  [fill={rgb, 255:red, 0; green, 0; blue, 0 }  ,fill opacity=1 ] (378,136.25) .. controls (378,133.9) and (379.9,132) .. (382.25,132) .. controls (384.6,132) and (386.5,133.9) .. (386.5,136.25) .. controls (386.5,138.6) and (384.6,140.5) .. (382.25,140.5) .. controls (379.9,140.5) and (378,138.6) .. (378,136.25) -- cycle ;
\draw  [fill={rgb, 255:red, 80; green, 227; blue, 194 }  ,fill opacity=1 ] (319,160.25) .. controls (319,157.9) and (320.9,156) .. (323.25,156) .. controls (325.6,156) and (327.5,157.9) .. (327.5,160.25) .. controls (327.5,162.6) and (325.6,164.5) .. (323.25,164.5) .. controls (320.9,164.5) and (319,162.6) .. (319,160.25) -- cycle ;
\draw  [fill={rgb, 255:red, 80; green, 227; blue, 194 }  ,fill opacity=1 ] (351,159.25) .. controls (351,156.9) and (352.9,155) .. (355.25,155) .. controls (357.6,155) and (359.5,156.9) .. (359.5,159.25) .. controls (359.5,161.6) and (357.6,163.5) .. (355.25,163.5) .. controls (352.9,163.5) and (351,161.6) .. (351,159.25) -- cycle ;

\draw (168,114.4) node [anchor=north west][inner sep=0.75pt] [font=\tiny]   {$\infty $};
\draw (225,108.4) node [anchor=north west][inner sep=0.75pt]   [font=\tiny]  {$I_{1}$};
\draw (299,148.4) node [anchor=north west][inner sep=0.75pt]  [font=\tiny]   {$\tilde{I}_{1}$};

\end{tikzpicture}

\end{center}
\caption{Global Coulomb  branch geometry with $I_1$ and $\tilde{I}_1$ type  singular fibers  at the bulk. The fiber at infinity gives the information for UV theory, i.e. which space-time dimension the UV theory lives.}
\label{global}
\end{figure}

Let's now construct some global Coulomb branch geometries, where all the fibers at the bulk 
are the $I_1$ $(I_{1-0-0}$ in table. \ref{parabolic3}) or $\tilde{I}_1$  ($I_0-I_0-1$ in table. \ref{elliptic2a}) type fibers. These would describe the singular fiber configuration of generic deformation
on Coulomb branch. The general structure is shown in figure. \ref{global}: there is a singular fiber at infinity which encodes 
the UV behavior of the theory,  i.e. which space time dimension the UV theory lives. 

There are several global constraints that a singular fiber configuration has to satisfy, for example, the total monodromy around 
the singularities should be trivial, and the holomorphic quantities (the photon couplings) should be consistently defined.
The study of those constraints would be left in \cite{Xie:ranktwob}. Here we'd like to discuss some topological constraints which 
are related to the local invariants of the singularity.

The first constraint comes from the geometry of the total space $X$ formed by genus two fibration over $\mathbb{P}^1$:
 the total space $X$ should be a \textbf{rational} surface so that  it can describe the Coulomb branch configuration for a local field theory.  
 The rational condition would put constraint on the choice  of singular fibers. 
 
 Let's first review some facts about complex surface. For a compact rational surface, the irregularity and geometric genus satisfies $q=p_g=0$ \footnote{The irregularity of a compact complex surface $X$
is defined as $q=dim H^{1,0}(X,C)$, and the geometric genus is given as $p_g= dim H^{2,0}$. The only nontrivial Hodge numbers for a connected rational surface are $h^{0,0}=h^{2,2}=1$, and $h^{1,1}$.}, and the 
Euler number for the rational surface is $\chi(X)=1$. For a compact complex surface, Norther's theorem relates the Euler number of the surface to the first  Chern class $c_1$ and second Chern class $c_2$ as follows:
\begin{equation}
12\chi=c_1^2+c_2.
\label{nother}
\end{equation}
Given a genus two fibered surface, $c_1^2$ and $c_2$ can be computed from the local data associated with singular fiber \cite{kenji1988discriminants}. 
The local contribution for a singular fiber $F$ is given by the local data $d_x, \delta_x$ as follows \cite{kenji1988discriminants}:
\begin{align*}
& c_1^2(F)= \frac{6}{5} d_x-\delta_x, \nonumber\\
& c_2(F)=\delta_x.
\end{align*}
The total Chern class of the surface $X$ are computed from the local data as:
\begin{align}
&c_1^2(X)=-8+\sum c_1^2(F)=-8+\sum (\frac{6}{5} d_x-\delta_x) \nonumber\\
&c_2(X)=-4+\sum c_2(F)=-4+\sum \delta_x
\label{chern}
\end{align}
Using Noether's theorem (see equation \ref{nother}), the local data for the singular fibers should satisfy following equation:
\begin{equation*}
12 \chi=-12+\sum ( c_1^2(F)+c_2(F))=-12+\sum (\frac{6}{5} d_x).
\end{equation*}
Since the Euler number  $\chi$ of a rational surface $X$ is one, one get from above equation
\begin{equation}
\boxed{24=\sum (\frac{6}{5} d_x) \to \sum d_x=20}.
\label{euler}
\end{equation}
This is the first important equation that a configuration of singular fibers  relevant for the Coulomb branch should satisfy. 
Notice only the assumption that the total space $X$ should be rational is used in deriving above equation.

The second constraint comes from the relation between  flavor symmetry and the Mordell-Weil lattice. One can define a Mordell-Weil 
lattice for higher genus fibration over $\mathbb{P}^1$, whose rank is given as \cite{shioda1992mordell}: 
\begin{equation}
rank(MW(X))=\rho(X)-2-\sum (n_t-1).
\end{equation}
Here $n_t$ is the number of components 
in the dual graph of the singular fiber; and $\rho(X)$ is the Picard number of $X$.  $11\leq \rho(X)\leq 14$ for a genus two fibration $X$. 

Let's now relate $\rho(X)$ to the local data of the singular fibers. First of all, there is the following formula for 
the Picard number:
\begin{equation*}
\rho(X)=c_2(X)-2=\sum \delta_x-6.
\end{equation*}
Formula \ref{chern} for $c_2(X)$ is used in deriving above equation.
Let's now recall that the number $n_x=2\delta_x-d_x, t_x=d_x-\delta_x$ ($n_x$ ($t_x$) is the number of $I_1$ ($\tilde{I}_1$) singularities in the deformation of the singularity):
and the Picard number becomes 
\begin{equation*}
\rho(X)=\sum (d_x-t_x)-6=14-\sum t_x.
\end{equation*}
Here we used the constraint \ref{euler}. The maximal value for Picard number is 14 when all the local data $t_x$ is zero!  

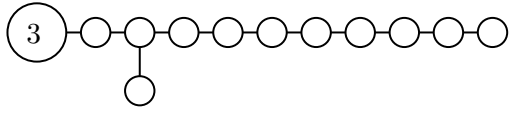
\begin{figure}[H]
\begin{center}
\tikzset{every picture/.style={line width=0.75pt}} 

\begin{tikzpicture}[x=0.55pt,y=0.55pt,yscale=-1,xscale=1]

\draw   (120,110) .. controls (120,98.95) and (128.95,90) .. (140,90) .. controls (151.05,90) and (160,98.95) .. (160,110) .. controls (160,121.05) and (151.05,130) .. (140,130) .. controls (128.95,130) and (120,121.05) .. (120,110) -- cycle ;
\draw   (170,110) .. controls (170,104.48) and (174.48,100) .. (180,100) .. controls (185.52,100) and (190,104.48) .. (190,110) .. controls (190,115.52) and (185.52,120) .. (180,120) .. controls (174.48,120) and (170,115.52) .. (170,110) -- cycle ;
\draw   (200,110) .. controls (200,104.48) and (204.48,100) .. (210,100) .. controls (215.52,100) and (220,104.48) .. (220,110) .. controls (220,115.52) and (215.52,120) .. (210,120) .. controls (204.48,120) and (200,115.52) .. (200,110) -- cycle ;
\draw   (230,110) .. controls (230,104.48) and (234.48,100) .. (240,100) .. controls (245.52,100) and (250,104.48) .. (250,110) .. controls (250,115.52) and (245.52,120) .. (240,120) .. controls (234.48,120) and (230,115.52) .. (230,110) -- cycle ;
\draw   (260,110) .. controls (260,104.48) and (264.48,100) .. (270,100) .. controls (275.52,100) and (280,104.48) .. (280,110) .. controls (280,115.52) and (275.52,120) .. (270,120) .. controls (264.48,120) and (260,115.52) .. (260,110) -- cycle ;
\draw   (290,110) .. controls (290,104.48) and (294.48,100) .. (300,100) .. controls (305.52,100) and (310,104.48) .. (310,110) .. controls (310,115.52) and (305.52,120) .. (300,120) .. controls (294.48,120) and (290,115.52) .. (290,110) -- cycle ;
\draw   (320,110) .. controls (320,104.48) and (324.48,100) .. (330,100) .. controls (335.52,100) and (340,104.48) .. (340,110) .. controls (340,115.52) and (335.52,120) .. (330,120) .. controls (324.48,120) and (320,115.52) .. (320,110) -- cycle ;
\draw   (350,110) .. controls (350,104.48) and (354.48,100) .. (360,100) .. controls (365.52,100) and (370,104.48) .. (370,110) .. controls (370,115.52) and (365.52,120) .. (360,120) .. controls (354.48,120) and (350,115.52) .. (350,110) -- cycle ;
\draw   (380,110) .. controls (380,104.48) and (384.48,100) .. (390,100) .. controls (395.52,100) and (400,104.48) .. (400,110) .. controls (400,115.52) and (395.52,120) .. (390,120) .. controls (384.48,120) and (380,115.52) .. (380,110) -- cycle ;
\draw   (410,110) .. controls (410,104.48) and (414.48,100) .. (420,100) .. controls (425.52,100) and (430,104.48) .. (430,110) .. controls (430,115.52) and (425.52,120) .. (420,120) .. controls (414.48,120) and (410,115.52) .. (410,110) -- cycle ;
\draw   (440,110) .. controls (440,104.48) and (444.48,100) .. (450,100) .. controls (455.52,100) and (460,104.48) .. (460,110) .. controls (460,115.52) and (455.52,120) .. (450,120) .. controls (444.48,120) and (440,115.52) .. (440,110) -- cycle ;
\draw   (200,150) .. controls (200,144.48) and (204.48,140) .. (210,140) .. controls (215.52,140) and (220,144.48) .. (220,150) .. controls (220,155.52) and (215.52,160) .. (210,160) .. controls (204.48,160) and (200,155.52) .. (200,150) -- cycle ;
\draw    (190,110) -- (200,110) ;
\draw    (220,110) -- (230,110) ;
\draw    (250,110) -- (260,110) ;
\draw    (280,110) -- (290,110) ;
\draw    (310,110) -- (320,110) ;
\draw    (340,110) -- (350,110) ;
\draw    (370,110) -- (380,110) ;
\draw    (400,110) -- (410,110) ;
\draw    (430,110) -- (440,110) ;
\draw    (160,110) -- (170,110) ;
\draw    (210,120) -- (210,140) ;

\draw (131,102.4) node [anchor=north west][inner sep=0.75pt]    {$3$};

\end{tikzpicture}

\end{center}
\caption{The maximal Mordell-Weil lattice for a genus two fibered rational surface. There is one  basis section with self-intersection number $-3$, and all  the other basis sections have self-intersection number $-2$. }
\label{mwgenus2}
\end{figure}

So finally, the rank of MW group is given by following formula:
\begin{equation}
\boxed{rank(MW(X))=12-\sum t_x-\sum (n_t-1)}.
\label{mw}
\end{equation}
Here the sum is over all the singular fibers. The Mordell-Weil group forms a lattice by introducing a pairing \cite{shioda1992mordell}.  The maximal 
Mordell-Weil lattice for genus two fibration is given in figure. \ref{mwgenus2}, which is a 12 dimensional lattice, see \cite{khac2002mordell}. This lattice is called
$D_{12}^+$ in \cite{conway2013sphere} and is the unique unimodular lattice in dimension 12.  

It was argued in \cite{Caorsi:2018ahl} that  the rank of the Mordell-Weil lattice is the same as that of the flavor symmetry of rank one theory.  
In rank two case, the above identification no longer holds. However, there is a more general formula relating the rank of Mordell-Weil group and the number 
of bulk singularities $\Gamma$:
 \begin{equation}
\Gamma=(MW(X)+4)+D-4 =MW(X)+D.
\label{charge}
\end{equation}
Here $D$ is the space-time dimension the UV theory lives. The difference comes from the fact there is one (two) extra KK charges for 5d (6d) theories.

Now assuming the local data of the singular fiber at $\infty$ is $d_\infty, \delta_\infty, n_\infty$, we have following two equations (using \ref{euler}, \ref{mw} and \ref{charge}):
\begin{align*}
& d_\infty+\Gamma+\sum_{F/\infty} t_x=20, \nonumber\\
&rank(MW)+D=[12-\sum_{F/\infty} t_x-t_\infty-(n_\infty -1)]+D=\Gamma.
\end{align*}
Here  $\Gamma$ is the total number of singular fibers at the bulk. Solving  above equations, one find:
\begin{equation}
\boxed{d_\infty-n_\infty=7+t_\infty-D}.
\label{infinity}
\end{equation}
This equation would tell us which kind of singular fiber can be put at infinity for a theory with given space-time dimension $D$.

\textbf{Remark 1}: We'd like to clarify an important subtly here. The rank of flavor symmetry is the same as the rank of Mordell-Weil lattice for  rank one theory \cite{Caorsi:2018ahl}; this is no longer true for rank two theories. 
For example, consider 4d theory defined by quiver $2-SU(2)-SU(2)-2$,  its Coulomb branch geometry 
is given by the configuration $(I_{0-0-0}^*,I_1^{10})$; the rank of MW lattice of above configuration is 6, but the rank of physical flavor symmetry is only $5$. 
So the real meaning of \ref{charge} is that the $MW(X)+D=\Gamma$, with  $\Gamma$ the number of $I_1$ and $\tilde{I}_1$ singularities in the bulk.

\textbf{Remark 2}: For a large class of 4d $\mathcal{N}=2$ theories defined by hypersurface singularity, the number of $I_1$ singularities in the generic deformation is given by the formula
$\mu=2r+f$. This is not true in general, as also illustrated by the theory $2-SU(2)-SU(2)-2$, here $2r+f=9$, but there are  10 $I_1$ singularities in the bulk (the Coulomb branch geometry is $(I_{0-0-0}^*,I_1^{10})$).

\subsection{4d theories}
Let's now consider Coulomb branch configurations for  4d theories. The constraint on the topological data of the fiber at $\infty$ is (see formula. \ref{infinity}):
\begin{equation}
\boxed{d_\infty-n_\infty=3+t_\infty}
\label{4dinfinity}
\end{equation}
One can search from tables of section. \ref{data} to find candidate singular fiber at $\infty$ for a 4d theory.

\subsubsection{Isotrivial family and SCFTs}
Let's first consider the so-called iso-trivial family over $\mathcal{P}^1$, which  has only two singular  fibers. Those
families are already classified in \cite{gong2016families}, see table. \ref{iso}.  Interestingly, all the singular fibers in table. \ref{iso}
satisfies  the condition \ref{4dinfinity} and so they would give the 4d $\mathcal{N}=2$ SCFT.

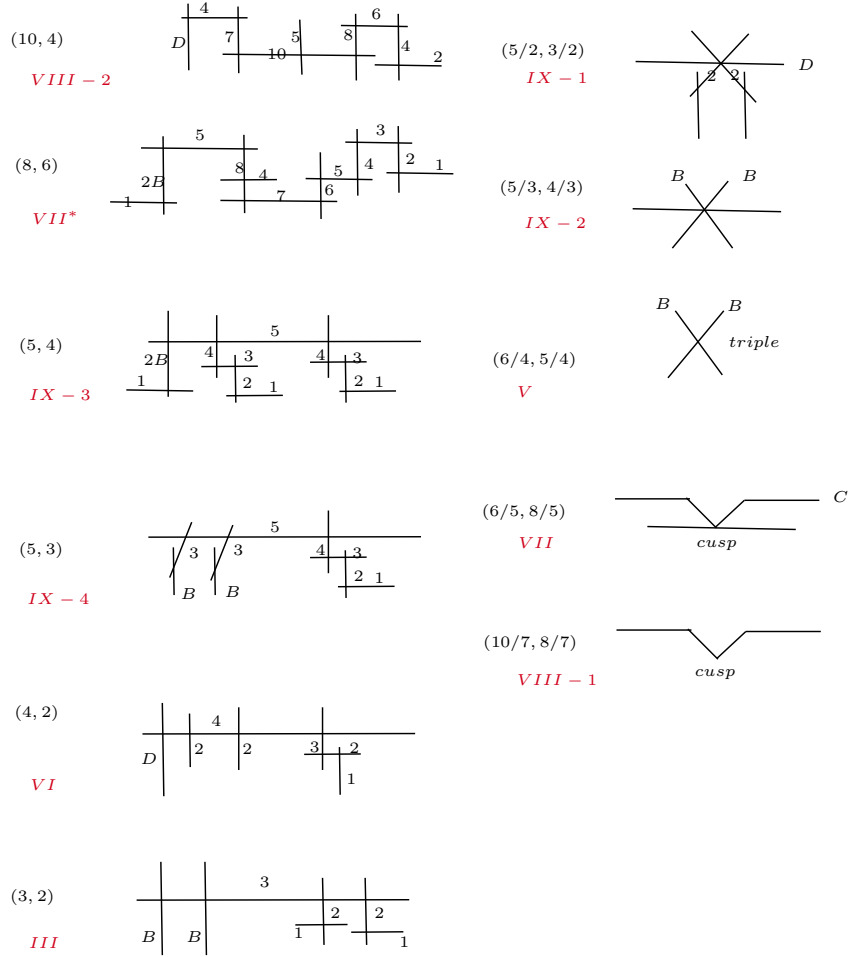
\begin{figure}
\begin{center}
\tikzset{every picture/.style={line width=0.55pt}} 

\begin{tikzpicture}[x=0.55pt,y=0.55pt,yscale=-1,xscale=1]

\draw    (177,110) -- (177.5,163.72) ;
\draw    (161,155) -- (240.5,155.72) ;
\draw    (229,122) -- (229.5,167.72) ;
\draw    (219,140) -- (264.5,140.72) ;
\draw    (254,106) -- (254.5,151.72) ;
\draw    (246,115) -- (291.5,115.72) ;
\draw    (282,104) -- (282.5,149.72) ;
\draw    (274,136) -- (319.5,136.72) ;
\draw    (107,119) -- (186.5,119.72) ;
\draw    (122,111) -- (122.5,164.72) ;
\draw    (86,156) -- (131.5,156.72) ;
\draw    (161,141) -- (199.5,140.72) ;
\draw    (112,252) -- (297.5,251.72) ;
\draw    (125.5,230.72) -- (125.5,289.72) ;
\draw    (158.5,233.72) -- (158.5,276.72) ;
\draw    (234.5,233.72) -- (234.5,276.72) ;
\draw    (148,269) -- (186.5,268.72) ;
\draw    (222,266) -- (260.5,265.72) ;
\draw    (171,261) -- (171.5,293.72) ;
\draw    (246,261) -- (246.5,293.72) ;
\draw    (242,286) -- (280.5,285.72) ;
\draw    (165,289) -- (203.5,288.72) ;
\draw    (97,285) -- (142.5,285.72) ;
\draw    (112,386) -- (297.5,385.72) ;
\draw    (169.5,377.72) -- (154.5,415.72) ;
\draw    (234.5,367.72) -- (234.5,410.72) ;
\draw    (222,400) -- (260.5,399.72) ;
\draw    (157,393) -- (157.5,425.72) ;
\draw    (246,395) -- (246.5,427.72) ;
\draw    (241,420) -- (279.5,419.72) ;
\draw    (141.5,375.72) -- (126.5,413.72) ;
\draw    (129,393) -- (129.5,425.72) ;
\draw    (162.5,54.72) -- (266.5,55.72) ;
\draw    (173,22) -- (173.5,67.72) ;
\draw    (215,31) -- (216.5,68.72) ;
\draw    (253,27) -- (253.5,72.72) ;
\draw    (243,35) -- (288.5,35.72) ;
\draw    (282,27) -- (282.5,72.72) ;
\draw    (134.5,29.72) -- (179.5,29.72) ;
\draw    (266,62) -- (311.5,62.72) ;
\draw    (139,20) -- (139.5,65.72) ;
\draw    (108,521) -- (293.5,520.72) ;
\draw    (121.5,499.72) -- (122.5,563.72) ;
\draw    (173.5,502.72) -- (173.5,545.72) ;
\draw    (230.5,502.72) -- (230.5,545.72) ;
\draw    (218,535) -- (256.5,534.72) ;
\draw    (140,507) -- (140.5,543.72) ;
\draw    (242,530) -- (242.5,562.72) ;
\draw    (104,635) -- (289.5,634.72) ;
\draw    (120.5,608.72) -- (121.5,672.72) ;
\draw    (150.5,608.72) -- (151.5,672.72) ;
\draw    (230.5,619.72) -- (231.5,665.72) ;
\draw    (259.5,619.72) -- (260.5,665.72) ;
\draw    (212,652) -- (247.5,651.72) ;
\draw    (250,657) -- (285.5,656.72) ;
\draw    (443.5,59.72) -- (544.5,61.72) ;
\draw    (481.5,38.72) -- (525.5,87.72) ;
\draw    (480.5,83.72) -- (520.5,40.72) ;
\draw    (486.5,112.72) -- (485.5,66.72) ;
\draw    (518.5,112.72) -- (517.5,66.72) ;
\draw    (441.5,160.72) -- (542.5,162.72) ;
\draw    (468.5,187.72) -- (506.5,141.72) ;
\draw    (509.5,187.72) -- (477.5,143.72) ;
\draw    (465.5,276.72) -- (503.5,230.72) ;
\draw    (502.5,274.72) -- (470.5,230.72) ;
\draw    (429.5,359.72) -- (480.5,359.72) ;
\draw    (478,359) -- (498.5,379.72) ;
\draw    (498,379) -- (517.5,360.72) ;
\draw    (517.5,360.72) -- (568.5,360.72) ;
\draw    (430.5,449.72) -- (481.5,449.72) ;
\draw    (479,449) -- (499.5,469.72) ;
\draw    (499,469) -- (518.5,450.72) ;
\draw    (518.5,450.72) -- (569.5,450.72) ;
\draw    (451.5,378.72) -- (552.5,380.72) ;

\draw (105,136.4) node [anchor=north west][inner sep=0.75pt]  [font=\tiny]  {$2B$};
\draw (169,127.4) node [anchor=north west][inner sep=0.75pt]  [font=\tiny]  {$8$};
\draw (142,104.4) node [anchor=north west][inner sep=0.75pt]  [font=\tiny]  {$5$};
\draw (185,132.4) node [anchor=north west][inner sep=0.75pt]  [font=\tiny]  {$4$};
\draw (197,146.4) node [anchor=north west][inner sep=0.75pt]  [font=\tiny]  {$7$};
\draw (230,142.4) node [anchor=north west][inner sep=0.75pt]  [font=\tiny]  {$6$};
\draw (236,129.4) node [anchor=north west][inner sep=0.75pt]  [font=\tiny]  {$5$};
\draw (257,124.4) node [anchor=north west][inner sep=0.75pt]  [font=\tiny]  {$4$};
\draw (265,101.4) node [anchor=north west][inner sep=0.75pt]  [font=\tiny]  {$3$};
\draw (285,121.4) node [anchor=north west][inner sep=0.75pt]  [font=\tiny]  {$2$};
\draw (305,125.4) node [anchor=north west][inner sep=0.75pt]  [font=\tiny]  {$1$};
\draw (93,150.4) node [anchor=north west][inner sep=0.75pt]  [font=\tiny]  {$1$};
\draw (19,124.4) node [anchor=north west][inner sep=0.75pt]    [font=\tiny]  {$( 8,6)$};
\draw (193,239.4) node [anchor=north west][inner sep=0.75pt]  [font=\tiny]  {$5$};
\draw (148,253.4) node [anchor=north west][inner sep=0.75pt]  [font=\tiny]  {$4$};
\draw (224,255.4) node [anchor=north west][inner sep=0.75pt]  [font=\tiny]  {$4$};
\draw (175,256.4) node [anchor=north west][inner sep=0.75pt]  [font=\tiny]  {$3$};
\draw (174.25,274.76) node [anchor=north west][inner sep=0.75pt]  [font=\tiny]  {$2$};
\draw (192,277.4) node [anchor=north west][inner sep=0.75pt]  [font=\tiny]  {$1$};
\draw (249,257.4) node [anchor=north west][inner sep=0.75pt]  [font=\tiny]  {$3$};
\draw (250,273.4) node [anchor=north west][inner sep=0.75pt]  [font=\tiny]  {$2$};
\draw (264,274.4) node [anchor=north west][inner sep=0.75pt]  [font=\tiny]  {$1$};
\draw (107,258.4) node [anchor=north west][inner sep=0.75pt]  [font=\tiny]  {$2B$};
\draw (102,273.4) node [anchor=north west][inner sep=0.75pt]  [font=\tiny]  {$1$};
\draw (22,247.4) node [anchor=north west][inner sep=0.75pt]    [font=\tiny]  {$( 5,4)$};
\draw (22,387.4) node [anchor=north west][inner sep=0.75pt]    [font=\tiny]  {$( 5,3)$};
\draw (193,373.4) node [anchor=north west][inner sep=0.75pt]  [font=\tiny]  {$5$};
\draw (224,389.4) node [anchor=north west][inner sep=0.75pt]  [font=\tiny]  {$4$};
\draw (168,389.4) node [anchor=north west][inner sep=0.75pt]  [font=\tiny]  {$3$};
\draw (162.25,417.76) node [anchor=north west][inner sep=0.75pt]  [font=\tiny]  {$B$};
\draw (249,391.4) node [anchor=north west][inner sep=0.75pt]  [font=\tiny]  {$3$};
\draw (250,407.4) node [anchor=north west][inner sep=0.75pt]  [font=\tiny]  {$2$};
\draw (264,408.4) node [anchor=north west][inner sep=0.75pt]  [font=\tiny]  {$1$};
\draw (138,391.4) node [anchor=north west][inner sep=0.75pt]  [font=\tiny]  {$3$};
\draw (132.25,418.76) node [anchor=north west][inner sep=0.75pt]  [font=\tiny]  {$B$};
\draw (19,499.4) node [anchor=north west][inner sep=0.75pt]    [font=\tiny]  {$( 4,2)$};
\draw (304,51.4) node [anchor=north west][inner sep=0.75pt]  [font=\tiny]  {$2$};
\draw (282,43.4) node [anchor=north west][inner sep=0.75pt]  [font=\tiny]  {$4$};
\draw (262,21.4) node [anchor=north west][inner sep=0.75pt]  [font=\tiny]  {$6$};
\draw (243,36.4) node [anchor=north west][inner sep=0.75pt]  [font=\tiny]  {$8$};
\draw (207,37.4) node [anchor=north west][inner sep=0.75pt]  [font=\tiny]  {$5$};
\draw (191,49.4) node [anchor=north west][inner sep=0.75pt]  [font=\tiny]  {$10$};
\draw (162,37.4) node [anchor=north west][inner sep=0.75pt]  [font=\tiny]  {$7$};
\draw (145,17.4) node [anchor=north west][inner sep=0.75pt]  [font=\tiny]  {$4$};
\draw (125,41.4) node [anchor=north west][inner sep=0.75pt]  [font=\tiny]  {$D$};
\draw (16,37.4) node [anchor=north west][inner sep=0.75pt]    [font=\tiny]  {$( 10,4)$};
\draw (153,506.4) node [anchor=north west][inner sep=0.75pt]  [font=\tiny]  {$4$};
\draw (220,524.4) node [anchor=north west][inner sep=0.75pt]  [font=\tiny]  {$3$};
\draw (174.25,525.76) node [anchor=north west][inner sep=0.75pt]  [font=\tiny]  {$2$};
\draw (247,525.4) node [anchor=north west][inner sep=0.75pt]  [font=\tiny]  {$2$};
\draw (245,546.4) node [anchor=north west][inner sep=0.75pt]  [font=\tiny]  {$1$};
\draw (105,532.4) node [anchor=north west][inner sep=0.75pt]  [font=\tiny]  {$D$};
\draw (141.25,525.76) node [anchor=north west][inner sep=0.75pt]  [font=\tiny]  {$2$};
\draw (16,625.4) node [anchor=north west][inner sep=0.75pt]   [font=\tiny]   {$( 3,2)$};
\draw (186,617.4) node [anchor=north west][inner sep=0.75pt]  [font=\tiny]  {$3$};
\draw (234,638.4) node [anchor=north west][inner sep=0.75pt]  [font=\tiny]  {$2$};
\draw (264,638.4) node [anchor=north west][inner sep=0.75pt]  [font=\tiny]  {$2$};
\draw (281,658.4) node [anchor=north west][inner sep=0.75pt]  [font=\tiny]  {$1$};
\draw (209,652.4) node [anchor=north west][inner sep=0.75pt]  [font=\tiny]  {$1$};
\draw (136,654.4) node [anchor=north west][inner sep=0.75pt]  [font=\tiny]  {$B$};
\draw (105,654.4) node [anchor=north west][inner sep=0.75pt]  [font=\tiny]  {$B$};
\draw (350,45.4) node [anchor=north west][inner sep=0.75pt]  [font=\tiny]  {$( 5/2,3/2)$};
\draw (490,64.4) node [anchor=north west][inner sep=0.75pt]  [font=\tiny]  {$2$};
\draw (506,63.4) node [anchor=north west][inner sep=0.75pt]  [font=\tiny]  {$2$};
\draw (349,139.4) node [anchor=north west][inner sep=0.75pt]  [font=\tiny]  {$( 5/3,4/3)$};
\draw (552,56.4) node [anchor=north west][inner sep=0.75pt]  [font=\tiny]  {$D$};
\draw (514,132.4) node [anchor=north west][inner sep=0.75pt]  [font=\tiny]  {$B$};
\draw (465,132.4) node [anchor=north west][inner sep=0.75pt]  [font=\tiny]  {$B$};
\draw (344,257.4) node [anchor=north west][inner sep=0.75pt]  [font=\tiny]  {$( 6/4,5/4)$};
\draw (504,221.4) node [anchor=north west][inner sep=0.75pt]  [font=\tiny]  {$B$};
\draw (455,220.4) node [anchor=north west][inner sep=0.75pt]  [font=\tiny]  {$B$};
\draw (505,246.4) node [anchor=north west][inner sep=0.75pt]     [font=\tiny] {$triple$};
\draw (337,361.4) node [anchor=north west][inner sep=0.75pt]  [font=\tiny]  {$( 6/5,8/5)$};
\draw (338,451.4) node [anchor=north west][inner sep=0.75pt]  [font=\tiny]  {$( 10/7,8/7)$};
\draw (576,352.4) node [anchor=north west][inner sep=0.75pt]  [font=\tiny]  {$C$};
\draw (483,386.4) node [anchor=north west][inner sep=0.75pt]     [font=\tiny] {$cusp$};
\draw (482,474.4) node [anchor=north west][inner sep=0.75pt]    [font=\tiny]  {$cusp$};
\draw (30,66.4) node [anchor=north west][inner sep=0.75pt]  [color={rgb, 255:red, 208; green, 2; blue, 27 }  ,opacity=1 ]   [font=\tiny] {$VIII-2$};
\draw (31,160.4) node [anchor=north west][inner sep=0.75pt]  [color={rgb, 255:red, 208; green, 2; blue, 27 }  ,opacity=1 ]   [font=\tiny] {$VII^*$};
\draw (29,282.4) node [anchor=north west][inner sep=0.75pt]  [color={rgb, 255:red, 208; green, 2; blue, 27 }  ,opacity=1 ]  [font=\tiny]  {$IX-3$};
\draw (28,422.4) node [anchor=north west][inner sep=0.75pt]  [color={rgb, 255:red, 208; green, 2; blue, 27 }  ,opacity=1 ]   [font=\tiny] {$IX-4$};
\draw (30,550.4) node [anchor=north west][inner sep=0.75pt]  [color={rgb, 255:red, 208; green, 2; blue, 27 }  ,opacity=1 ]  [font=\tiny]  {$VI$};
\draw (29,658.4) node [anchor=north west][inner sep=0.75pt]  [color={rgb, 255:red, 208; green, 2; blue, 27 }  ,opacity=1 ]  [font=\tiny]  {$III$};
\draw (367,65.4) node [anchor=north west][inner sep=0.75pt]  [color={rgb, 255:red, 208; green, 2; blue, 27 }  ,opacity=1 ]   [font=\tiny] {$IX-1$};
\draw (366,164.4) node [anchor=north west][inner sep=0.75pt]  [color={rgb, 255:red, 208; green, 2; blue, 27 }  ,opacity=1 ]  [font=\tiny]  {$IX-2$};
\draw (361,280.4) node [anchor=north west][inner sep=0.75pt]  [color={rgb, 255:red, 208; green, 2; blue, 27 }  ,opacity=1 ]  [font=\tiny]  {$V$};
\draw (362,384.4) node [anchor=north west][inner sep=0.75pt]  [color={rgb, 255:red, 208; green, 2; blue, 27 }  ,opacity=1 ]   [font=\tiny] {$VII$};
\draw (361,478.4) node [anchor=north west][inner sep=0.75pt]  [color={rgb, 255:red, 208; green, 2; blue, 27 }  ,opacity=1 ]  [font=\tiny]  {$VIII-1$};

\end{tikzpicture}
  \caption{The dual graph for the singular fibers relevant for 4d  rank two SCFTs admitting an isotrivial family.}
  \label{dualscft}
  \end{center}
\end{figure}

The configuration in table. \ref{iso} should give the SW geometry for a
 rank two 4d SCFTs with only $t$ deformation (without the $u$ deformation and mass deformation). Here one use $t, u$ to denote the expectation values for two Coulomb branch
 operators (the scaling dimension satisfies $[v]\geq [u]$) \footnote{See \cite{Argyres:2022lah} for the  curves with two 
 Coulomb branch deformations (one can easily check that if one set the $u$ parameters of their curves to be zero, one can get the same curve as listed in table. \ref{iso}}.
Notice that all the singular fibers are of the elliptic type [1], and the dual graphs for them are listed in figure. \ref{dualscft}.

 \begin{table}
\begin{center}
\begin{tabular}{|c|c|c|}
\hline
No& $F_0,F_\infty$ & Families  \\ \hline
$1$ & $(I_{0-0-0}^*, I_{0-0-0}^*)$ & $y^2=t(x^5+ax^4+bx^3+cx^2+x)$ \\ \hline
$2$ &$ II,II$ & $y^2=t^6+dx^4t+ex^2t^2+t^3$ \\ \hline
$3$ &$ III,III$ & $y^2=t^6+dx^3t+t^2$ \\ \hline
$2$ &$ IV,IV$ & $y^2=(t^6+dx^3t+t^2)t$ \\ \hline
$2$ &$ V,V^*$ & $y^2=t^6+t$ \\ \hline
$2$ &$ VI,VI$ & $y^2=t^5+dtx^3+t^2x$ \\ \hline
$2$ &$ VII,VII^*$ & $y^2=x^5+xt$ \\ \hline
$2$ &$ VIII-1,VIII-4$ & $y^2=x^5+t$ \\ \hline
$2$ &$ VIII-2,VIII-3$ & $y^2=x^5+t^3$ \\ \hline
$2$ &$ IX-1,IX-4$ & $y^2=t^5+t^2$ \\ \hline
$2$ &$ IX-2,IX-3$ & $y^2=t(t+x^5)$ \\ \hline
\end{tabular}
\end{center}
\caption{The isotrivial families of genus two fibration with two singular fibers.}
\label{iso}
\end{table}%

\textbf{Scaling dimension from the curve}: The eigenvalues of monodromy group for the singular fibers can be used to constrain the possible scaling dimension, see table. \ref{scalingdimension}.
We also use the relation between the dual graph and the 3d mirror  to get the precise scaling dimension in section. \ref{data}. On the other hand, one can find 
the scaling dimension of coordinate $t$ by using  following formula satisfied by SW differential:
\begin{equation*}
\partial_t \lambda \subset \Omega^{1,0}.
\end{equation*}
Here $\lambda$ is the SW differential, and $\Omega^{1,0}$ is the space of holomorphic differential for the genus two curve. $\Omega$ is generated by ${dx\over y}, {xdx\over y}$ for 
genus two hypelliptic curve defined by the equation $y^2=f(x)$. 
Now $\lambda$ has scaling dimension one, so the following equations for scaling dimensions hold (using above equation, so $[\partial_t \lambda ]=[\omega_i]$ (with $\omega$ one of the basis of the holomorphic differential):
\begin{equation*}
1-[t]=[x]-[y],~~~or~~~1-t=2[x]-[y].
\end{equation*}
On the other hand, the SW curve is weighted homogenous, and from which one can get two more equations relating the scaling dimensions of $t, x,y$. Using above two three equations, one can find out the scaling dimension of $t$. 
The result is consistent with the data listed in section. \ref{data}.

\textit{Example}: Let's take the configuration $(V,V^*)$, and the curve takes the form (the singular fiber of $V$ is put at $t=0$):
\begin{equation*}
y^2=x^6+t.
\end{equation*}
There are following equations for the scaling dimensions:
\begin{equation*}
1-[t]=[x]-[y],~~2[y]=6[x],~~2[y]=[t],
\end{equation*}
and the solution is $[x]=\frac{1}{4}~,[y]=\frac{3}{4},~[t]=\frac{3}{2}$.   Here one need to assume $\partial_t\Omega \propto {dx\over y}$.

Let's now instead put singular fiber $V^*$ at $t=0$, and the curve changes as $y^2=x^6+t^5$ (since $y$ ($x$) is a section of  $O(-3)$ ($O(-1)$) bundle on $\mathbb{P}^1$), and the equations
for the scaling dimension change as
\begin{equation*}
1-[t]=2[x]-[y],~~2[y]=6[x],~~2[y]=5[t]
\end{equation*}
Notice that here we have to use $\partial_t \lambda \subset {xdx\over y}$ to get  a consistent solution. The solution for above equation is $[x]=5,[y]=15,[t]=6$.

\textbf{Singular configuration for generic deformation}:  Using the result for the deformation of singularity, one can easily write down the generic deformation patter for the SCFT. See table. \ref{4ddeform} for details.
We get the result by keeping the singular fiber at infinity intact, but split the singular fiber at $t=0$ into $I_1$ and $\tilde{I}_1$ fibers, whose number is given by the local invariant  $d_x, \delta_x$.
We notice that several of them do not correspond to known theories.

 \begin{table}[H]
\begin{center}
\begin{tabular}{|c|c|c|}
\hline
Singular configuration & Scaling dimension &Flavor   \\ \hline
$( II,I_1^{10})$ & $(2,2)$ & $SU(2)^5$ \\ \hline
$(III,I_1^{10})$ & $(3,2)$ & $U(6)$ \\ \hline
$(V^*, I_1^5)$ & $(\frac{6}{4}, \frac{5}{4})$ & $U(1)$ \\ \hline
$(VI, I_1^{10})$ &  $(4,2)$ & $SO(12)$ \\ \hline
$(VII^*, I_1^{5})$ &$ (\frac{8}{5}, \frac{6}{5})$ & $SU(2)$ \\ \hline
$(VII, I_1^{15})$ &$ (8, 6)$ & $SO(20)$ \\ \hline
$(VIII-4, I_1^4)$ &$(\frac{10}{7},\frac{8}{7})$ & $\emptyset$ \\ \hline
$(VIII-3, I_1^{12})$ &$(10,4)$ & $E_8$ \\ \hline
$(IX-4, I_1^{8})$ &$ (\frac{5}{2},\frac{3}{2})$ & $SU(5)$ \\ \hline
$(IX-3, I_1^{6})$ &$ (\frac{5}{3},\frac{4}{3})$ & $SU(2)\times U(1)$ \\ \hline
$(IX-2, I_1^{14})$ &$ (5,4)$ & $SU(10)$ \\ \hline
$(IX-1, I_1^{12})$ &$ (5,3)$ & $SO(14)\times U(1)$ \\ \hline
\end{tabular}
\begin{tabular}{|c|c|c|}
\hline
Singular configuration & Scaling dimension &Theory   \\ \hline
$( IV,I_1^{8}\tilde{I}_1)$ & $(6,2)$ & $G_2-4(\textbf{7})$? \\ \hline
$(V, I_1^{15})$ & $(6, 5)$ & $?$ \\ \hline
$(VIII-1, I_1^{16})$ &  $(10,8)$ & $?$ \\ \hline
$(VIII-2, I_1^{6}\tilde{I}_1)$ &$ (\frac{10}{3}, \frac{4}{3})$ & $?$ \\ \hline
\end{tabular}

\end{center}
\caption{The generic deformations for 4d rank two SCFTs which admit an isotrivial family. The status for the entries in the second table is not clear.}
\label{4ddeform}
\end{table}%

\begin{figure}
\begin{center}
\tikzset{every picture/.style={line width=0.75pt}} 

\begin{tikzpicture}[x=0.45pt,y=0.45pt,yscale=-1,xscale=1]

\draw    (80,80) -- (260,80) ;
\draw    (220,40) -- (220,140) ;
\draw    (200,120) -- (280,120) ;
\draw    (240,100) -- (240,160) ;
\draw    (227,151) -- (307,151) ;
\draw    (128,48) -- (128,148) ;
\draw    (108,128) -- (188,128) ;
\draw    (148,108) -- (148,168) ;
\draw    (135,159) -- (215,159) ;
\draw    (91,62) -- (91,162) ;
\draw    (28,152) -- (108,152) ;
\draw    (70,360) -- (210,360) ;
\draw    (99,319) -- (179,399) ;
\draw    (99,399) -- (179,319) ;
\draw    (111,373) -- (111,433) ;
\draw    (167,372) -- (167,432) ;
\draw   (461,80.75) .. controls (461,69.84) and (469.84,61) .. (480.75,61) .. controls (491.66,61) and (500.5,69.84) .. (500.5,80.75) .. controls (500.5,91.66) and (491.66,100.5) .. (480.75,100.5) .. controls (469.84,100.5) and (461,91.66) .. (461,80.75) -- cycle ;
\draw   (506,113.75) .. controls (506,106.16) and (512.16,100) .. (519.75,100) .. controls (527.34,100) and (533.5,106.16) .. (533.5,113.75) .. controls (533.5,121.34) and (527.34,127.5) .. (519.75,127.5) .. controls (512.16,127.5) and (506,121.34) .. (506,113.75) -- cycle ;
\draw   (536,141.75) .. controls (536,134.16) and (542.16,128) .. (549.75,128) .. controls (557.34,128) and (563.5,134.16) .. (563.5,141.75) .. controls (563.5,149.34) and (557.34,155.5) .. (549.75,155.5) .. controls (542.16,155.5) and (536,149.34) .. (536,141.75) -- cycle ;
\draw   (570,171.75) .. controls (570,164.16) and (576.16,158) .. (583.75,158) .. controls (591.34,158) and (597.5,164.16) .. (597.5,171.75) .. controls (597.5,179.34) and (591.34,185.5) .. (583.75,185.5) .. controls (576.16,185.5) and (570,179.34) .. (570,171.75) -- cycle ;
\draw   (602,198.75) .. controls (602,191.16) and (608.16,185) .. (615.75,185) .. controls (623.34,185) and (629.5,191.16) .. (629.5,198.75) .. controls (629.5,206.34) and (623.34,212.5) .. (615.75,212.5) .. controls (608.16,212.5) and (602,206.34) .. (602,198.75) -- cycle ;
\draw   (433,128.75) .. controls (433,121.16) and (439.16,115) .. (446.75,115) .. controls (454.34,115) and (460.5,121.16) .. (460.5,128.75) .. controls (460.5,136.34) and (454.34,142.5) .. (446.75,142.5) .. controls (439.16,142.5) and (433,136.34) .. (433,128.75) -- cycle ;
\draw   (405,167.75) .. controls (405,160.16) and (411.16,154) .. (418.75,154) .. controls (426.34,154) and (432.5,160.16) .. (432.5,167.75) .. controls (432.5,175.34) and (426.34,181.5) .. (418.75,181.5) .. controls (411.16,181.5) and (405,175.34) .. (405,167.75) -- cycle ;
\draw   (376,202.75) .. controls (376,195.16) and (382.16,189) .. (389.75,189) .. controls (397.34,189) and (403.5,195.16) .. (403.5,202.75) .. controls (403.5,210.34) and (397.34,216.5) .. (389.75,216.5) .. controls (382.16,216.5) and (376,210.34) .. (376,202.75) -- cycle ;
\draw   (348,236.75) .. controls (348,229.16) and (354.16,223) .. (361.75,223) .. controls (369.34,223) and (375.5,229.16) .. (375.5,236.75) .. controls (375.5,244.34) and (369.34,250.5) .. (361.75,250.5) .. controls (354.16,250.5) and (348,244.34) .. (348,236.75) -- cycle ;
\draw   (467,29.75) .. controls (467,22.16) and (473.16,16) .. (480.75,16) .. controls (488.34,16) and (494.5,22.16) .. (494.5,29.75) .. controls (494.5,37.34) and (488.34,43.5) .. (480.75,43.5) .. controls (473.16,43.5) and (467,37.34) .. (467,29.75) -- cycle ;
\draw    (481.5,42.22) -- (481.5,60.22) ;
\draw    (497.5,93.22) -- (509.5,104.22) ;
\draw    (529.5,123.22) -- (540.5,133.22) ;
\draw    (561.5,151.22) -- (573.5,162.22) ;
\draw    (595.5,180.22) -- (607.5,190.22) ;
\draw    (471.5,98.22) -- (455.5,119.22) ;
\draw    (439.5,141.22) -- (427.5,157.22) ;
\draw    (410.5,178.22) -- (398.5,192.22) ;
\draw    (379.5,212.22) -- (368.5,226.22) ;
\draw    (99,577) -- (179,497) ;
\draw    (101.5,506.22) -- (191.5,579.22) ;
\draw   (382,344.75) .. controls (382,333.84) and (390.84,325) .. (401.75,325) .. controls (412.66,325) and (421.5,333.84) .. (421.5,344.75) .. controls (421.5,355.66) and (412.66,364.5) .. (401.75,364.5) .. controls (390.84,364.5) and (382,355.66) .. (382,344.75) -- cycle ;
\draw   (481,344.75) .. controls (481,333.84) and (489.84,325) .. (500.75,325) .. controls (511.66,325) and (520.5,333.84) .. (520.5,344.75) .. controls (520.5,355.66) and (511.66,364.5) .. (500.75,364.5) .. controls (489.84,364.5) and (481,355.66) .. (481,344.75) -- cycle ;
\draw   (439,302.75) .. controls (439,295.16) and (445.16,289) .. (452.75,289) .. controls (460.34,289) and (466.5,295.16) .. (466.5,302.75) .. controls (466.5,310.34) and (460.34,316.5) .. (452.75,316.5) .. controls (445.16,316.5) and (439,310.34) .. (439,302.75) -- cycle ;
\draw   (346,388.75) .. controls (346,381.16) and (352.16,375) .. (359.75,375) .. controls (367.34,375) and (373.5,381.16) .. (373.5,388.75) .. controls (373.5,396.34) and (367.34,402.5) .. (359.75,402.5) .. controls (352.16,402.5) and (346,396.34) .. (346,388.75) -- cycle ;
\draw   (524,397.75) .. controls (524,390.16) and (530.16,384) .. (537.75,384) .. controls (545.34,384) and (551.5,390.16) .. (551.5,397.75) .. controls (551.5,405.34) and (545.34,411.5) .. (537.75,411.5) .. controls (530.16,411.5) and (524,405.34) .. (524,397.75) -- cycle ;
\draw    (369,378) -- (387.5,359.22) ;
\draw    (532,386) -- (512.5,362.22) ;
\draw    (481.5,347.22) -- (422.5,347.22) ;
\draw    (485.5,331.22) -- (464.5,311.22) ;
\draw    (440.5,310.22) -- (417.5,331.22) ;
\draw   (361,516.75) .. controls (361,505.84) and (369.84,497) .. (380.75,497) .. controls (391.66,497) and (400.5,505.84) .. (400.5,516.75) .. controls (400.5,527.66) and (391.66,536.5) .. (380.75,536.5) .. controls (369.84,536.5) and (361,527.66) .. (361,516.75) -- cycle ;
\draw   (482,519.75) .. controls (482,508.84) and (490.84,500) .. (501.75,500) .. controls (512.66,500) and (521.5,508.84) .. (521.5,519.75) .. controls (521.5,530.66) and (512.66,539.5) .. (501.75,539.5) .. controls (490.84,539.5) and (482,530.66) .. (482,519.75) -- cycle ;
\draw    (486.5,507.22) -- (399.5,507.22) ;
\draw    (483.5,518.22) -- (401.5,518.22) ;
\draw    (483.5,528.22) -- (397.5,528.22) ;

\draw (67,101) node [anchor=north west][inner sep=0.75pt]   [align=left] {2B};
\draw (160,62) node [anchor=north west][inner sep=0.75pt]   [align=left] {5};
\draw (208,85) node [anchor=north west][inner sep=0.75pt]   [align=left] {4};
\draw (228,105) node [anchor=north west][inner sep=0.75pt]   [align=left] {3};
\draw (248,125) node [anchor=north west][inner sep=0.75pt]   [align=left] {2};
\draw (268,153) node [anchor=north west][inner sep=0.75pt]   [align=left] {1};
\draw (114,91) node [anchor=north west][inner sep=0.75pt]   [align=left] {4};
\draw (134,111) node [anchor=north west][inner sep=0.75pt]   [align=left] {3};
\draw (153,131) node [anchor=north west][inner sep=0.75pt]   [align=left] {2};
\draw (167,157) node [anchor=north west][inner sep=0.75pt]   [align=left] {1};
\draw (49,134) node [anchor=north west][inner sep=0.75pt]   [align=left] {1};
\draw (357,228) node [anchor=north west][inner sep=0.75pt]   [align=left] {1};
\draw (440,119) node [anchor=north west][inner sep=0.75pt]   [align=left] {4};
\draw (514,105) node [anchor=north west][inner sep=0.75pt]   [align=left] {4};
\draw (412,160) node [anchor=north west][inner sep=0.75pt]   [align=left] {3};
\draw (545,133) node [anchor=north west][inner sep=0.75pt]   [align=left] {3};
\draw (578,164) node [anchor=north west][inner sep=0.75pt]   [align=left] {2};
\draw (610,188) node [anchor=north west][inner sep=0.75pt]   [align=left] {1};
\draw (384,192) node [anchor=north west][inner sep=0.75pt]   [align=left] {2};
\draw (475,72) node [anchor=north west][inner sep=0.75pt]   [align=left] {5};
\draw (476,20) node [anchor=north west][inner sep=0.75pt]   [align=left] {2};
\draw (7,180.4) node [anchor=north west][inner sep=0.75pt]  [font=\tiny]  {$( 5,4)$};
\draw (4,384.4) node [anchor=north west][inner sep=0.75pt]  [font=\tiny]  {$( 5/2,3/2)$};
\draw (210,352.62) node [anchor=north west][inner sep=0.75pt]    {$D$};
\draw (123,370) node [anchor=north west][inner sep=0.75pt]   [align=left] {2};
\draw (150,371) node [anchor=north west][inner sep=0.75pt]   [align=left] {2};
\draw (179,477.62) node [anchor=north west][inner sep=0.75pt]    {$B$};
\draw (91,479.62) node [anchor=north west][inner sep=0.75pt]    {$B$};
\draw (8,526.4) node [anchor=north west][inner sep=0.75pt]  [font=\tiny]  {$( 3/2,5/4)$};
\draw (448,294) node [anchor=north west][inner sep=0.75pt]   [align=left] {1};
\draw (493,336) node [anchor=north west][inner sep=0.75pt]   [align=left] {2};
\draw (533,388) node [anchor=north west][inner sep=0.75pt]   [align=left] {1};
\draw (395,337) node [anchor=north west][inner sep=0.75pt]   [align=left] {2};
\draw (354,380) node [anchor=north west][inner sep=0.75pt]   [align=left] {1};
\draw (161,526.62) node [anchor=north west][inner sep=0.75pt]    {$triple$};
\draw (375,509) node [anchor=north west][inner sep=0.75pt]   [align=left] {1};
\draw (495,513) node [anchor=north west][inner sep=0.75pt]   [align=left] {1};

\end{tikzpicture}
  \caption{3d mirror  for 4d SCFT defined using the genus two singular fiber. The 3d mirror is read from the dual graph of the singular fiber.}
  \label{figmirror}

\end{center}
\end{figure}
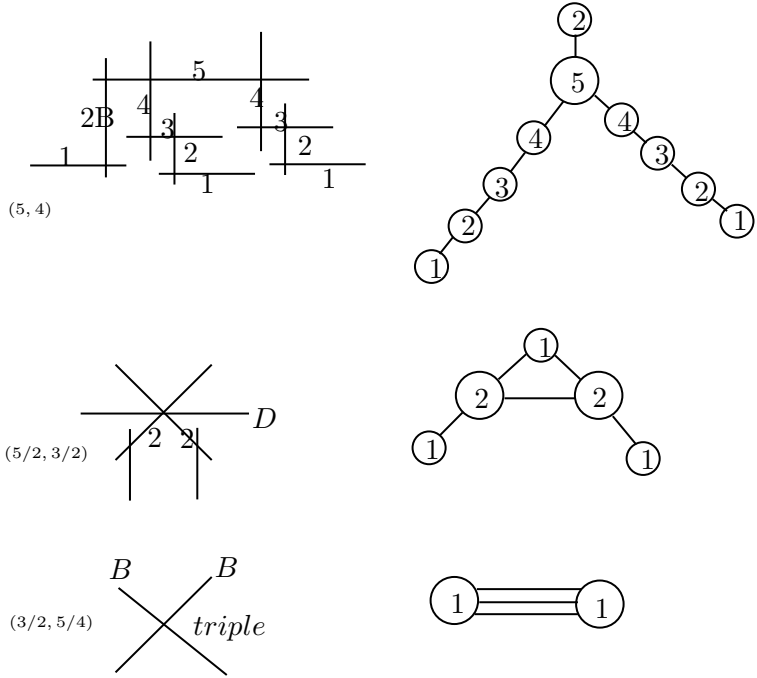

\textbf{3d Mirror}: If the 4d theory is put on a circle, then one can get a 3d $\mathcal{N}=4$ SCFT in the IR. 3d $\mathcal{N}=4$ theory has interesting 
mirror symmetry. In our case, the 3d mirror theory could be read from the dual graph of the singular fiber $X_0$, see figure. \ref{figmirror}. The rule is the following: Let's take 
$X_0=\sum n_i C_i$, the 3d mirror is found as follows:
\begin{enumerate}
\item If $X_0$ is of $D$ or $(B,B)$ type, then the mirror quiver is formed as follows: there is a $U(n_i)$ quiver node, and the number of bi-fundamental hypers between them is $C_i\cdot C_j$.
\item If $X_0$ is $2B$ type, one need to first modify the dual graph as follows: if  there are two $-2$ curve intersecting with the $2B$ curve, and one need to remove one of them. The new dual 
graph is denoted as $X_0^{'}$, and one can find a quiver by the same rule as above.
\end{enumerate}
One can check that the mirror quiver is the same as found in the literature, see \cite{xie:2022lowrank}.

\subsubsection{Other rank two SCFTs}
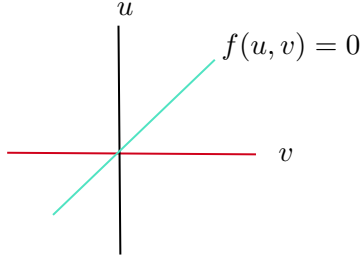
\begin{figure}
\begin{center}

\tikzset{every picture/.style={line width=0.75pt}} 

\begin{tikzpicture}[x=0.35pt,y=0.35pt,yscale=-1,xscale=1]

\draw [color={rgb, 255:red, 208; green, 2; blue, 27 }  ,draw opacity=1 ]   (161,238) -- (427,239.72) ;
\draw    (280,101) -- (282,347.72) ;
\draw [color={rgb, 255:red, 80; green, 227; blue, 194 }  ,draw opacity=1 ]   (210,305) -- (383,137.72) ;

\draw (448,230.4) node [anchor=north west][inner sep=0.75pt]    {$v$};
\draw (275,72.4) node [anchor=north west][inner sep=0.75pt]    {$u$};
\draw (387,104.4) node [anchor=north west][inner sep=0.75pt]    {$f( u,v) =0$};

\end{tikzpicture}

\end{center}
\caption{The two dimensional Coulomb branch for a rank two theory. On the $u=0$ plane ($v$ flane), 
the genus two fibration is degenerated into a (singular) genus one fibration. To get a genus two fibration, one need to find a slice defined by an equation $f(u,v)=0$. }
\label{twobase}
\end{figure}

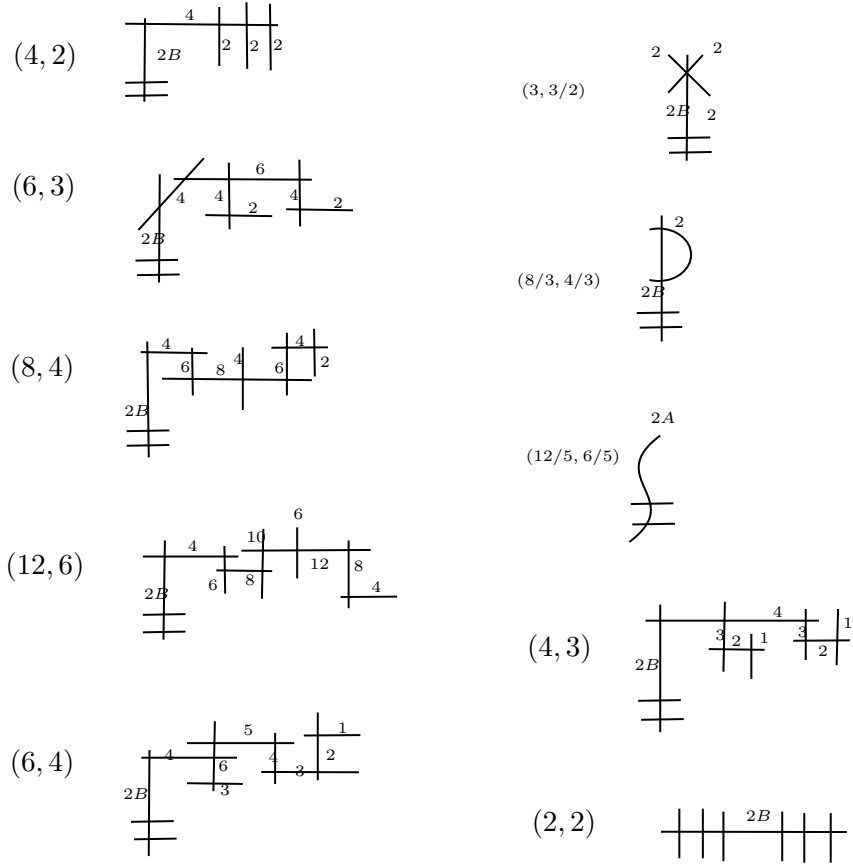
\begin{figure}[H]
\begin{center}
 \tikzset{every picture/.style={line width=0.75pt}} 

\begin{tikzpicture}[x=0.55pt,y=0.55pt,yscale=-1,xscale=1]

\draw    (148,119) -- (103.5,168.22) ;
\draw    (165,122) -- (165.5,167.72) ;
\draw    (149,158) -- (194.5,158.72) ;
\draw    (127.5,133.22) -- (221.5,133.72) ;
\draw    (213,120) -- (213.5,165.72) ;
\draw    (204,154) -- (249.5,154.72) ;
\draw    (118,130) -- (117.5,204.22) ;
\draw    (109.5,244.72) -- (110.5,324.22) ;
\draw    (174.5,248.72) -- (174.5,291.72) ;
\draw    (204.5,238.72) -- (204.5,281.72) ;
\draw    (194,249) -- (232.5,248.72) ;
\draw    (140,249) -- (140.5,281.72) ;
\draw    (223,236) -- (223.5,268.72) ;
\draw    (119.5,270.22) -- (221.5,271.22) ;
\draw    (105,252) -- (150.5,252.72) ;
\draw    (173.5,388.22) -- (261.5,387.72) ;
\draw    (194.5,402.22) -- (156.5,401.72) ;
\draw    (171.5,392.22) -- (106.5,392.22) ;
\draw    (211.5,372.22) -- (211.5,407.22) ;
\draw    (246.5,381.22) -- (246.5,427.72) ;
\draw    (241,420) -- (279.5,419.72) ;
\draw    (188.5,373.22) -- (187.5,421.22) ;
\draw    (162,385) -- (162.5,417.72) ;
\draw    (94.5,26.72) -- (198.5,27.72) ;
\draw    (108,23) -- (107.5,80.22) ;
\draw    (158.5,15.22) -- (158.5,55.72) ;
\draw    (177,12) -- (177.5,57.72) ;
\draw    (193,13) -- (193.5,58.72) ;
\draw    (459.5,158.22) -- (459.5,244.72) ;
\draw    (121.5,381.22) -- (120.5,449.22) ;
\draw  [draw opacity=0] (451.22,202.47) .. controls (453.21,202.96) and (455.32,203.22) .. (457.5,203.22) .. controls (469.65,203.22) and (479.5,195.16) .. (479.5,185.22) .. controls (479.5,175.28) and (469.65,167.22) .. (457.5,167.22) .. controls (455.32,167.22) and (453.21,167.48) .. (451.22,167.96) -- (457.5,185.22) -- cycle ; \draw   (451.22,202.47) .. controls (453.21,202.96) and (455.32,203.22) .. (457.5,203.22) .. controls (469.65,203.22) and (479.5,195.16) .. (479.5,185.22) .. controls (479.5,175.28) and (469.65,167.22) .. (457.5,167.22) .. controls (455.32,167.22) and (453.21,167.48) .. (451.22,167.96) ;  
\draw    (437.5,382.22) .. controls (477.5,352.22) and (418.5,339.22) .. (458.5,309.22) ;
\draw    (464,48) -- (492.5,76.22) ;
\draw    (464,74) -- (487.5,48.22) ;
\draw    (477,48) -- (476.5,120.72) ;
\draw    (136.5,520.22) -- (209.5,520.22) ;
\draw    (174.5,548.22) -- (136.5,547.72) ;
\draw    (170.5,530.22) -- (105.5,530.22) ;
\draw    (196.5,513.22) -- (196.5,548.22) ;
\draw    (225.5,499.22) -- (225.5,545.72) ;
\draw    (216,515) -- (254.5,514.72) ;
\draw    (155.5,505.22) -- (154.5,553.22) ;
\draw    (111,525) -- (110.5,597.72) ;
\draw    (187,540) -- (253.5,540.22) ;
\draw    (448.5,436.22) -- (566.5,436.22) ;
\draw    (529.5,456.22) -- (491.5,455.72) ;
\draw    (520.5,445.22) -- (520.5,476.22) ;
\draw    (557.5,428.22) -- (557.5,463.22) ;
\draw    (579.5,428.22) -- (578.5,466.72) ;
\draw    (549,450) -- (587.5,449.72) ;
\draw    (502.5,423.22) -- (501.5,471.22) ;
\draw    (458.5,425.22) -- (458.5,512.72) ;
\draw    (459,581) -- (585.5,581.22) ;
\draw    (471.5,599.22) -- (471.5,565.22) ;
\draw    (487.5,599.22) -- (487.5,565.22) ;
\draw    (501.5,601.22) -- (501.5,567.22) ;
\draw    (541.5,601.22) -- (541.5,567.22) ;
\draw    (556.5,602.22) -- (556.5,568.22) ;
\draw    (574.5,602.22) -- (574.5,568.22) ;
\draw    (123.5,66.72) -- (94.5,67.22) ;
\draw    (123.5,75.72) -- (94.5,76.22) ;
\draw    (130.5,188.72) -- (101.5,189.22) ;
\draw    (131.5,198.72) -- (102.5,199.22) ;
\draw    (124.5,305.72) -- (95.5,306.22) ;
\draw    (124.5,315.72) -- (95.5,316.22) ;
\draw    (135.5,431.72) -- (106.5,432.22) ;
\draw    (135.5,443.72) -- (106.5,444.22) ;
\draw    (127.5,573.72) -- (98.5,574.22) ;
\draw    (129.5,585.72) -- (100.5,586.22) ;
\draw    (492.5,104.72) -- (463.5,105.22) ;
\draw    (493.5,114.72) -- (464.5,115.22) ;
\draw    (471.5,224.72) -- (442.5,225.22) ;
\draw    (473.5,234.72) -- (444.5,235.22) ;
\draw    (467.5,355.72) -- (438.5,356.22) ;
\draw    (468.5,369.72) -- (439.5,370.22) ;
\draw    (472.5,490.72) -- (443.5,491.22) ;
\draw    (474.5,503.72) -- (445.5,504.22) ;

\draw (103,168.4) node [anchor=north west][inner sep=0.75pt]  [font=\tiny]  {$2B$};
\draw (127,140.4) node [anchor=north west][inner sep=0.75pt]  [font=\tiny]  {$4$};
\draw (153,139.4) node [anchor=north west][inner sep=0.75pt]  [font=\tiny]  {$4$};
\draw (176,147.4) node [anchor=north west][inner sep=0.75pt]  [font=\tiny]  {$2$};
\draw (204,138.4) node [anchor=north west][inner sep=0.75pt]  [font=\tiny]  {$4$};
\draw (181,120.4) node [anchor=north west][inner sep=0.75pt]  [font=\tiny]  {$6$};
\draw (234,144.4) node [anchor=north west][inner sep=0.75pt]  [font=\tiny]  {$2$};
\draw (15,127.4) node [anchor=north west][inner sep=0.75pt]    {$( 6,3)$};
\draw (194,257.4) node [anchor=north west][inner sep=0.75pt]  [font=\tiny]  {$6$};
\draw (166,251.4) node [anchor=north west][inner sep=0.75pt]  [font=\tiny]  {$4$};
\draw (208,238.4) node [anchor=north west][inner sep=0.75pt]  [font=\tiny]  {$4$};
\draw (117.25,240.76) node [anchor=north west][inner sep=0.75pt]  [font=\tiny]  {$4$};
\draw (154,258.4) node [anchor=north west][inner sep=0.75pt]  [font=\tiny]  {$8$};
\draw (225,253.4) node [anchor=north west][inner sep=0.75pt]  [font=\tiny]  {$2$};
\draw (92,286.4) node [anchor=north west][inner sep=0.75pt]  [font=\tiny]  {$2B$};
\draw (14,251.4) node [anchor=north west][inner sep=0.75pt]    {$( 8,4)$};
\draw (11,387.4) node [anchor=north west][inner sep=0.75pt]    {$( 12,6)$};
\draw (207,357.4) node [anchor=north west][inner sep=0.75pt]  [font=\tiny]  {$6$};
\draw (218,391.4) node [anchor=north west][inner sep=0.75pt]  [font=\tiny]  {$12$};
\draw (174,402.4) node [anchor=north west][inner sep=0.75pt]  [font=\tiny]  {$8$};
\draw (105.25,411.76) node [anchor=north west][inner sep=0.75pt]  [font=\tiny]  {$2B$};
\draw (248,393.4) node [anchor=north west][inner sep=0.75pt]  [font=\tiny]  {$8$};
\draw (260,407.4) node [anchor=north west][inner sep=0.75pt]  [font=\tiny]  {$4$};
\draw (175,373.4) node [anchor=north west][inner sep=0.75pt]  [font=\tiny]  {$10$};
\draw (149,406.4) node [anchor=north west][inner sep=0.75pt]  [font=\tiny]  {$6$};
\draw (362,65.4) node [anchor=north west][inner sep=0.75pt]  [font=\tiny]  {$( 3,3/2)$};
\draw (193,35.4) node [anchor=north west][inner sep=0.75pt]  [font=\tiny]  {$2$};
\draw (114,42.4) node [anchor=north west][inner sep=0.75pt]  [font=\tiny]  {$2B$};
\draw (133,15.4) node [anchor=north west][inner sep=0.75pt]  [font=\tiny]  {$4$};
\draw (16,37.4) node [anchor=north west][inner sep=0.75pt]    {$( 4,2)$};
\draw (359,195.4) node [anchor=north west][inner sep=0.75pt]  [font=\tiny]  {$( 8/3,4/3)$};
\draw (365,316.4) node [anchor=north west][inner sep=0.75pt]  [font=\tiny]  {$( 12/5,6/5)$};
\draw (158,35.4) node [anchor=north west][inner sep=0.75pt]  [font=\tiny]  {$2$};
\draw (177,36.4) node [anchor=north west][inner sep=0.75pt]  [font=\tiny]  {$2$};
\draw (130,256.4) node [anchor=north west][inner sep=0.75pt]  [font=\tiny]  {$6$};
\draw (135,379.4) node [anchor=north west][inner sep=0.75pt]  [font=\tiny]  {$4$};
\draw (443.25,204.76) node [anchor=north west][inner sep=0.75pt]  [font=\tiny]  {$2B$};
\draw (466.25,156.76) node [anchor=north west][inner sep=0.75pt]  [font=\tiny]  {$2$};
\draw (450,291.4) node [anchor=north west][inner sep=0.75pt]  [font=\tiny]  {$2A$};
\draw (488.25,83.76) node [anchor=north west][inner sep=0.75pt]  [font=\tiny]  {$2$};
\draw (460.25,80.76) node [anchor=north west][inner sep=0.75pt]  [font=\tiny]  {$2B$};
\draw (492.25,37.76) node [anchor=north west][inner sep=0.75pt]  [font=\tiny]  {$2$};
\draw (450.25,40.76) node [anchor=north west][inner sep=0.75pt]  [font=\tiny]  {$2$};
\draw (14,522.4) node [anchor=north west][inner sep=0.75pt]    {$( 6,4)$};
\draw (173,505.4) node [anchor=north west][inner sep=0.75pt]  [font=\tiny]  {$5$};
\draw (229,522.4) node [anchor=north west][inner sep=0.75pt]  [font=\tiny]  {$2$};
\draw (157,546.4) node [anchor=north west][inner sep=0.75pt]  [font=\tiny]  {$3$};
\draw (91.25,548.76) node [anchor=north west][inner sep=0.75pt]  [font=\tiny]  {$2B$};
\draw (190,524.4) node [anchor=north west][inner sep=0.75pt]  [font=\tiny]  {$4$};
\draw (208,533.4) node [anchor=north west][inner sep=0.75pt]  [font=\tiny]  {$3$};
\draw (156,530.4) node [anchor=north west][inner sep=0.75pt]  [font=\tiny]  {$6$};
\draw (119,522.4) node [anchor=north west][inner sep=0.75pt]  [font=\tiny]  {$4$};
\draw (237,504.4) node [anchor=north west][inner sep=0.75pt]  [font=\tiny]  {$1$};
\draw (366,443.4) node [anchor=north west][inner sep=0.75pt]    {$( 4,3)$};
\draw (524,442.4) node [anchor=north west][inner sep=0.75pt]  [font=\tiny]  {$1$};
\draw (564,451.4) node [anchor=north west][inner sep=0.75pt]  [font=\tiny]  {$2$};
\draw (494,440.4) node [anchor=north west][inner sep=0.75pt]  [font=\tiny]  {$3$};
\draw (439.25,460.76) node [anchor=north west][inner sep=0.75pt]  [font=\tiny]  {$2B$};
\draw (533,424.4) node [anchor=north west][inner sep=0.75pt]  [font=\tiny]  {$4$};
\draw (550,438.4) node [anchor=north west][inner sep=0.75pt]  [font=\tiny]  {$3$};
\draw (505,444.4) node [anchor=north west][inner sep=0.75pt]  [font=\tiny]  {$2$};
\draw (581,432.4) node [anchor=north west][inner sep=0.75pt]  [font=\tiny]  {$1$};
\draw (369,563.4) node [anchor=north west][inner sep=0.75pt]    {$( 2,2)$};
\draw (515,564.4) node [anchor=north west][inner sep=0.75pt]  [font=\tiny]  {$2B$};

\end{tikzpicture}

  \caption{The dual graph for singularities relevant for 4d $\mathcal{N}=2$ SCFT. Notice that for the Coulomb branch configuration $(F, I_1^b)$, the property of the UV theory 
  should be derived from the dual fiber  $F^*$. The dual pairs in table.\ref{4ddeform1} are $(2II-0,2II^*-0),(2III-0,2III^*-0),(2IV-0,2IV^*-0)$. $2I_0^*-0$ is self-dual.  }
  \label{dualforscft}
  \end{center}
\end{figure}

There are other rank two SCFTs which does not admit an isotrivial family. The reason is: if only 
Coulomb branch operator  $v$ with maximal scaling dimension is turned on,  the genus two curve would be degenerated into a singular genus one curve along the $v$ direction.
To use the genus two fibration studied in this paper, one have to turn on other deformations (such as the other Coulomb branch operator $u$
or the mass deformation), so that a genus two fibration over $\mathbb{P}^1$ can be formed, see figure. \ref{twobase}. Now the singular fiber at $\infty$ will still reflect the UV properties, but one can not find 
a genus two fibration with just two singular fibers.  One might still define these SCFTs by specifying the singular fiber at $\infty$ and the singular fibers for generic deformations, see table. \ref{4ddeform}.
The dual graph for the singularity at infinity is shown in figure. \ref{dualforscft}, from which one can also find the 3d mirror. Those 3d mirrors match the known results.

\textit{Example}: Consider  configuration $(2II-0, I_1^{13})$, which should give rank two $E_8$ theories. Notice that $2II-0$ fiber has $t_\infty=1$, and so (see formula. \ref{mw}):
$rank(MW)=11-n_t(2II-0)+1=9$, which is the same as the rank of flavor symmetry.

 \begin{table}[H]
\begin{center}
\begin{tabular}{|c|c|c|}
\hline
Singular configuration & Scaling dimension &Flavor   \\ \hline
$( 2I_0^*-0,I_1^{9})$ & $(4,2)$ & $SO(8)\times SU(2)$ \\ \hline
$(2II-0, I_1^{13})$ & $(12, 6)$ & $E_8\times SU(2)$ \\ \hline
$(2II^*-0, I_1^5)$ & $(\frac{12}{5}, \frac{6}{5})$ & $SU(2)$ \\ \hline
$(2IV-0, I_1^{11})$ & $(6, 3)$ & $E_6\times SU(2)$ \\ \hline
$(2IV^*-0, I_1^7)$ & $(3, \frac{3}{2})$ & $SU(3)\times SU(2)$ \\ \hline
$(2III-0, I_1^{12})$ & $(8, 4)$ & $E_7\times SU(2)$ \\ \hline
$(2III^*-0, I_1^6)$ & $(\frac{8}{3}, \frac{4}{3})$ & $SU(2)\times SU(2)$ \\ \hline
\end{tabular}
\begin{tabular}{|c|c|c|}
\hline
Singular configuration & Scaling dimension &Flavor   \\ \hline
$(II-II^*_0,I_1^{13})$ & $(6,4)$ & $SO(16)\times SU(2)$ \\ \hline
$(III-II_0^*,I_1^{12})$ & $(4,3)$ & $SU(8)\times SU(2)$ \\ \hline
\end{tabular}

\end{center}
\caption{The generic deformations for 4d  rank two SCFTs which do not  have an isotrivial limit. The entries in first table are rank two version 
of  rank one $E_n$ and $H_n$ type theories.}
\label{4ddeform1}
\end{table}%

\subsubsection{Asymptotic free theories}
The rank two gauge groups are $A_2, B_2, C_2, G_2, A_1\times A_1$. One can couple the gauge group with matter fields in 
various kinds of representations, so that the theory is asymptotic free. The typical example is $SU(3)$ gauge group coupled with $N_f\leq 6$ 
fundamental hypermultiplets.  In our framework, the SW solutions are represented by a configuration of  genus two singular fibers on $\mathbb{P}_1$:
Given the classification of singular fiber in last section, it is easy to find the SW geometry for various asymptotic free theories. See table .\ref{af}. 

Another possibility is to consider $SU(2)$ gauge group coupled with rank one SCFT and various free matter. The possibilities are: a) $n-SU(2)-H_1,~n=0,1,2$, b) $n-SU(2)-H_2, n=0,1,2$.
 The SW geometries for these theories are also listed in table. \ref{af}.

\begin{table}
\begin{center}
\begin{tabular}{|c|c|}
\hline
Configuration & Theory  \\ \hline
$(III_n, I_1^{10-n}),~~n=0,1\ldots, 6$ & $SU(3)-(6-n)$ \\ \hline
$(\tilde{III}_n, I_1^{10-n}),~~n=0,1\ldots, 6$ & $Sp(4)-(6-n)$ \\ \hline
$(2I_{n}^*-0, I_1^{9-n} ),~~n=0,1,\ldots, 4$ & $V-Sp(4)-(4-n)$ \\ \hline
$(III_n^*, I_1^{8-n}\tilde{I}_1),~~n=0,1\ldots, 4$ & $G_2-(4-n)(\textbf{7})$ \\ \hline
$(I^*_{n-p-0}, I_1^{10-n-p}),~~n=0,1, 2,~~p=0,1, 2$ & $(2-n)-SU(2)-SU(2)-(2-p)$    \\ \hline
$(II^*-II_n^*,I_1^{6-n})$ & $(2-n)-SU(2)-H_1,~n=0,1,2$ \\ \hline
$(III^*-II_n^*,I_1^{7-n})$ & $(2-n)-SU(2)-H_2,~n=0,1,2$ \\ \hline
\end{tabular}
\end{center}
\caption{The singular configuration for 4d asymptotic free theories.}
\label{af}
\end{table}%

\textbf{Example}: Let's consider  $SU(3)$ gauge theory coupled with $6-n$ hypermultiplets in fundamental representation. 
The fiber at $\infty$ is  $III_n$, and the bulk has $(10-n)$ $I_{1-0-0}$ (which will be denoted as $I_1$ singularity). 
This would give the SW geometry for $SU(3)$ with $6-n$ fundamental flavors.

\subsubsection{Other possibilities}
There are other singular fibers satisfying the condition \ref{4dinfinity}. Potentially, they would 
define 4d $\mathcal{N}=2$ UV complete theories. The basic data for these singular fibers are listed in table. \ref{other}, here one assume 
that all the eigenvalues are different from one. Whether those theories give rise to SCFT or just asymptotical free theories would be 
discussed in \cite{Xie:ranktwob}.

\begin{table}
\begin{center}
\resizebox{3.5in}{!}{\begin{tabular}{|c|c|c|c|c|c|c|c|c|c|c|}
\hline
Type &  Monodromy & $d_x$ & $\delta_x$ & $l$ & Eigenvalue  & $n_t$ (components)\\  \hline 
$I^*_0-I_0^*-0$ &$ \left(\begin{array}{cccc} -1&0&0&0\\0&-1&0&0\\0&0&-1&0\\0&0&0&-1\end{array}\right)$& $12$ & $12$ & -1 & $(-1,-1,-1,-1)$    &9 \\ \hline
$I_0^*-II-0$ & $ \left(\begin{array}{cccc}1&0&1&0\\0&-1&0&0\\-1&0&0&0\\0&0&0&-1\end{array}\right)$& $8$ & $8$ & $-1$  & $(-1,-1, \exp(2\pi i/6),\exp(10\pi/6))$ &5 \\ \hline
\textcolor{red}{$I_0^*-II^*-\alpha$} & $ ~$& $14$ & $14$ & $-1$ & $(-1,-1, \exp(2\pi i/6),\exp(10\pi/6))$  &11  \\ \hline
$I_0^*-IV-0$ &$ \left(\begin{array}{cccc} 0&0&1&0\\0&-1&0&0\\-1&0&-1&0\\0&0&0&-1\end{array}\right)$& $10$ & $10$ & $-1$  &  $(-1,-1, \exp(2\pi i/3),\exp(4\pi/3))$  &7\\ \hline
\textcolor{red}{$I_0^*-IV^*-\alpha$ }&  $ ~$ &$12$ & $12$ & $-1$  & $(-1,-1, \exp(2\pi i/3),\exp(4\pi/3))$  &9 \\ \hline
$I_0^*-III-0$ &$ \left(\begin{array}{cccc} 0&0&1&0\\0&-1&0&0\\-1&0&0&0\\0&0&0&-1\end{array}\right)$& $9$ & $9$ & $-1$  & $(-1,-1, \exp(2\pi i/4),\exp(6\pi/4))$  & 6 \\ \hline
$I_0^*-III^*-0$ &$ \left(\begin{array}{cccc} 0&0&-1&0\\0&-1&0&0\\1&0&0&0\\0&0&0&-1\end{array}\right)$& $15$ & $15$ & $-1$  & $(-1,-1, \exp(2\pi i/4),\exp(6\pi/4))$    &12\\ \hline
\textcolor{red}{$I_0^*-III^*-\alpha$} &$ ~$& $13$ & $13$ & $-1$  & $(-1,-1, \exp(2\pi i/4),\exp(6\pi/4))$ &10  \\ \hline
$II-IV-0$ & $ \left(\begin{array}{cccc} 0&0&1&0\\0&1&0&1\\-1&0&-1&0\\0&-1&0&0\end{array}\right)$& $6$ & $6$ & $-1$  & $(\exp(2\pi i/6),\exp(10\pi/6), \exp(2\pi i/3),\exp(4\pi/3))$  &3 \\ \hline
$II^*-IV-0$ & $ \left(\begin{array}{cccc} -1&0&-1&0\\0&1&0&0\\-1&0&-1&0\\0&0&0&1\end{array}\right)$ & $14$ & $14$ & $-1$  & $(\exp(2\pi i/6),\exp(10\pi/6), \exp(2\pi i/3),\exp(4\pi/3))$  &11\\ \hline
\textcolor{red}{$II^*-IV^*-\alpha$ }& $ ~$& $16$ & $16$ & $-1$  & $(\exp(2\pi i/6),\exp(10\pi/6), \exp(2\pi i/3),\exp(4\pi/3))$ & 13\\ \hline
\textcolor{red}{$IV^*-IV^*-\alpha$} &~& $14$ & $14$ & $-1$ & $(\exp(2\pi i/6),\exp(10\pi/6), \exp(2\pi i/3),\exp(4\pi/3))$ &11 \\ \hline
$II-III-0$ &$ \left(\begin{array}{cccc} 1&0&1&0\\0&0&0&-1\\-1&0&0&0\\0&1&0&0\end{array}\right)$& $5$ & $5$ & $-1$ & $(\exp(2\pi i/6),\exp(10\pi/6), \exp(2\pi i/4),\exp(6\pi/4))$ &2 \\ \hline
\textcolor{red}{$II^*-III-\alpha$ }&$ ~$& $11$ & $11$ & $-1$ & $(\exp(2\pi i/6),\exp(10\pi/6), \exp(2\pi i/4),\exp(6\pi/4))$ &8 \\ \hline
$IV-III-0$  &$ \left(\begin{array}{cccc} 0&0&1&0\\0&0&0&1\\-1&0&-1&0\\0&-1&0&0\end{array}\right)$& $7$ & $7$ & $-1$ & $(\exp(2\pi i/6),\exp(10\pi/6), \exp(2\pi i/3),\exp(4\pi/3))$&4  \\ \hline
\textcolor{red}{$IV-III^*-\alpha$}& $ ~$& $11$ & $11$ & $0$ &$(\exp(2\pi i/4),\exp(6\pi/4), \exp(2\pi i/3),\exp(4\pi/3))$ & 8 \\ \hline
\textcolor{red}{$IV^*-III^*-\alpha$} &$~$& $15$ & $15$ & $-1$ & $(\exp(2\pi i/4),\exp(6\pi/4), \exp(2\pi i/3),\exp(4\pi/3))$& 12  \\ \hline
$III-III-0$ & $ \left(\begin{array}{cccc} 0&0&1&0\\0&0&0&1\\-1&0&0&0\\0&-1&0&0\end{array}\right)$& $6$ & $6$ & $-1$ & $(\exp(2\pi i/4),\exp(6\pi/4), \exp(2\pi i/4),\exp(6\pi/4))$ & 3\\ \hline
\textcolor{red}{$III^*-III^*-\alpha$ }&$~$& $16$ & $16$ & $-1$ & $(\exp(2\pi i/4),\exp(6\pi/4), \exp(2\pi i/4),\exp(6\pi/4))$& 13 \\ \hline
\end{tabular}}
\end{center}
\caption{Singular fibers which may give rise to UV complete 4d $\mathcal{N}=2$ theories. }
\label{other}
\end{table}%

\newpage
\subsection{5d KK theories}
Let's now discuss the singular fiber configuration for 5d KK theories. The constraint on the topological data of the fiber at $\infty$ is 
\begin{equation}
\boxed{d_\infty-n_\infty=2+t_\infty}.
\label{5dinfinity}
\end{equation}
One can search from tables listed in section. \ref{data} for the singular fiber satisfying above condition, and there are  many candidates.  
Of course, one would like to find more constraints, i.e. the eigenvalues for the monodromy group at $\infty$ of 
the SW curve of 5d theory.  In the following, we will use the known results of 5d rank two theory to get some insights.

In fact, one can engineer 5d $\mathcal{N}=1$ theory by putting M theory on a 3d canonical singularity \cite{Xie:2017pfl}. 
A particular interesting class is the toric singularity, and the rank two cases are classified in \cite{Xie:2017pfl}. The SW geometry for  those 5d KK theories
are easy to find \cite{Aharony:1997bh}: they are just given by the monomials associated with the toric diagram. The eigenvalues at infinity can be computed 
by the combinatorial formula found in \cite{libgober1995zeta}.  

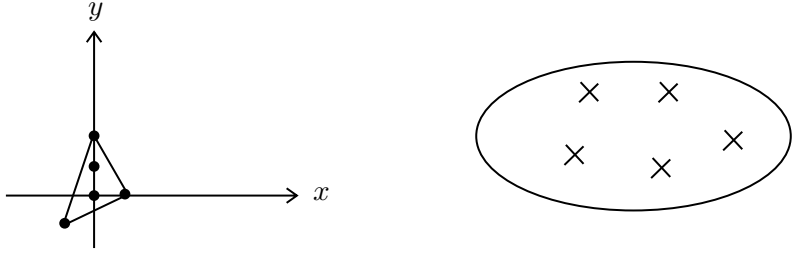
\begin{figure}
\begin{center}

\tikzset{every picture/.style={line width=0.75pt}} 

\begin{tikzpicture}[x=0.55pt,y=0.55pt,yscale=-1,xscale=1]

\draw  (23,192) -- (221,192)(83,79.78) -- (83,228) (214,187) -- (221,192) -- (214,197) (78,86.78) -- (83,79.78) -- (88,86.78)  ;
\draw    (83,151) -- (106,191.22) ;
\draw    (63,212) -- (104,192.22) ;
\draw    (82,152) -- (62,212) ;
\draw  [fill={rgb, 255:red, 0; green, 0; blue, 0 }  ,fill opacity=1 ] (80,151) .. controls (80,149.34) and (81.34,148) .. (83,148) .. controls (84.66,148) and (86,149.34) .. (86,151) .. controls (86,152.66) and (84.66,154) .. (83,154) .. controls (81.34,154) and (80,152.66) .. (80,151) -- cycle ;
\draw  [fill={rgb, 255:red, 0; green, 0; blue, 0 }  ,fill opacity=1 ] (80,172) .. controls (80,170.34) and (81.34,169) .. (83,169) .. controls (84.66,169) and (86,170.34) .. (86,172) .. controls (86,173.66) and (84.66,175) .. (83,175) .. controls (81.34,175) and (80,173.66) .. (80,172) -- cycle ;
\draw  [fill={rgb, 255:red, 0; green, 0; blue, 0 }  ,fill opacity=1 ] (101,191) .. controls (101,189.34) and (102.34,188) .. (104,188) .. controls (105.66,188) and (107,189.34) .. (107,191) .. controls (107,192.66) and (105.66,194) .. (104,194) .. controls (102.34,194) and (101,192.66) .. (101,191) -- cycle ;
\draw  [fill={rgb, 255:red, 0; green, 0; blue, 0 }  ,fill opacity=1 ] (60,211) .. controls (60,209.34) and (61.34,208) .. (63,208) .. controls (64.66,208) and (66,209.34) .. (66,211) .. controls (66,212.66) and (64.66,214) .. (63,214) .. controls (61.34,214) and (60,212.66) .. (60,211) -- cycle ;
\draw  [fill={rgb, 255:red, 0; green, 0; blue, 0 }  ,fill opacity=1 ] (80,192) .. controls (80,190.34) and (81.34,189) .. (83,189) .. controls (84.66,189) and (86,190.34) .. (86,192) .. controls (86,193.66) and (84.66,195) .. (83,195) .. controls (81.34,195) and (80,193.66) .. (80,192) -- cycle ;
\draw   (343,151.22) .. controls (343,123.05) and (390.91,100.22) .. (450,100.22) .. controls (509.09,100.22) and (557,123.05) .. (557,151.22) .. controls (557,179.39) and (509.09,202.22) .. (450,202.22) .. controls (390.91,202.22) and (343,179.39) .. (343,151.22) -- cycle ;
\draw    (414,114.22) -- (426,127.22) ;
\draw    (413,128) -- (426,115.22) ;

\draw    (468,114.22) -- (480,127.22) ;
\draw    (467,128) -- (480,115.22) ;

\draw    (404,157.22) -- (416,170.22) ;
\draw    (403,171) -- (416,158.22) ;

\draw    (463,166.22) -- (475,179.22) ;
\draw    (462,180) -- (475,167.22) ;

\draw    (512,147.22) -- (524,160.22) ;
\draw    (511,161) -- (524,148.22) ;

\draw (77,57.4) node [anchor=north west][inner sep=0.75pt]    {$y$};
\draw (230,184.4) node [anchor=north west][inner sep=0.75pt]    {$x$};

\end{tikzpicture}

\end{center}
\caption{Left:The toric diagram for a 3d canonical singularity. M theory on it gives rise to a 5d $\mathcal{N}=1$ SCFT without any flavor symmetry; Right: The Coulomb branch geometry for the 5d KK theory defined using 
the toric diagram on the left. There are a total of five $I_1$ singularities at the bulk.}
\label{5dexample}
\end{figure}

Let's give a review for how to compute the eigenvalues at infinity. A toric diagram is given as a convex polygon $P$, see an example in figure. \ref{5dexample}.
The SW geometry is given as 
\begin{equation*}
f=\sum_{(a,b)\in \Gamma} \lambda_{a,b} x^a y^b.
\end{equation*}
Here $\Gamma$ contains lattice points inside or at the boundary of the convex polygon $P$, $x$ and $y$ are $\mathbb{C}^*$ variable. 
We further assume that one of the interior point of $P$ is at the origin. So there is a one parameter family of maps $f: \mathbb{C}^* \to C$, and 
the generic fiber $X_t=f^{-1}(t)$ is a smooth curve. When $t\to \infty$, there is monodromy group $M$ action acting on the homology group $H_i(X_t)$. One define 
a Zeta function by using the monodromy group action:
\begin{equation*}
\zeta(s)=\prod det((I-sM)|H_i(X_t))^{(-1)^{i}}.
\end{equation*}
In our particular case, only $H_0$ and $H_1$ are non-vanishing, and so the $\zeta$ function takes the form
\begin{equation}
\zeta(s)={ (1-s)\over (1-s \lambda_1)\ldots (1-s\lambda_n)}.
\label{zeta}
\end{equation}
Here $n$ is the dimension of $H_1$, and $\lambda_1, \ldots, \lambda_n$ are the eigenvalues of the monodromy group acting on $H_1$. On the other hand, $\zeta(s)$ has the following combinatorial description
\begin{equation*}
\zeta(s)=\prod (1-s^{m_\sigma})^{-Vol(\sigma)}.
\end{equation*}
Here the product is over the one dimensional boundary $\sigma$ of the convex polygon, and $\sigma$ has the equation $\sum a_i^\sigma x_i=m_\sigma, m_\sigma>0$. $Vol(\sigma)$ is 
equal to interior lattice points of $\sigma$ plus one.

\textbf{Example}: Let's look at the convex polygon in figure. \ref{5dexample}. The SW geometry is given by:
\begin{equation*}
f=y^2+x+{1\over xy}+y,
\end{equation*}
and there are three one dimensional boundaries with equations: $y+2x=2,~~y-3x=2,~~x-2y=1$, and there is no interior lattice points for those boundaries. So we have 
\begin{equation*}
\zeta(s)={1\over (1-s^2)(1-s^2)(1-s)}={1\over (1-s)(1-s)(1-s)(1+s)(1+s)}
\end{equation*}
Comparing with formula $\zeta(s)$ in \ref{zeta}, one can see that the eigenvalue for $M$ on $H_1$  are $(1,1,1,1,-1,-1)$.  The weight one part of $H_1$ is just four dimensional, 
and the eigenvalues of the monodromy acting on weight one part is $(1,1,-1,-1)$. 

One quick comment is that there are always two eigenvalues $1$ if one put an interior point of the convex polygon at the origin, 
and the other pair of eigenvalues could be $(-1,-1)$ for most cases. However, it is also possible that 
the other two eigenvalues are $(\exp(i 2\pi/3), \exp(4\pi/3))$.

All possible singular fibers satisfying equation \ref{5dinfinity} are listed in table. \ref{5ddata}, and the eigenvalues are $(1,1,-1,-1)$ or $(1,1,\exp(i 2\pi/3), \exp(4\pi/3))$. 
\begin{table}[H]
\begin{center}
\resizebox{4.5in}{!}{\begin{tabular}{|c|c|c|c|c|c|c|c|}
\hline
Type &  Monodromy & $d_x$ & $\delta_x$ & $l$ & Eigenvalue  & $n_t$ (components) & Gauge algebra\\  \hline 
$IV-I_n-0$ & $ \left(\begin{array}{cccc}0&0&1&0\\0&1&0&n\\-1&0&-1&0\\0&0&0&1\end{array}\right)$& $n+4$ & $n+4$ & $-1$  & $(1, 1,\exp(2i\pi/3),\exp(4i\pi/3))$  & $n+2$ & $A_2\oplus A_{n-1}$ \\ \hline
$IV-II_n$ &$ \left(\begin{array}{cccc} 0&0&1&0\\0&1&-1&n+1\\-1&0&-1&1\\0&0&0&1\end{array}\right)$& $n+4$ & $n+4$ & $-1$  & $(1, 1,\exp(2i\pi/3),\exp(4i\pi/3))$  & $n+2$ & $A_1\oplus A_{n-1}$ \\ \hline
$IV^*-II_0$ &~& $7$ & $6$ & $-1$  & $(1, 1,\exp(2i\pi/3),\exp(4i\pi/3))$  & $4$ & $D_5$ \\ \hline
$IV^*-II_n$ &$ \left(\begin{array}{cccc} -1&0&-1&-1\\-1&1&0&n\\1&0&0&0\\0&0&0&1\end{array}\right)$& $n+7$ & $n+7$ & $-1$  & $(1,1,\exp(2i\pi/3),\exp(4i\pi/3))$  & $n+5$ & $D_5\oplus A_{n-2}$ \\ \hline
$I_n-I_p^*-0$ & $ \left(\begin{array}{cccc} -1&0&-p&0\\0&1&0&n\\0&0&-1&0\\0&0&0&1\end{array}\right)$& $n+p+6$ & $n+p+6$ & $-1$  & $(1,1,-1,-1)$ & $n+p+4$ & $A_{n-1}\oplus D_{p+4}$\\ \hline
$II_{n-p}~(n>0,~p\geq 0)$&$ \left(\begin{array}{cccc} -1&0&-p&-1\\0&1&1&n\\0&0&-1&0\\0&0&0&1\end{array}\right)$& $n+p+5$ & $n+p+5$ & $-1$ & $(1,1,-1,-1)$  & $n+p+3$ &$ A_{n-2}\oplus D_{p+3}$ \\ \hline
$\tilde{II}_{n-p},~(n> 0,p\geq 0)$ & $ \left(\begin{array}{cccc} -1&0&-p&0\\1&1&p&n\\0&0&-1&1\\0&0&0&1\end{array}\right)$& $n+p+5$ & $n+p+5$ & $-1$  & $(1,1,-1,-1)$  & $n+p+3$ & $A_{n-2}\oplus A_{p+3}$ \\ \hline
$II_{n-p}^*,~~p>0$ &$ \left(\begin{array}{cccc} 1&0&-p&0\\-1&-1&p&-n\\0&0&1&-1\\0&0&0&-1\end{array}\right)$& $n+p+5$ & $n+p+4$ & $-1$  & $(1,1,-1,-1)$  & $n+p+2$ & $A_{p-1}\oplus D_{n+2}$\\ \hline
\end{tabular}}
\end{center}
\caption{List of singular fibers which would give the Coulomb branch geometry for the 5d KK theories.}
\label{5ddata}
\end{table}%

Once the singular fiber at $\infty$ is given, it is possible to construct some candidate Coulomb branch geometry of 5d KK theories (see also \cite{Martone:2021drm} for related studies.),
the results are listed in table. \ref{5ddeform}. One interesting observation is that there are many possibilities for a given rank of flavor symmetry, 
one can compare our findings with the geometric classification in \cite{Xie:2017pfl,Jefferson:2018irk,Hayashi:2018lyv,Bhardwaj:2018yhy,Apruzzi:2019opn,Apruzzi:2019enx,Bhardwaj:2019jtr,Saxena:2020ltf}, and it seems that our theory can cover all of them (comparing our table with that in \cite{Jefferson:2018irk,Apruzzi:2019opn}). Moreover, 
it suggests that there are more. A detailed studies of those theories will be left in \cite{Xie:ranktwob}.

\begin{table}[H]
\begin{center}
\begin{tabular}{|l|c|}
\hline
Configuration & Flavor  \\ \hline
$(IV-II_0, I_1^{16})$ & $SO(20)$ \\ \hline
$(II_{0-1}^*, I_1^{14})$ & $E_8\times SU(2)$ \\ \hline
$(II_{n-p}, I_1^{15-(n+p)}),~~1\leq (n+p)\leq 10$ & $rank=10-(n+p)$ \\ \hline
$(\tilde{II}_{n-p}, I_1^{15-(n+p)}),~~1\leq (n+p)\leq10$ & $rank=10-(n+p)$ \\ \hline
$(II_{n-p}^*, I_1^{15-(n+p)}),~~1\leq (n+p)\leq10$ & $rank=10-(n+p)$ \\ \hline
\end{tabular}
\end{center}
\caption{Possible singular configurations for 5d $\mathcal{N}=1$ KK theories.}
\label{5ddeform}
\end{table}%

\subsection{6d KK theories}
Let's now give the singular fiber configuration for 6d KK theories. The constraint on the topological data of the fiber at $\infty$ is 
\begin{equation}
\boxed{d_\infty-n_\infty=1+t_\infty}
\end{equation}
The list of singularities satisfying the above equation is given in table. \ref{6ddata}. Here we further assume that at least two eigenvalues of the monodromy group is identity.
Some global Coulomb branch geometries are listed in table. \ref{6ddeform}.

\begin{table}[H]
\begin{center}
\resizebox{4.5in}{!}{\begin{tabular}{|c|c|c|c|c|c|c|c|c|c|}
\hline
Type &  Monodromy & $d_x$ & $\delta_x$ & $l$ & Eigenvalue  & $n_t$ (components) & Gauge algebra\\  \hline 
$I_0-I_n^*-0$  &$ \left(\begin{array}{cccc} 1&0&0&0\\0&-1&0&-n\\0&0&1&0\\0&0&0&-1\end{array}\right)$& $n+6$ & $n+6$ & $-1$ & $(1,1,-1,-1)$  & $n+5$ & $D_{n+4}$  \\ \hline
$II_{n-0}^*$ & $ \left(\begin{array}{cccc} 1&0&0&0\\1&-1&0&-n\\0&0&1&1\\0&0&0&-1\end{array}\right)$& $n+5$ & $n+4$ & $-1$& $(1,1,-1,-1)$  & $n+3$ & $D_{n+2}$ \\ \hline
$IV^*-I_n-0$ & $ \left(\begin{array}{cccc} -1&0&-1&0\\0&1&0&n\\1&0&0&0\\0&0&0&1\end{array}\right)$& $n+8$ & $n+8$ & $-1$  & $(1,1,\exp(2\pi/3),\exp(4\pi/3))$  & $n+7$ & $E_6\oplus A_{n-1}$\\ \hline
$I_{n-p-q},~n>0,p>0$ & $ \left(\begin{array}{cccc}1&0&p+q&-q\\0&1&-q&n+q\\0&0&1&0\\0&0&0&1\end{array}\right)$& $n+p+q$ & $n+p+q$ & $-1$  & $(1, 1,1,1)$  & $n+p+q-1$  & $A_{n-1}\oplus A_{p-1}\oplus A_{q-1}$ \\ \hline
\end{tabular}}
\end{center}
\caption{List of singular fibers which would give the Coulomb branch geometry for the 6d KK theories.}
\label{6ddata}
\end{table}%

\begin{table}[H]
\begin{center}
\begin{tabular}{|c|c|}
\hline
Configuration & Flavor  \\ \hline
$(I_{1-1-1},I_1^{17})$ & $SO(20)$ \\ \hline
$(II_{0-0}^*,I_1^{15})$ & $E_8\times SU(2)$ \\ \hline
$(I_{n-p-q}, I_1^{20-n-p-q}),~n>0,p>0,q>0$& Rank=$14-(n+p+q)$ \\ \hline
$(II_{n-0}^*,I_1^{15-n})$ & rank= $9-n$ \\ \hline
\end{tabular}
\end{center}
\caption{Possible singular configuration for 6d $(1,0)$ KK theories.}
\label{6ddeform}
\end{table}%

\section{Conclusion}
We have studied the local singularities for rank two  $\mathcal{N}=2$ Coulomb branch geometries, and 
the local invariants $d_x, \delta_x, n_t$, gauge algebra are  listed in table. \ref{elliptic1},\ref{elliptic2a},\ref{elliptic2b},\ref{parabolic3}, \ref{parabolic4}.
The low energy theories for most  of them are also included there, which could be 
SCFT, IR free gauge theory, or the sum of the above components, etc.  

We then construct some  potential global Coulomb branch geometries for 4d SCFTs (table. \ref{4ddeform} and \ref{4ddeform1}) and asymptotical 
free theories (table. \ref{af}), 5d $\mathcal{N}=1$ KK (table. \ref{5ddeform}) and 6d $(1,0)$ KK theories (table. \ref{6ddeform}). Here we assume 
that the bulk singularities are just $I_1$ type (the low energy theory is $U(1)$ coupled with one massless hypermultiplet plus another free $U(1)$ vector-multiplet).
The condition on the singular fiber at $\infty$ is given for a UV theory in space time dimension $D\geq 4$.
It is quite a remarkable fact that our approach can essentially cover all the known results in the literature.

The next  question is the systematical study of global genus two fiberation, 
and the details will appear in \cite{Xie:ranktwob}, where one can construct other type of undeformable singularities, such as $I_{n-0-0}$ singularities.
To study general undeformable singularities, one need to study the base change of genus two fiberation, and the details will appear in \cite{Xie:ranktwoc}.

It is possible to generalize the study of the local singularities to  rank $g\geq 3$ theory. Similarly, locally one has a one dimensional family of abelian varieties 
whose complex structure is determined by $\tau_{ij}(v)$ (the photon couplings). One might first study the case where 
there is an associate genus $g$ fibration and so one should study a genus $g$ pencils. A complete classification for genus three
degeneration is given in \cite{ashikaga2002classification}, and it seems not difficult to get the low energy theory by following the method proposed in this paper.
A even small subset would be the study of genus $g$ hypelliptic  pencils, which would give some of the well-known theories (such as $SU(N)$ gauge theory with $N_f$ fundamental flavors).
The details will appear elsewhere.

\section*{Acklowledgement}
DX would like to thank D.X Zhang for helpful discussions. DX is supported by Yau mathematical science center at Tsinghua University. 

\bibliographystyle{JHEP}
\bibliography{ADhigher}

\end{document}